\documentclass[twocolumn,nofootinbib, notitlepage,superscriptaddress,showpacs,amsmath,amssymb,pra,longbibliography]{revtex4-1}

\makeatletter
\renewcommand{\p@subsection}{}
\renewcommand{\p@subsubsection}{}
\makeatother

\usepackage[pdftex]{graphicx} 
\usepackage{dcolumn}  
\usepackage{bm}       
\usepackage[usenames,dvipsnames]{color}
\usepackage{epstopdf}
\usepackage{braket}
\usepackage{dsfont}
\usepackage{enumitem}   
\usepackage{tikz}

\usepackage[pdftex,bookmarks,breaklinks,bookmarksopen,bookmarksnumbered,linktocpage,pdfpagemode=UseOutline,pdftex]{hyperref}

\usepackage{booktabs}
\usepackage{comment}
\usepackage{natbib}

\def\ben{\begin{equation}}
\def\een{\end{equation}}
\def\benu{\begin{enumerate}}
\def\enu{\end{enumerate}}

\def\LP{_{\rm LP}}
\def\UP{_{\rm UP}}

\def\nonum{\nonumber \\}
\def\mbf{\mathbf}
\def\t{\text}

\begin{document}
\title{Enhanced optical nonlinearities under collective strong light-matter coupling}
\author{Raphael F. Ribeiro} 
\affiliation{Department of Chemistry and Biochemistry, University of California San Diego, La Jolla, CA 92093}
\author{Jorge A. Campos-Gonzales-Angulo}
\affiliation{Department of Chemistry and Biochemistry, University of California San Diego, La Jolla, CA 92093}
\author{Noel C. Giebink}
\affiliation{Department of Electrical Engineering, The Pennsylvania State University, University Park, PA,
16802}
\author{Wei Xiong}
\affiliation{Department of Chemistry and Biochemistry, University of California San Diego, La Jolla, CA 92093}
\affiliation{Materials Science and Engineering Program, University of California San Diego, La Jolla, CA 92093}
\author{Joel Yuen-Zhou}
\affiliation{Department of Chemistry and Biochemistry, University of California San Diego, La Jolla, CA 92093}
\date{\today}

\begin{abstract} Optical microcavities and metallic nanostructures have been shown to significantly modulate the  dynamics and spectroscopic response of molecular systems. We present a study of the nonlinear optics of a model consisting of $N$ anharmonic multilevel systems (e.g., Morse oscillators) undergoing collective strong coupling with a resonant infrared microcavity. We find that, under experimentally accessible conditions,  molecular systems in microcavities may have nonlinear phenomena significantly intensified due to the high quality of polariton resonances and the enhanced microcavity electromagnetic energy density relative to free space. Particularly large enhancement of multiphoton absorption happens when multipolariton states are resonant with bare molecule multiphoton transitions. In particular, our model predicts two-photon absorption cross section enhancements by several orders of magnitude relative to free space when the Rabi splitting $\Omega_R$ is approximately equal to the molecular anharmonic shift $2\Delta$.  Our results provide rough upper bounds to resonant nonlinear response enhancement factors as relaxation to dark states is treated phenomenologically. Notably, ensembles of two-level systems undergoing strong coupling with a cavity (described by the Tavis-Cummings model) show no such optical nonlinearity enhancements, highlighting the rich phenomenology afforded by multilevel anharmonic systems. Similar conclusions are expected to hold for excitonic systems that share features with our model (e.g., molecular dyes with accessible $S_0 \rightarrow S_1 \rightarrow S_2$ transitions) and strongly interact with a UV-visible cavity.
\end{abstract}	
\maketitle
\section{Introduction}\label{sec:intro}
Light-induced nonequilibrium phenomena is a topic of great contemporary interest due to its relevance to the energy, biochemical, and material sciences. Nonlinear spectroscopy provides tools for probing and controlling nonequilibrium quantum dynamics \cite{mukamel1999principles,yuen2014ultrafast} driven by external radiation. Applications of nonlinear optics to chemistry include investigations of the dynamics of energy and charge transport in light-harvesting complexes \cite{brixner2005, thyrhaug2018}, organic electronics \cite{xiang2017}, and other excitonic systems \cite{breen2017}. Nonlinear optical processes are also basic to various developing technologies including all-optical devices \cite{sanvitto2016, zasedatelev2019}, quantum information processors \cite{monroe2002, braunstein2005}, and enhanced sensors \cite{chulhai2016}. 

Unfortunately, the nonlinearities of molecular systems are generally weak \cite{boyd2008nonlinear}. Recently, hybrid materials consisting of a molecular ensemble hosted by a photonic (or plasmonic) device (e.g., optical microcavities and metallic nanostructures) have been explored as potential sources of magnified nonlinear optical response \cite{kavokin2017microcavities, ebbesen_hybrid_2016, sukharev_optics_2017, ribeiro2018d}. Under accessible experimental conditions (room temperature and atmospheric pressure) the light-matter interaction in photonic materials can become strong enough that excited states corresponding to superposition of (collective) material polarization and cavity excitations emerge \cite{sukharev_ultrafast_2014, vasa2018, s.dovzhenko2018, ribeiro2018d, herrera2020}. The corresponding hybrid quasiparticles (modes) are commonly denoted by \textit{(cavity)-polaritons}\cite{agranovich2009excitations}. They show controllable coherence and relaxation dynamics that allow modulation of various physicochemical properties. Molecular phenomena significantly influenced by strong light-matter interactions include: energy transfer\cite{coles2014b,zhong2017,du2018}, charge and exciton transport \cite{orgiu2015, feist_extraordinary_2015, hagenmuller2017a}, and chemical kinetics \cite{thomas2016, thomas2019, campos-gonzalez-angulo2019a, li2020}.
\begin{figure}[h]
\includegraphics[width=\columnwidth]{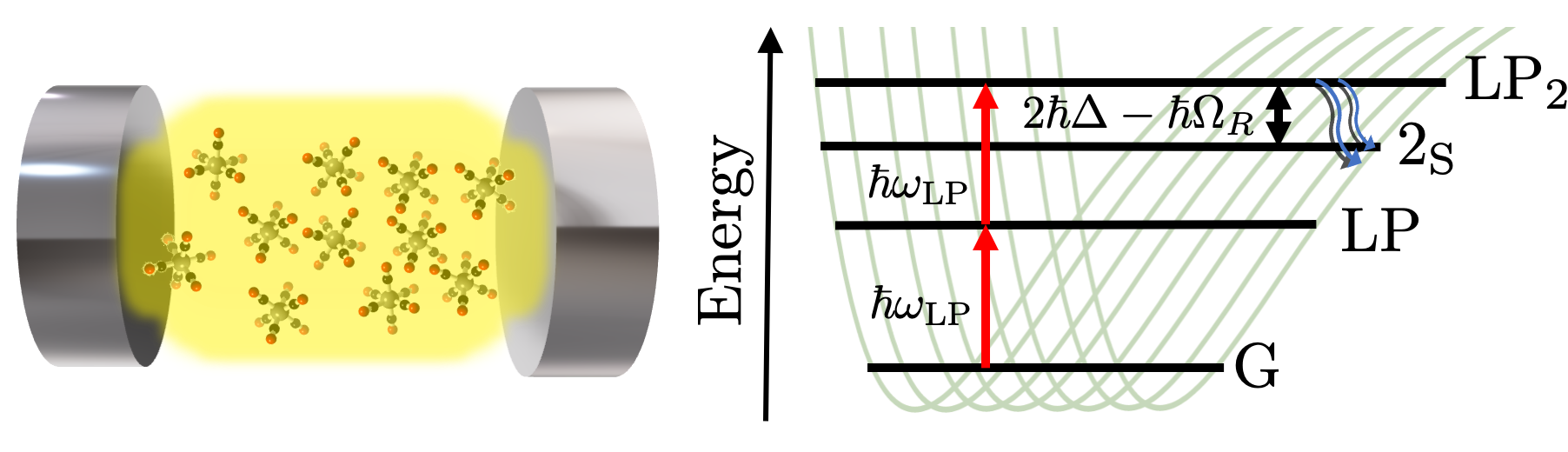}
\caption{Left: Planar microcavity consisting of two highly reflective mirrors filled with a molecular ensemble (e.g., W(CO)$_6$ in solution) with sufficiently large collective oscillator strength that hybrid polaritonic states are formed. Right: Mechanism for enhancement of two-photon absorption by an ensemble of Morse oscillators (represented by the various illustrative Morse potentials) under strong coupling with an optical cavity. An external field resonant with the lower-polariton (LP) drives the hybrid cavity and excites two-LP states which can be tuned to be near-resonant with the anharmonically shifted doubly-excited molecular states forming the totally-symmetric $2_S$ state. This polariton-mediated absorption channel allows enhancement of several orders of magnitude of the molecular two-photon absorption cross-section.} \label{fig:cavpes}
\end{figure}

Recent experiments \cite{fofang2011, virgili2011, vasa2013, chervy2016, dunkelberger_modified_2016, xiang2018, dunkelberger2018a, avramenko2020, delpo2020, finkelstein-shapiro2020} have surveyed the nonlinear optics of polaritonic systems to gain further insight into the relaxation kinetics and optical response of strongly coupled devices. In Refs. \cite{dunkelberger_modified_2016, dunkelberger2018a, xiang2018, xiang2019a}, the transient response and relaxation to equilibrium of vibrational polaritons (those arising from the strong coupling of molecular infrared polarization with a resonant microcavity) were investigated with pump-probe and two-dimensional infrared spectroscopy. These studies demonstrated how vibrational anharmonicity is manifested in the pump-probe polariton response \cite{ribeiro2018c}. However, the observed time-resolved spectra were sensitive to various system-dependent effects arising from the small Rabi splittings of the studied materials, and significant static and dynamical disorder which induces ultrafast polariton decay into the weakly-coupled (dark) molecular modes.

In this work, we focus on universal (system-independent) features of molecular polariton nonlinear optics. Our aim is to provide qualitative and quantitative insight on the potential to achieve giant optical nonlinearities with molecular polaritons in the \textit{collective} regime (which is the case in most experiments) with a large number of molecules in a microcavity (nonlinear optical effects of \textit{single-molecule} polaritonic systems have been studied within a non-adiabatic model of the dynamical Casimir effect in Ref. \cite{perez-sanchez2020}, as well as in  vibrational polariton spectra in Ref. \cite{hernandez2019}). 

In Sec. \ref{sec:mnre}, we describe our model, provide an analytical expression for the nonlinear optical susceptibility of an ideal molecular ensemble under strong interaction with a microcavity (the full derivation is in the SI Sec. 2), and discuss its main features. In Sec. \ref{sec:pmtpa}, we compare the free space and the polariton-mediated two-photon absorption (TPA) rates, and show that, especially when overtone polariton transitions are resonant with multiphoton molecular transitions, nonlinearity enhancements of several orders of magnitude may be achieved with currently available optical cavities (Fig. \ref{fig:cavpes}) as a result of three main effects: increased electromagnetic energy density in the optical microcavity relative to free space \cite{steck2017, kavokin2017microcavities}, creation of new optical resonances, and strong coupling induced suppression of lineshape broadening \cite{houdre_vacuum-field_1996}.
 A discussion of our main results and conclusions are given in Sec. \ref{sec:dac}. Our article is accompanied by Supporting Information (SI) containing detailed derivations of the molecular nonlinear susceptibility, and rate of nonlinear absorption in free space and under strong coupling with an optical microcavity.

\section{Molecular nonlinear response}\label{sec:mnre}
\subsection{Effective Hamiltonian}
The physical system of interest consists of a molecular ensemble containing $N$ molecules uniformly distributed in a region enclosed by two highly-reflective planar mirrors separated by a distance $L_c$ of the order of the wavelength of a specific material infrared excitation ($L_c$ is usually between 0.1 and 20 $\mu$m) \cite{george_liquid-phase_2015,long_coherent_2015,casey_vibrational_2016}. This setup corresponds to a Fabry-Perot (FP) microcavity \cite{kavokin2017microcavities, steck2017} filled with a homogeneous molecular system. Our description of the molecular subsystem will include explicitly only the modes which are nearly-resonant with the optical cavity. The effects of all other molecular degrees of freedom will be treated phenomenologically by introduction of damping to the molecular polarization (see below). 

We suppose that the interaction between the cavity field and the molecular polarization $\sum_{i=1}^N \braket{1_i|p_i|0_i}$ (where $p_i$ is the effective dipole operator of the $i$th molecule and $0_i$ and $1_i$ denotes states where the $i$th molecule is in the ground and first excited-state, respectively, whereas all other molecules are in the ground-state) is significantly stronger than the coupling of either subsystem to external (bath) degrees of freedom, but still only a tenth or less of the bare vibrational and cavity frequencies (so considerations exclusive to ultrastrong coupling can be ignored \cite{ciuti_quantum_2005, todorov_ultrastrong_2010, strashko2016}). 

The total Hamiltonian of the composite material is given by
$H_T(t) = H_{\t{L}}(t) + H_{\t{M}} + H_{\t{LM}},$ 	
where $H_{\t{L}}(t)$ and $H_{\t{M}}$ are the bare cavity (driven by an external time-dependent field) and molecular Hamiltonians and $H_{\t{LM}}$ contains the interaction between the cavity EM field and matter. The cavity Hamiltonian is given by:
\begin{align}
H_{\t{L}}(t) = & \sum_{\mbf{k}} \hbar \omega_{\mbf{k}} b^\dagger_{\mbf{k}} b_{\mbf{k}} \nonum
& + i\hbar \sqrt{\frac{\kappa}{2}}\sum_{\mbf{k}}\left\{\left[b_{\mbf{k}\t{in}}^{\t{L}}(t)\right]^\dagger b_{\mbf{k}} - b^\dagger_{\mbf{k}}b_{\mbf{k}\t{in}}^{\t{L}}(t)\right\}, \label{eq:hem}
\end{align}
where this effective Hamiltonian can be obtained from input-output theory \cite{gardiner_input_1985, steck2007quantum, li2018} which describes the interaction of the optical cavity with left input and right output flux operators $b_{\mbf{k}\t{in}}^{\t{L}}(t)$ and $b_{\mbf{k}\t{out}}^{\t{R}}(t)$, respectively (SI Sec. I), and we include only a single cavity band and EM field polarization (as the cavity band gaps are much larger than the cavity and molecular linewidths, due to the smallness of the cavity's longitudinal length $L_c$, and electric field polarization conversion gives a tiny perturbation on the results presented here especially as we consider isotropic molecular ensembles \cite{litinskaya_loss_2006}). The frequency of the mode with (in-plane) wave-vector $\mbf{k} = (k_x, k_y)$ is  $\omega_{\mbf{k}} = c\sqrt{\mbf{k}^2+m^2\pi^2/L_c^2}/n$ ($m \in \mathbb{Z}$ is the index of the cavity band; $n$ is the index of refraction of the cavity interior; hereafter $n=1$), and $b_{\mbf{k}}$ is its annihilation operator. The cavity leakage (decay) rate is $\kappa$. The Heisenberg equations of motion generated by Eq. \ref{eq:hem} are turned into the Heisenberg-Langevin equations when the replacement $\omega_{\mbf{k}} \rightarrow \tilde{\omega}_\mbf{k} \equiv \omega_{\mbf{k}} - i\kappa/2$ is performed (SI Sec. 1). 

The bare vibrational dynamics is generated by the Hamiltonian $H_M$ given by
\begin{align}
H_{\t{M}} = \sum_{i=1}^{N} \hbar\omega_0a_i^\dagger a_i -\hbar \Delta \sum_{i=1}^N a_i^\dagger a_i^\dagger a_i a_i,   \label{eq:hbmol}
\end{align}
where the vibrational creation and annihilation operators of the $i$th molecule are $a_i^\dagger$ and $a_i$, respectively. The fundamental frequency of each molecule is $\omega_0$, and the anharmonic coupling is $\Delta > 0$. We neglect intermolecular interactions as they are too weak relative to light-matter coupling (the situation could be different in other situations, e.g., molecular crystals and liquid-solid interfaces \cite{agranovich2009excitations, kraack2017}). We treat the relaxation of the molecular subsystem phenomenologically by converting the Heisenberg equations of motion (EOMs) of molecular operators into Heisenberg-Langevin EOMs via the substitution $\omega_0 \rightarrow \tilde{\omega}_0 = \omega_0 - i\gamma_m/2$, where $\gamma_m$ is the bare molecule fundamental transition (homogeneous) linewidth. 

The light-matter interaction is treated with the multipolar gauge \cite{craig1998molecular} in the long-wavelength limit within the rotating wave approximation\cite{steck2007quantum} (see next paragraph for a discussion of these and other approximations): \begin{align}
H_{\t{LM}} = & -\sum_{\mbf{k}} \sum_{i=1}^N \left(g_{i\mbf{k}}a_i^\dagger b_{\mbf{k}}+\bar{g}_{i\mbf{k}}a_i b_{\mbf{k}}^\dagger\right)+ H_{\t{P}^2}, \label{eq:hlm}
\end{align}
 where $g_{j\mbf{k}} = \mu_j \cdot E_{j\mbf{k}}^c$ is the coupling constant for the interaction between the $j$th molecular vibration (with effective transition dipole moment  $\mu_j$) and the cavity mode $\mbf{k}$, with mode profile evaluated at the position $\mbf{r}_j$ of the $j$th molecule, i.e.,  $E_{j\mbf{k}}^c = i\sqrt{\hbar \omega_{\mbf{k}}/(2\epsilon_0 V_c)}e^{i\mbf{k}\cdot \mbf{r}_j}\t{sin}(m\pi z_j/L_z)$  
   ($\epsilon_0$ is the electrical permittivity of free space, $V_c$  is the cavity quantization volume and $z_j$ is the position of the molecule along the cavity longitudinal axis), $\bar{f}$ denotes the complex conjugate of $f$, and $H_{\t{P}^2}$ is the molecular self-polarization energy \cite{craig1998molecular}. Although this term ensures the existence of a ground-state for the composite system \cite{rokaj2018} and it becomes essential for an appropriate treatment of a system with total light-matter interaction energy approaching or surpassing the bare cavity and molecular frequencies \cite{deliberato2014}, $H_{\t{P}^2}$ can be neglected under the strong coupling conditions assumed here. Therefore, we will disregard this term onward.
            
The length scale over which the cavity mode profile varies substantially (of order 0.1-20 $\mu$m) is much larger than typical molecular diameters (of order $0.5-5\t{nm}$). Thus, under strong coupling, the $\mbf{k}\approx 0$ cavity modes interact coherently with material polarization consisting of a macroscopic number of molecules. This notion forms the basis for neglecting spatial, orientational and energetic dispersion of the molecular excitations, since fluctuations of these quantities are necessarily weak effects compared to the collective light-matter interactions from which polaritons emerge. Fluctuations about the mean values of the molecular transition frequency and dipole moment can lead to dephasing-induced polariton decay \cite{pino2015, zhang2019}, weak-coupling of light to states which are dark according to Eq. \ref{eq:hlm}, as well as polariton \cite{agranovich2003, agranovich_nature_2007} and dark-state localization \cite{agranovich2003, botzung2020}. Since we are not concerned with transport phenomena, we will not include them in our model, although Sec. \ref{sec:dac}  qualitatively analyzes their implications to our main results. Despite neglecting these effects, we highlight that our input-output treatment of the material and photonic components naturally accounts for polariton dissipation via cavity leakage and molecular homogeneous dephasing \cite{gardiner_input_1985, gardiner1991quantum, steck2007quantum, ribeiro2018c}. In Eq. \ref{eq:hlm}, we also assumed validity of the so-called rotating-wave-approximation: only light-matter interactions preserving the total number of cavity and molecular excitations are retained. This approximation is justified by $ \sqrt{\sum_{i=1}^N|g_{i\mbf{k}}|^2} \ll \omega_0,~~\forall~~\mbf{k}$.
 
We aim to investigate the nonlinear response of the hybrid system to an input radiation field with $\mbf{k} \in \mathbb{R}^2$ centered at $\mbf{k}_0 \approx 0$, with a small width $\delta \mbf{k}$. The frequency $\omega_{\mbf{k}_0}$ is nearly resonant with the bare molecule fundamental frequency $\omega_0$. Therefore, we shall retain only a single cavity-mode corresponding to $\mbf{k}_0 \approx 0$. This assumes there is no variation in the polariton nonlinear response with respect to changes of magnitude $|\delta \mbf{k}|$ in the incident wave-vector $\mbf{k}_0$.

Based on the above, we simplify $H_L(t)$ and $H_{\t{LM}}$ and employ the following effective Hamiltonian for the hybrid cavity-matter system
 \begin{align}
 	H_T(t) = & \hbar \omega_{c}b^\dagger b + \sum_{i=1}^{N} \hbar\omega_0 a_i^\dagger a_i -\hbar \Delta \sum_{i=1}^N a_i^\dagger a_i^\dagger a_i a_i \nonum
 	&- \sum_{i=1}^N \mu \left(E_{0}^c a_i^\dagger b+\bar{E_0^c} b^\dagger a_i\right)\nonum
 	& -i\hbar\sqrt{\frac{\kappa}{2}}\left\{\left[b_{\t{in}}^{\t{L}}(t)\right]^\dagger b - b^\dagger b_{\t{in}}^{\t{L}}(t)\right\} \label{eq:heffinal},
 \end{align}
 where $\omega_c \equiv \omega_{\mbf{k}_0}$, $b = b_{\mbf{k}_0}$, $b_{\mbf{k}_0\t{in}}^{\t{L}} = b_{\t{in}}^{\t{L}}$, and $\mu E_{0}^c \equiv g_{j\mbf{k}_0}  = i\mu\sqrt{\hbar \omega_{c}/(2\epsilon_0V_c)}$. 

\subsection{Nonlinear molecular polarization under strong light-matter coupling}\label{sec:nlmolpol}
The optical response of a hybrid microcavity can be investigated by measuring the transmission, reflection or absorption spectrum of light input into the system. For instance, transmission and reflection spectra can be obtained by applying the input-output relations to the steady-state cavity field $b(t) = \sum_{\omega > 0} b(\omega) e^{-i\omega t}$. Because the cavity is weakly-coupled to the external fields, the expectation value $\braket{b(t)}$ admits a perturbative expansion in powers of the input amplitude $\braket{b(t)} = \sum_{p =1}^{\infty} \braket{b(t)}^{(2p-1)}$, where $\braket{b(t)}^{(2p-1)} = O\left[|b_{\t{in}}^{\t{L}}|^{2p-1}\right]$ (only odd powers of the input field appear in the cavity response because the material is assumed homogeneous and symmetric with respect to spatial inversion \cite{mukamel1999principles, boyd2008nonlinear}). The material polarization $P(t) = \mu \sum_{i=1}^N a_i(t)$ is strongly coupled to the optical cavity. Therefore, molecular observables also admit a perturbative expansion in powers of $\left|b_{\t{in}}^{\t{L}}\right|$. Note that the empty cavity is a linear system, and thus, the source of the nonlinear part of $\braket{b(t)}$ is the molecular subsystem (specifically, the source of $\braket{b(t)}^{(3)}$ is $\braket{P(t)}^{(3)} = \sum_{i=1}^N \mu \braket{a_i(t)}^{(3)}$; see SI Sec. 2). Therefore, $\braket{P(t)}^{(3)}$ directly determines the amplitude of the nonlinear optical response of a strongly coupled system as measured by the output transmitted and reflected light.

Neglecting quantum fluctuations of the input field, $b_{\t{in}}(t)$ is a complex number that we express as: \begin{align} b_{\t{in}}(t) = i\sum_{\omega >0}\sqrt{\frac{\mathcal{P}(\omega)}{\hbar \omega}} e^{i\theta_{\t{in}}(\omega)} e^{-i \omega t},\end{align}
where $\mathcal{P}(\omega)$ is the power of the free space mode with frequency $\omega$ driving the  cavity, and $\theta_{\t{in}}(\omega)$ is its phase (SI Sec. 1).

It follows (SI Sec. 2) that the third-order polarization in the frequency domain $\braket{P}^{(3)}(\omega_s)$ can be written in terms of the input electric fields as follows
\begin{align}
	\braket{P}^{(3)}(\omega_s) = & \sum_{\omega_u,\omega_v,\omega_w} \chi^{(3)}(-\omega_s; \omega_v,-\omega_w,\omega_u) E_{\t{in}}^{(+)}(\omega_v)	\times \nonum & E_{\t{in}}^{(-)}(\omega_w) E_{\t{in}}^{(+)}(\omega_u)+ \t{h.c.}, \label{eq:tmpol}
\end{align}
where $\omega_s > 0$ is the signal frequency, the brackets denote expectation values, the driving frequencies $\omega_u,\omega_v,\omega_w$  are all positive, the input fields $E_{\t{in}}^{(+)}(\omega_u)$ are directly proportional to the $b_{\t{in}}(\omega_u)$ (see SI Sec. 1), and $\chi^{(3)}$ is the nonlinear molecular susceptibility \cite{boyd2008nonlinear} under strong light-matter interaction conditions. The ratio between $\chi^{(3)}$  and the bare molecular system nonlinear susceptibility $\chi_0^{(3)}$ provides an external-field independent measure of strong light-matter coupling effects on the optical nonlinearities of an arbitrary molecular system.

  To obtain $\chi^{(3)}$ \cite{mukamel1999principles} for the system described by Eq. \ref{eq:heffinal}, we solve perturbatively the Heisenberg-Langevin equations of motion (EOM) for the molecular polarization to third-order in the driving field $b_{\t{in}}^{\t{L}}$ \cite{leegwater1992, wang1994a, chernyak1995}. The EOMs for the cavity and material polarization expectation values admit simple solutions since the initial condition (ground-state) and the time-evolution of our system ensures that it remains in a \textit{pure state} at all times. The result is (SI Sec. 2): 
\begin{widetext}
\begin{align}
\chi^{(3)}(-\omega_s;\omega_v,-\omega_w,\omega_u) =  & N \mu (2\hbar\Delta) G_{mm}(\omega_s)\bar{G}_{mm}(\omega_w)\Gamma_{mm,mm}(\omega_u+\omega_v)G_{mm}(\omega_u)G_{mm}(\omega_v)\times
\nonum
& \left[\mu \sqrt{\frac{2\mathcal{F}}{\pi}} G_{pp}^{(0)}(\omega_v)\frac{\hbar\kappa}{2}\right]\left[\mu \sqrt{\frac{2\mathcal{F}}{\pi}} \frac{\hbar \kappa}{2} \bar{G}_{pp}^{(0)}(\omega_w)\right] \left[\mu \frac{\hbar \kappa}{2}\sqrt{\frac{2\mathcal{F}}{\pi}} G_{pp}^{(0)}(\omega_u)\right]\delta_{\omega_s,\omega_v-\omega_w+\omega_u}, \label{eq:chi3_ne}
\end{align}
\end{widetext}
where $\mathcal{F}$ is the cavity finesse (the electromagnetic field intensity in a resonant cavity is stronger than in free space by the factor $2\mathcal{F}/\pi$, or alternatively, $\mathcal{F} = Q/m$, where $m$ is the aforementioned band index, and $Q=\omega_c/\kappa$ is the quality factor; SI Sec. I and Ref. \cite{steck2017}), and $G_{mm}(\omega)$ is the Fourier transform (FT) of the retarded single-molecule  propagator
\begin{align}
 G_{mm}(\omega) = \frac{1}{\hbar\omega -\hbar\omega_0+i\hbar\gamma_m/2- \frac{\left|\mu E_0\right|^2 N}{\hbar\omega-\hbar\omega_c+i\kappa/2}}.
\end{align}
Note the real part of the poles of $G_{mm}(\omega)$ are the fundamental polariton resonance frequencies
\begin{align}
& \omega_{\t{LP}} = \frac{\omega_c+\omega_0}{2} -\frac{\sqrt{(\omega_c-\omega_0)^2+\Omega_R^2}}{2}, \\
&\omega_{\t{UP}} = \frac{\omega_c+\omega_0}{2} + \frac{\sqrt{(\omega_c-\omega_0)^2+\Omega_R^2}}{2}.
\end{align}
where $\Omega_R = 2 |\mu E_0^c|\sqrt{N}/\hbar$ is the Rabi frequency (splitting). The imaginary parts of the polariton poles in $G_{mm}(\omega)$ correspond to their (linear) absorption linewidths. Under weak-coupling conditions, we can neglect the cavity-induced self-energy $\frac{\left|\mu E_0\right|^2 N}{\hbar\omega-\hbar\omega_c+i\kappa/2}$ to obtain the bare molecule propagator  $G_{mm}^{(0)}(\omega) = 1/(\hbar\omega -\hbar\omega_0+i\hbar\gamma_m/2)$. Similarly, the photon-photon correlator $G_{pp}(\omega)$ has resonances at the polariton frequencies, as is clear from its explicit form:
\begin{align}
G_{pp}(\omega) = \frac{1}{\hbar \omega-\hbar\omega_c+i\hbar \kappa/2 - \frac{\left|\mu E_0^c\right|^2 N}{\hbar\omega-\hbar \omega_0+i\hbar\gamma_m/2}}.
\end{align} 
In the weak-coupling limit, $G_{pp}(\omega)$ approaches the empty cavity frequency-domain propagator $G_{pp}^{(0)}(\omega) = 1/(\hbar \omega-\hbar\omega_c+i\hbar \kappa/2)$.

\begin{figure}[t]
\includegraphics[width=\columnwidth]{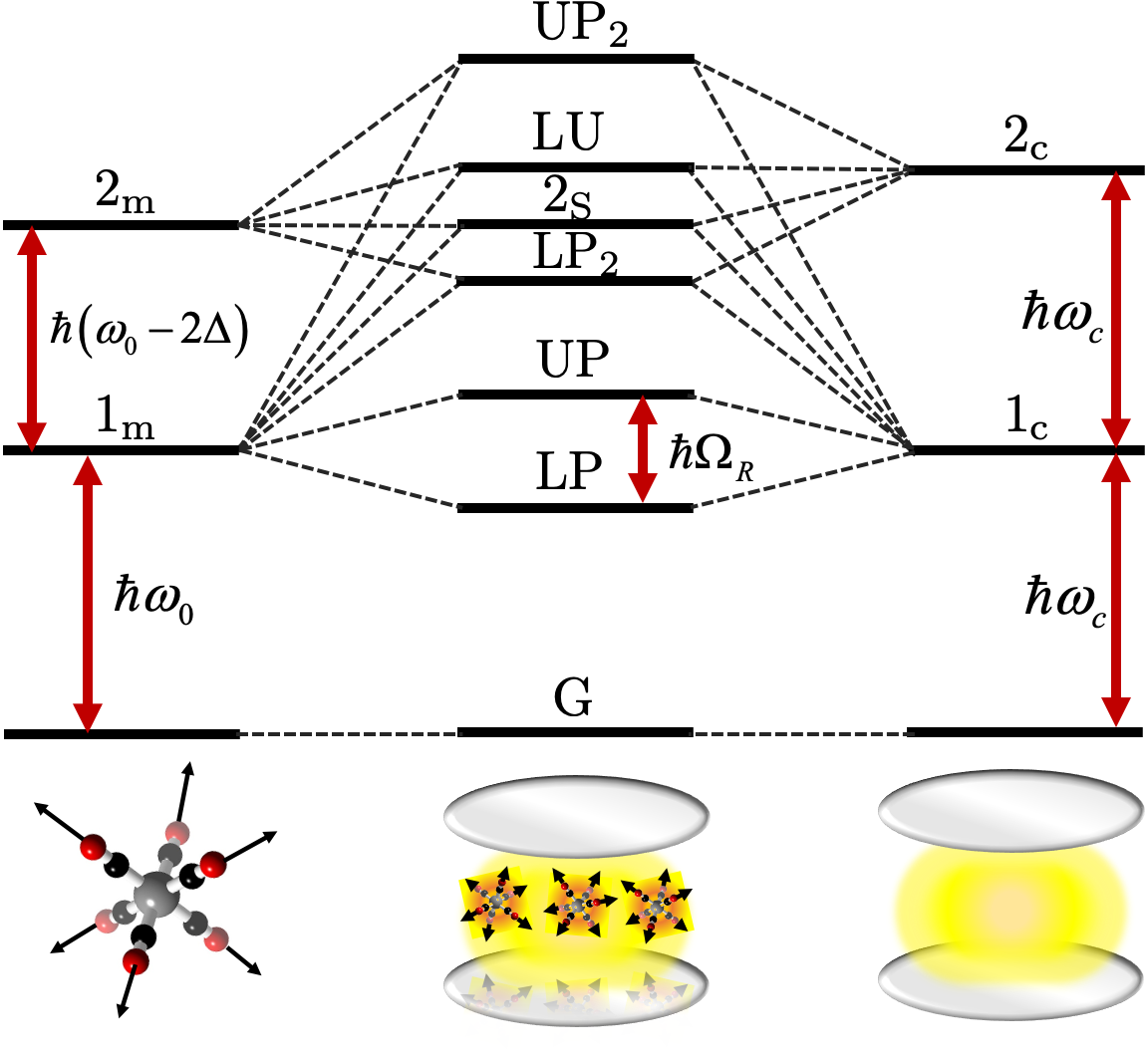}
\caption{Energy level diagram for a model system with zero-detuning $(\omega_c = \omega_0)$ and Rabi splitting $\Omega_R >  ~\t{anharmonicity}~~2\Delta$, including only bright excitations of the single and two-polariton manifolds (SI Sec. 7).} \label{fig:enlev}
\end{figure}

 The function $\Gamma_{mm,mm}(\omega_u+\omega_v)$ is the two-particle elastic scattering matrix element given by:
\begin{widetext}
\begin{align}
\Gamma_{mm,mm}(\omega) = \frac{(\hbar\omega-2\hbar\tilde{\omega}_0)(\hbar\omega-\hbar\tilde{\omega}_0-\hbar\tilde{\omega}_c)\left[(\hbar\omega-2\hbar\tilde{\omega}_0)(\hbar\omega-2\hbar\tilde{\omega}_c)-4g^2 N \right]}{D(\omega)}, \label{eq:gammm}
\end{align}
\end{widetext}
where $D(\omega)$ is a 4th-order polynomial of $\omega$ given by:
\begin{align}
D(\omega) = & D^{(0)}(\omega)   -2g^2N(\hbar\omega-2\hbar\tilde{\omega}_0)(\hbar\omega-2\hbar\tilde{\omega}_0+2\hbar\Delta) \nonum
& -2g^2(N-1)(\hbar\omega-2\hbar\tilde{\omega}_0+2\hbar\Delta)(\hbar\omega-2\hbar\tilde{\omega}_c)  \nonum & -2g^2(\hbar\omega-2\hbar\tilde{\omega}_c)(\hbar\omega-2\hbar\tilde{\omega}_0),
\end{align}
where $g = |\mu E_0^c|$ is the single-molecule light-matter coupling and  $D^{(0)}(\omega) = (\hbar\omega-\hbar\tilde{\omega}_c-\hbar\tilde{\omega}_0)(\hbar\omega-2\hbar\tilde{\omega}_0+2\hbar\Delta)\times$ $(\hbar\omega-2\hbar\tilde{\omega}_c)(\hbar\omega-2\hbar\tilde{\omega}_0)$.
The roots of $D(\omega)$ correspond to the \textit{bright} two-particle resonances of the hybrid system, as can be verified by comparison to the eigenvalues of the doubly-excited block of the Hamiltonian in Eq. \ref{eq:heffinal} with $b_{\t{in}}^{\t{L}} = 0$ (see SI Sec. 7 for a discussion and analytical results for the $\omega_c = \omega_0$ case). In the large $N$ limit appropriate to almost all experimental studies of infrared strong coupling, the real parts of the two-particle resonances in the zero-detuning case are given by $\omega_{2\t{UP}} = 2\omega_{\t{UP}} +O(g/\sqrt{N})$, $\omega_{\t{LU}} = \omega_c + \omega_0$, $\omega_{2_m} = 2\omega_0 -2\Delta + O(\Delta/N)$, and  $\omega_{2\t{LP}} = 2\omega_{\t{LP}} + O(g/\sqrt{N})$, where the subscripts label the dominant character of each state, e.g., the highest-frequency resonance is dominated by the component with a doubly-excited UP mode while the resonance with frequency $\omega_{\t{LU}}$ corresponds to that containing an LP,UP pair (see Fig. \ref{fig:enlev}). 

 Physically, $\Gamma_{mm}(\omega_u+\omega_v) \propto G_{mm,mm}(\omega_u,\omega_v)$ (SI Sec. 2), where $G_{mm,mm}(\omega)$ is the frequency-domain single-molecule two-excitation propagator, i.e.,  the FT of the probability amplitude that a molecule initially in its doubly excited-state remains in the same state after time $t$. 

Importantly, when $N \rightarrow \infty$, Eq. \ref{eq:gammm} becomes:
\begin{align}
\Gamma_{mm,mm}(\omega) \approx \Gamma_{mm,mm}^{(0)}(\omega) \equiv \frac{\omega-2\tilde{\omega}_0}{\omega-2\tilde{\omega}_0+2\Delta},~N \rightarrow \infty, \label{eq:gammamm}
\end{align}
where $\Gamma^{(0)}(\omega)$ is the bare single-molecule two-particle (elastic) scattering matrix (see next subsection and SI Sec. 4). This result is expected, since $\Gamma_{mm,mm}(t)$ describes the time-dependent propagation of single-molecule doubly excited-states under interaction with the optical cavity, and as we show in SI Sec. 7, the totally-symmetric doubly-excited molecular state $\ket{2_m} = \frac{1}{\sqrt{N}}\sum_{i=1}^N \ket{2_i}$ (where $\ket{2_i}$ is the state where the $i$th molecule is in the 2nd excited-state while the cavity and all other molecules are in the ground-state) is only \textit{weakly-coupled} to two-polariton states via an interaction that is proportional to the single-molecule light-matter coupling $g$. Therefore, while polaritons play an essential role as intermediate states for TPA by the molecular subsystem, Eq. \ref{eq:gammamm} indicates the dynamics of molecular doubly excited-states is almost insensitive to their coupling to the cavity electromagnetic field in the ensemble strong coupling limit.

To gain further insight into the molecular nonlinear polarization in the strong light-matter coupling regime, we now compare Eq. \ref {eq:chi3_ne} to the nonlinear susceptibility of the bare molecules in free space (under the rotating-wave approximation) given by
\begin{widetext}
\begin{align}
\chi_0^{(3)}(-\omega_s; \omega_v, -\omega_w, \omega_u)=  N  \mu (2\hbar\Delta) \mu^3 G_{mm}^{(0)}(\omega_s) \bar{G}_{mm}^{(0)}(\omega_w) \Gamma_{mm,mm}^{(0)}(\omega_u+\omega_v) G_{mm}^{(0)}(\omega_v)G_{mm}^{(0)}(\omega_u)\delta_{\omega_s,\omega_v-\omega_w+\omega_u}.\label{eq:chi3_0}\end{align}\end{widetext}
\par By contrasting Eqs. \ref{eq:chi3_0} and \ref{eq:chi3_ne}, we find that the nonlinear optical response of a molecular system (e.g., solution \cite{muallem2016, xiang2018}, polymer \cite{menghrajani2019, shalabney_coherent_2015}, etc) in an optical microcavity is significantly distinct from that in free space mainly because of: (i) near-resonant intracavity field intensity enhancement [which renormalizes the induced molecular transition dipole moments $\mu \rightarrow \tilde{\mu}(\omega) =  \mu \sqrt{2\mathcal{F}/\pi} i\hbar\sqrt{\kappa/2} G_{pp}^{(0)}(\omega)]$, and (ii) the appearance of new (polariton) resonances corresponding to hybrid superpositions of molecular polarization and cavity modes. In other words, the molecular nonlinear response under strong coupling can be written entirely in terms of \textit{cavity-renormalized} single-particle [$G_{mm}^{(0)}(\omega) \rightarrow G_{mm}(\omega)$] and two-particle molecular response functions [$\Gamma_{mm,mm}^{(0)}(\omega) \rightarrow \Gamma_{mm,mm}(\omega)$] which are non-perturbatively dressed by the interaction with the cavity field, as well as (ii) molecular transition dipoles $\mu$ which are renormalized by factors that depend on the cavity finesse $\mathcal{F}$ and the bare photon propagator [$\mu \rightarrow \tilde{\mu}(\omega)]$. The renormalization of the induced molecular dipoles is a result of the well-known enhancement of the intracavity electric field relative to free space \cite{born2013principles, steck2017}. In the weak coupling regime, a Purcell-like result follows where the molecular nonlinear susceptibility in a microcavity (Eq. \ref{eq:chi3_ne}) is simply related to that of the bare system (Eq. \ref{eq:chi3_0}), $\chi^{(3)} \rightarrow \chi_0^{(3)} \times$ intracavity field enhancement factors. 

As discussed in Sec. \ref{sec:pmtpa}, by virtue of the cavity-matter strong coupling, the  nonlinear polarization contribution to the energy absorbed by the molecular subsystem is \textit{not} directly proportional to the imaginary part of  $\chi^{(3)}(-\omega;\omega,-\omega,\omega)$ (in contrast to the nonlinear absorption of bare molecules; see SI Sec. 5). Therefore, we provide additional comments and a numerical comparison of the real and imaginary parts of Eqs. \ref{eq:chi3_ne} and \ref{eq:chi3_0} as a function of cavity detuning and Rabi splitting in SI Sec. 6, and focus below on the nonlinear absorption spectrum of the strongly coupled material.

\section{Polariton-enhanced two-photon absorption} \label{sec:pmtpa}
The steady-state rate of excitation of a molecular system driven by the electromagnetic field can be written as (SI Secs. 3 and 5)
\begin{align}
W_T(\omega) \equiv \frac{2}{\hbar}\left[\t{Im}\braket{\left[E(\omega)\right]^\dagger P(\omega)}\right],
\end{align}
where $E(\omega)$ is the frequency-domain representation of the free space or cavity Heisenberg electric field operator. For a molecular system in free space interacting weakly with a classical monochromatic EM field with (positive-frequency) amplitude $E_{\t{in}}^{(+)}(\omega)$, it follows that the photon absorption rate (in the rotating-wave approximation) is \cite{mukamel1999principles}
\begin{align}
W_0(\omega) \equiv & \frac{2}{\hbar} \t{Im}\left[\chi_0^{(1)}(-\omega;\omega)\right]|E_{\t{in}}^{(+)}(\omega)|^2 \nonum
& + \frac{2}{\hbar}\t{Im}\left[\chi_0^{(3)}(-\omega; \omega,-\omega,\omega)\right]|E_{\t{in}}^{(+)}(\omega)|^4  + ... 
\end{align}
This expression is clearly invalid when the molecular ensemble interacts strongly with a cavity, since in this instance, the cavity field and the material electrical polarization are correlated, and therefore $\braket{E(\omega)P(\omega)}$ cannot (in general) be factorized into $\braket{E}(\omega)\braket{P}(\omega)$ (where $E$ refers to the cavity EM field). Nevertheless, the external input field interacts weakly with the cavity, and the rate of absorption by the strongly coupled molecular system admits the following perturbative expansion in powers of the input field amplitude 
\begin{align}
W(\omega) = \frac{2}{\hbar}\t{Im}\left[\braket{\left[E_c(\omega)\right]^\dagger P(\omega)}^{(2)}+\braket{\left[E_c(\omega)\right]^\dagger P(\omega)}^{(4)}\right]+...
\end{align} 
where $E_c(\omega) = E_0^c b(\omega)$. The contribution to the absorption spectrum dependent on the nonlinear response of the molecular subsystem is given by $W^{\t{NL}}(\omega) = \frac{2}{\hbar}\t{Im}\left[\braket{E_c^\dagger(\omega)P(\omega)}^{(4)}\right]$. In SI Sec. 3, we obtain $W^{\t{NL}}(\omega)$ by employing the Heisenberg-Langevin EOMs following the same approach taken to derive Eq. \ref{eq:chi3_ne}. 

For simplicity, we restrict our analysis of the nonlinear absorption spectrum to the zero-detuning case where $\omega_c=\omega_0$. We also simplify $W^{\t{NL}}(\omega)$ by using the following conditions necessarily valid at strong coupling: $\Omega_R \gg \hbar\eta_s \equiv \hbar (\kappa +\gamma_m)$, and $\Omega_R \gg \hbar \eta \equiv \hbar \kappa \gamma_m/(\kappa+\gamma_m)$. Under these conditions, the nonlinear component of molecular absorption under strong coupling with a cavity can be expressed as $W^{\t{NL}}(\omega) = \sum_{\alpha=1}^4 W^{\t{NL}_\alpha}(\omega) \left|E_{\t{in}}^{(+)}(\omega)\right|^4$, where 
\begin{align}
 W^{\t{NL}_1} (\omega) \approx &-\frac{2\eta\kappa}{\hbar}\left[\frac{1}{4(\omega-\omega_0)^2+\kappa^2}+\frac{1}{(\Omega_R/\hbar)^2} \right] 
\nonum
& \times \t{Re}\left[\sqrt{\frac{2\mathcal{F}}{\pi}}  \chi^{(3)}(\omega)\right], \label{eq:w1} 
\end{align}
\begin{align}
& W^{\t{NL}_2}(\omega) \approx \frac{4\eta}{\hbar}
\frac{\omega-\omega_0}{4(\omega-\omega_0)^2+\kappa^2}\t{Im}\left[\sqrt{\frac{2\mathcal{F}}{\pi}}  \chi^{(3)}(\omega)\right], \label{eq:w2}\\
& W^{\t{NL}_3} (\omega) \approx \eta \frac{4\Delta^2 N}{(\omega-\omega_0)^2+\gamma_m^2/4}\frac{\left|\braket{a_i a_i}^{(2)}(2\omega)\right|^2}{|E_{\t{in}}^{(+)}(\omega)|^4} \label{eq:w3}, \\
& W^{\t{NL}_4} (\omega) \approx -\eta \frac{(2\omega -\omega_{20})2\Delta N }{(\Omega_R/2\hbar)^2} \frac{\left|\braket{a_i a_i}^{(2)}(2\omega)\right|^2}{|E_{\t{in}}^{(+)}(\omega)|^4}\label{eq:w4},
\end{align}
where $\chi^{(3)}(\omega) \equiv 6\chi^{(3)}(-\omega; \omega,-\omega,\omega)$, $\omega_{20} \equiv 2\omega_0 -2\Delta$, and
\begin{align}
\braket{a_i a_i}^{(2)}(2\omega) & = -\Gamma_{mm,mm}(2\omega)G_{mm}(\omega) G_{mm}(\omega)\nonum
& \times \left[\mu\sqrt{\frac{2\mathcal{F}}{\pi}}G_{pp}^{(0)}(\omega)\frac{\hbar\kappa}{2} E_{\t{in}}^{(+)}(\omega)\right]^2. 
\end{align}
The expression for the nonlinear absorption by the molecular system under strong coupling with a cavity is more complicated relative to the bare system given by $W^{\t{NL}}_0(\omega)|E_{\t{in}}^{(+)}(\omega)|^4 = \frac{2}{\hbar}\t{Im}\left[\chi_0^{(3)}(-\omega;\omega,-\omega,\omega)\right] |E_{\t{in}}^{(+)}(\omega)|^4$. For example, $W^{\t{NL}}(\omega)$ shows dependence on both the real and imaginary parts of $\chi^{(3)}$ (Eqs. \ref{eq:w1} and \ref{eq:w2}) in addition to the steady-state population of molecular doubly excited-states $P_{2m}^T(\omega) = \sum_{i=1}^N |\braket{a_i a_i}^{(2)}(2\omega)|^2/2$ (Eqs. \ref{eq:w3} and \ref{eq:w4}). This additional complexity of nonlinear absorption under strong coupling conditions is expected, since while external fields acting on the bare system drives transitions between three molecular states (ground, first and second excited-state), at least seven energy levels (Fig. \ref{fig:enlev}) may play a role in the nonlinear response of a material strongly coupled to an optical cavity.
     \begin{figure}[t]
     \includegraphics[width=\columnwidth]{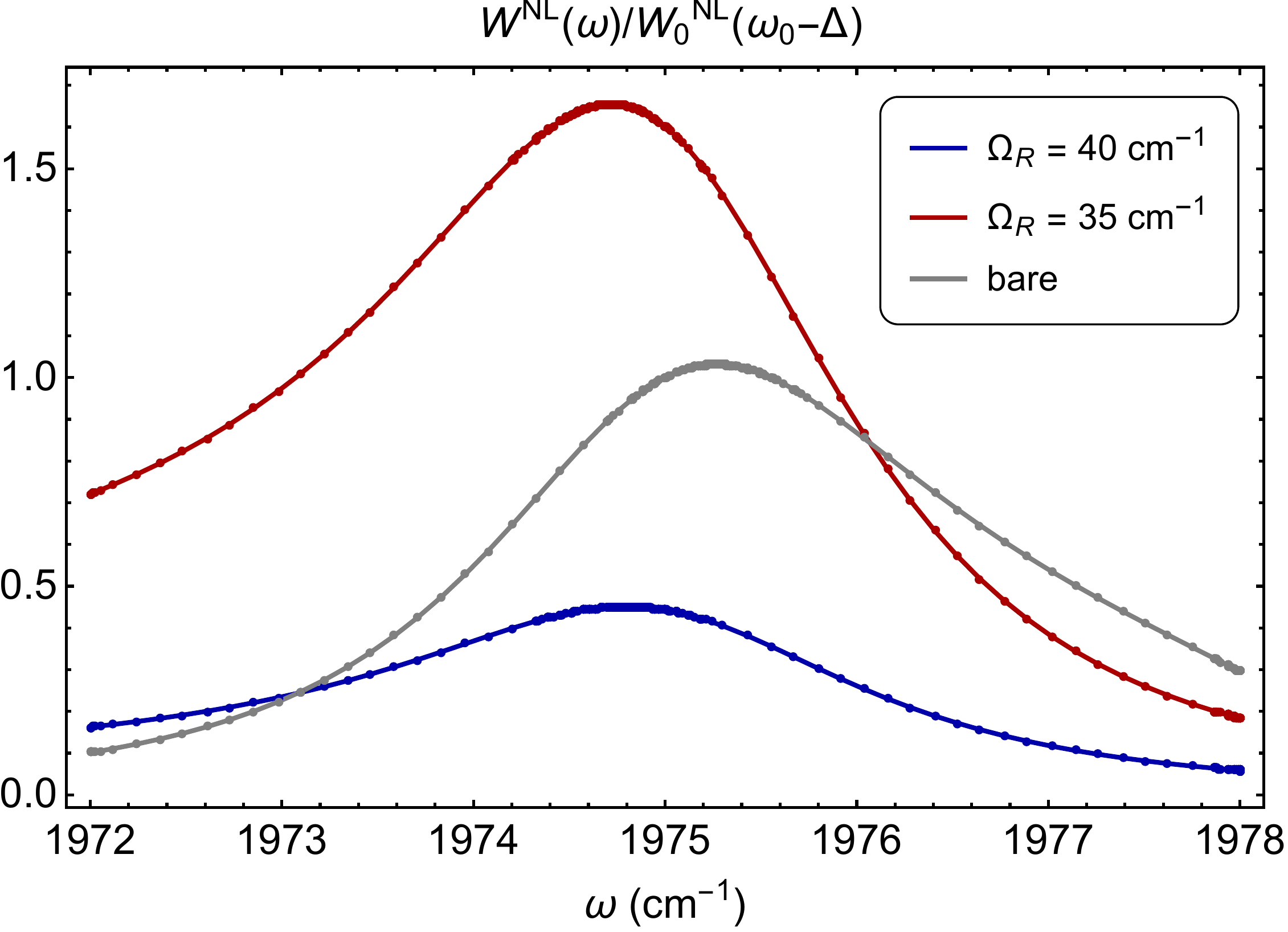}
     \caption{Ratio of two-photon absorption rate of two strongly coupled ($\omega_0 =  \omega_c = 1983~\t{cm}^{-1}, \gamma_m = \kappa = 3~\t{cm}^{-1}, \Delta = 8~\t{cm}^{-1}$, and $\Omega_R = 40~\t{cm}^{-1}$ or $\Omega_R = 35~\t{cm}^{-1}$) systems relative to that of the molecular ensemble in free space normalized by the maximum of the latter.}
     \label{fig:tpa_or}
    \end{figure}
Nevertheless, the main features of $W^{\t{NL}}(\omega)$ can be obtained from Eqs. \ref{eq:w1}$-$\ref{eq:w4}. Specifically,
\begin{enumerate}
\item The nonlinear absorption intensity is largest when the input field is nearly-resonant with either the LP or UP, since this maximizes $|G_{mm}(\omega)|^4$ which appears in all of Eqs. \ref{eq:w1}$-
$\ref{eq:w4}. Physically, the polariton resonance condition for maximal photon absorption is a consequence of the optical filtering performed by a microcavity (off-resonant external fields are suppressed relative to the resonant).
\item $W^{\t{NL}_3}(\omega)$ is the only component of $W^{\t{NL}}(\omega)$ which is positive for all values of the input frequency. Therefore, it necessarily gives molecular excited-state absorption contributions to $W^{\t{NL}}(\omega)$. Further evidence is given by the fact that $W^{\t{NL}_3}(\omega)$ is proportional to the steady-state population of molecules in the doubly-excited state, and thus,
\begin{align}
~~~~W^{\t{NL}_3}(\omega) \approx  \frac{8\eta \Delta^2}{(\omega-\omega_0)^2+\gamma_m^2/4}\frac{P_{2_m}^{\t{T}}(2\omega)}{\left|E_{\t{in}}^{(+)}(\omega)\right|^4}.
\end{align}
  All other contributions to the nonlinear absorption can be either positive (when excited-state absorption processes dominate) or negative (when ground-state bleach and stimulated emission processes dominate \cite{mukamel1999principles}) depending on $\omega$.

\item Based on items 1 and 2, we expect the TPA rate will be largely enhanced relative to free space when the doubly-excited molecular states are approximately resonant with either one of the available two-polariton transitions (see Fig. \ref{fig:cavpes}), i.e.,
\begin{align}
 	&2\omega_0-2\Delta = 2\omega\LP, ~~\Delta > 0, ~\t{or} \nonum
 	&2\omega_0 -2\Delta = 2\omega\UP, ~~\Delta < 0 \label{eq:rescon},
 \end{align}
since in this case, all response functions showing up in Eqs. \ref{eq:chi3_ne} and $19-23$ (namely, $G_{mm}(\omega)$, and the scattering amplitude $\Gamma_{mm,mm}(2\omega)$) become resonant at $\omega = \omega_{\t{LP}} ~(\t{if}~\Delta > 0)$ or $\omega = \omega_{\t{UP}}~(\t{if}~\Delta < 0)$. Physically, the polaritons provide the resonant optical window to efficiently drive the transitions of interest. In the studied case of zero cavity detuning, the conditions described in Eq. \ref{eq:rescon} can be summarized as the Rabi splitting being equal to the anharmonic shift, that is, $\Omega_R  = \pm 2\Delta$. 

When the criteria in Eq. \ref{eq:rescon} are satisfied, we expect strong enhancement of nonlinear absorption based on the following argument: if the input field consists of photons with $\omega = \omega_0 - \Delta$ and Eq. \ref{eq:rescon} is satisfied, polaritons will be efficiently pumped, and a fraction of those will subsequently decay by populating molecular doubly-excited states. In other words, when the two-polariton resonance condition is satisfied, the molecular doubly excited-states provides an efficient sink for the energy stored in two-polariton modes. This effect was indeed reported in a recent experiment \cite{xiang2019a}, where evidence was given that (for systems with weak system-bath interactions and slow molecular polarization dephasing) the second excited vibrational state was preferentially populated over the first when the pump (input) field was resonant with LP.

\item Conversely, in the limit where two-polariton states are highly off-resonant with the molecular TPA ($|2\Delta - \Omega_R| \gg 0$, Fig. \ref{fig:tpa_or}), the nonlinear response substantially weakens. In this limit, the studied model approaches the Tavis-Cummings \cite{tavis_exact_1968}, where a collection of two-level systems interact strongly with a single-mode cavity. The nonlinear response given by this system is known to become negligible in the large $N$ limit \cite{varada1987} (e.g., as we show in the SI Sec. 7, in the Tavis-Cummings model, the two-level system nonlinearity produces a large $N$ limit anharmonic shift proportional to $|\mu E_{0}^c|/\sqrt{N}$). 

It follows, therefore, that the condition given in Eq. \ref{eq:rescon} allows the harnessing of the enhanced electromagnetic field of optical cavities to enhance TPA. 

\end{enumerate}
Points 3 and 4 are the main conclusions of our work. We will now quantitatively illustrate that under experimentally accessible conditions, it is possible to employ cavity-strong coupling to substantially enhance the TPA cross section of a resonant molecular system. 

In Fig. \ref{fig:tpa_or}, we present the infrared TPA spectrum for a molecular system in free space  (we take representative parameters for W(CO)$_6$ in solution \cite{xiang2018, xiang2019a}, $\omega_0 = 1983 ~\t{cm}^{-1}, \Delta = 8~\t{cm}^{-1},\gamma = 3~\t{cm}^{-1}$), and under strong coupling with a microcavity  ($\omega_c = \omega_0, \kappa = \gamma$)  for $\Omega_R = 40~\t{cm}^{-1}$ and $\Omega_R = 35 ~\t{cm}^{-1}$. The curves are normalized by the maximum of the bare system TPA. Figure \ref{fig:tpa_or} shows the strong dependence of the TPA cross-section on the light-matter interaction: when $\Omega_R = 40~\t{cm}^{-1}$ ($\Omega_R - 2\Delta = 24~\t{cm}^{-1}$), the nonlinear absorption is suppressed relative to that given by the bare system. However, a slight decrease of $\Omega_R$ to $35~\t{cm}^{-1}$ leads to enhanced TPA due to a stronger spectral overlap between the LP$_2$ mode and the molecular doubly excited-state transition from the ground-state. 

\begin{figure}[t]
     \includegraphics[width=\columnwidth]{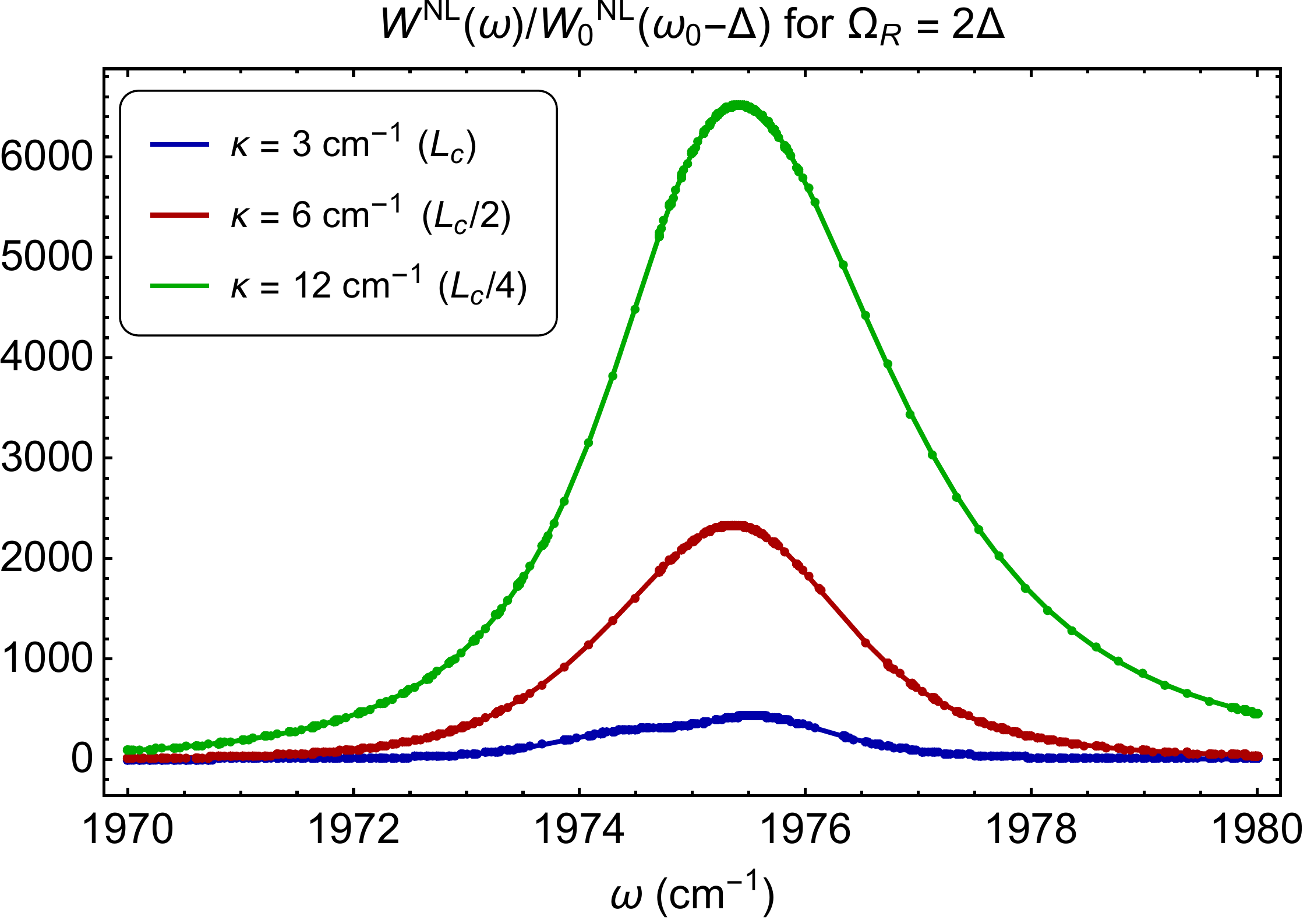}
     \caption{Enhancement of TPA rate of strongly coupled system when $\Omega_R = 2\Delta$, with $\omega_0 = \omega_c, \gamma = 3~\t{cm}^{-1}, \Delta = 8~\t{cm}^{-1}$ for optical microcavities with different cavity lengths $(L_c, L_c/2, L_c/4)$  and corresponding decay rates.}
     \label{fig:tpa_k}
    \end{figure}

In Fig. \ref{fig:tpa_k}, we explore the great potential for obtaining polariton-enhanced TPA with optical microcavities of varying longitudinal lengths $L_c, L_c/2$ and $L_c/4$ (with $L_c = 10~\mu m$, and cavity-mode indices $m=4, m =2$ and $m=1$, respectively, which would require cavity mirrors with transmissivity $|t|^2 \approx 0.01\%$). We assume the cavities are resonant with the molecular fundamental transition, and the TPA condition $\Omega_R = 2\Delta = 16~\t{cm}^{-1}$ is valid (the remaining bare molecule parameters are the same as in Fig. \ref{fig:tpa_or}). The two main conclusions from Fig. \ref{fig:tpa_k} are that: (a) the polariton-mediated TPA cross section predicted by our model can be larger than the bare one by close to 4 orders of magnitude for accessible parameters, and (b) a decrease in cavity length leads to stronger nonlinear signals, so that the cavity-mediated TPA will be maximally efficient when the strongly coupled cavity mode has the lowest possible longitudinal quantum number and mirrors with highest available reflectivity. These conditions, in fact, maximize the intracavity electromagnetic field enhancement relative to free space (see SI Sec. 1).

A similar increase in  nonlinear response signal strength with decreasing molecular concentration (Fig. \ref{fig:tpa_or}) or cavity longitudinal length (Fig. \ref{fig:tpa_k}), was observed and qualitatively analyzed in a different context in Ref. \cite{xiang2019b}.

\section{Discussion and Conclusions}\label{sec:dac}

The computed enhanced polariton-mediated TPA provides an upper bound estimate to future measurements of TPA under strong coupling conditions. Experiments performed on analogous systems could give reduced enhancements relative to those presented here for at least two reasons: (a) the intracavity enhancement factor (represented by the cavity finesse) varies spatially according to the cavity mode profile  ($\t{sin}(\pi z/L)$ in the simplest case), whereas we assumed that all molecules are within a small region around an antinode of the cavity field (so that the cavity field enhancement factor is maximal), and (b) inhomogeneous broadening of the molecular subsystem which allows for potentially fast polariton relaxation into reservoir (dark) modes, as well as reduction in efficiently of polariton pumping due to photonic intensity borrowing. Although we recognize the importance of these approximations, we disregard them in our explorations, since the inhomogeneity of the cavity mode profile is expected to change the nonlinear response properties by factors of order 1 (alternatively, spacers may be introduced between the molecular system and the optical cavity so that the molecules occupy only a small region around the antinode of the cavity mode profile), while (lower) polariton decay can be slowed down by increasing the Rabi splitting and (or) lowering the temperature. Moreover, polariton transitions are well-known to be homogeneously broadened within their lifetimes \cite{houdre_vacuum-field_1996}, and therefore, for molecular systems with significant inhomogeneously broadened transitions, we expect polariton lineshapes to be significantly narrower than that of the bare system (given the polariton ``hole-burning" effect \cite{cheng2018} yielding subnatural linewidths). In this instance, the mechanism for polariton-mediated TPA presented in our article would become even more efficient than in the model considered here.

While our study focused on infrared polaritonics, we note that the phenomenology observed in molecular vibrational and electronic strong coupling can be very similar depending on the system. For instance, ultrafast pump-probe transmission recorded for a microcavity strongly coupled to an organic semiconductor in Ref. \cite{virgili2011} showed qualitative features identical to the first reported vibrational polariton pump-probe data \cite{dunkelberger_modified_2016}. The exciton-biexciton ladder described in in Ref. \cite{virgili2011} is also notable in the studied context because the energy of the corresponding biexciton is more than twice of the excitonic, which would enable verification of the $\Delta < 0$ case of Eq. \ref{eq:rescon}. In fact, whenever electronic transitions are only weakly coupled to high-frequency vibrational modes, and the electronic $S_1 \rightarrow S_2$ (first to second excited-state) transition is dipole-allowed and slightly red- or blue-shifted from the $S_0 \rightarrow S_1$ (ground to first excited-state), we expect electronic TPA rates to have similarly appealing potential for enhancement in optical cavities under the strong coupling regime as discussed in Sec. \ref{sec:pmtpa}. 

It is also notable that, although our model and expressions allow us to derive quantitative nonlinear properties of molecular systems described as three-level system ensembles, the enhancement of molecular nonlinear polarization by intracavity field effects and subnatural polariton linewidths thoroughly discussed in Sec. \ref{sec:nlmolpol} are universal features of the nonlinear susceptibilities of molecular ensembles under strong coupling. In fact, our work qualitatively corroborates recently reported nonlinear response enhancement induced by strong coupling of microcavities with organic semiconductor materials \cite{barachati2018, wang2020} whose effective Hamiltonian is not given by Eq. \ref{eq:heffinal}. Specifically, Barachati and coworkers \cite{barachati2018} ascribed third-harmonic generation efficiency gains under cavity strong coupling to intracavity field energy density enhancement, whereas the recent Z-scan measurements reported by Wang et al. \cite{wang2020} showed that polariton resonance effects were also essential to obtain increases in the magnitude of the nonlinear index of refraction and absorption.

	In summary, we have derived and analyzed the nonlinear optical susceptibility and TPA rates for a molecular system under strong coupling with an infrared microcavity. By contrasting the polaritonic response with that of bare molecules in free space, we found that enhanced nonlinearities in the strong coupling regime may emerge due to intracavity field enhancement, creation of suitable optical resonances, and subnatural polaritonic linewidths. Our results suggest an increase of several orders of magnitude can potentially be achieved for the polaritonic nonlinear optical response, especially, when a multipolariton transition is resonant with a multiphonon (or multi-electronic state) absorption of the molecular system. Our work suggests new application of molecular polaritonics in two-photon imaging \cite{so2000}, and efficient generation of hot molecular excited-state distributions via (polariton) ladder climbing \cite{kraack2016a, morichika2019, juraschek2019}, possibly bypassing deleterious intramolecular vibrational relaxation.\\

\textbf{Acknowledgments $-$} R.F.R., J.C.G., N.C.G, and J.Y.Z. acknowledge funding support from the Defense Advanced Research Projects Agency under Award No. D19AC00011 for the calculations of nonlinear susceptibilities under strong coupling and the Air Force Office of Scientific Research award FA9550-18-1-0289 for the analysis of optimal conditions for multiphoton absorption under strong light-matter coupling. W.X. thanks the support of NSF CAREER Award DMR1848215.
\bibliography{lib}

\begin{thebibliography}{84}%
\makeatletter
\providecommand \@ifxundefined [1]{%
 \@ifx{#1\undefined}
}%
\providecommand \@ifnum [1]{%
 \ifnum #1\expandafter \@firstoftwo
 \else \expandafter \@secondoftwo
 \fi
}%
\providecommand \@ifx [1]{%
 \ifx #1\expandafter \@firstoftwo
 \else \expandafter \@secondoftwo
 \fi
}%
\providecommand \natexlab [1]{#1}%
\providecommand \enquote  [1]{``#1''}%
\providecommand \bibnamefont  [1]{#1}%
\providecommand \bibfnamefont [1]{#1}%
\providecommand \citenamefont [1]{#1}%
\providecommand \href@noop [0]{\@secondoftwo}%
\providecommand \href [0]{\begingroup \@sanitize@url \@href}%
\providecommand \@href[1]{\@@startlink{#1}\@@href}%
\providecommand \@@href[1]{\endgroup#1\@@endlink}%
\providecommand \@sanitize@url [0]{\catcode `\\12\catcode `\$12\catcode
  `\&12\catcode `\#12\catcode `\^12\catcode `\_12\catcode `\%12\relax}%
\providecommand \@@startlink[1]{}%
\providecommand \@@endlink[0]{}%
\providecommand \url  [0]{\begingroup\@sanitize@url \@url }%
\providecommand \@url [1]{\endgroup\@href {#1}{\urlprefix }}%
\providecommand \urlprefix  [0]{URL }%
\providecommand \Eprint [0]{\href }%
\providecommand \doibase [0]{http://dx.doi.org/}%
\providecommand \selectlanguage [0]{\@gobble}%
\providecommand \bibinfo  [0]{\@secondoftwo}%
\providecommand \bibfield  [0]{\@secondoftwo}%
\providecommand \translation [1]{[#1]}%
\providecommand \BibitemOpen [0]{}%
\providecommand \bibitemStop [0]{}%
\providecommand \bibitemNoStop [0]{.\EOS\space}%
\providecommand \EOS [0]{\spacefactor3000\relax}%
\providecommand \BibitemShut  [1]{\csname bibitem#1\endcsname}%
\let\auto@bib@innerbib\@empty
\bibitem [{\citenamefont {Mukamel}(1999)}]{mukamel1999principles}%
  \BibitemOpen
  \bibfield  {author} {\bibinfo {author} {\bibfnamefont {Shaul}\ \bibnamefont
  {Mukamel}},\ }\href@noop {} {\emph {\bibinfo {title} {Principles of Nonlinear
  Optical Spectroscopy}}}\ (\bibinfo  {publisher} {{Oxford University Press on
  Demand}},\ \bibinfo {year} {1999})\BibitemShut {NoStop}%
\bibitem [{\citenamefont {{Yuen-Zhou}}\ \emph {et~al.}(2014)\citenamefont
  {{Yuen-Zhou}}, \citenamefont {Krich}, \citenamefont {{Aspuru-Guzik}},
  \citenamefont {Kassal},\ and\ \citenamefont {Johnson}}]{yuen2014ultrafast}%
  \BibitemOpen
  \bibfield  {author} {\bibinfo {author} {\bibfnamefont {J.}~\bibnamefont
  {{Yuen-Zhou}}}, \bibinfo {author} {\bibfnamefont {J.J.}\ \bibnamefont
  {Krich}}, \bibinfo {author} {\bibfnamefont {A.}~\bibnamefont
  {{Aspuru-Guzik}}}, \bibinfo {author} {\bibfnamefont {I.}~\bibnamefont
  {Kassal}}, \ and\ \bibinfo {author} {\bibfnamefont {A.S.}\ \bibnamefont
  {Johnson}},\ }\href@noop {} {\emph {\bibinfo {title} {Ultrafast
  {{Spectroscopy}}: {{Quantum Information}} and {{Wavepackets}}}}},\ {{IOP}}
  Expanding Physics\ (\bibinfo  {publisher} {{Institute of Physics
  Publishing}},\ \bibinfo {year} {2014})\BibitemShut {NoStop}%
\bibitem [{\citenamefont {Brixner}\ \emph {et~al.}(2005)\citenamefont
  {Brixner}, \citenamefont {Stenger}, \citenamefont {Vaswani}, \citenamefont
  {Cho}, \citenamefont {Blankenship},\ and\ \citenamefont
  {Fleming}}]{brixner2005}%
  \BibitemOpen
  \bibfield  {author} {\bibinfo {author} {\bibfnamefont {Tobias}\ \bibnamefont
  {Brixner}}, \bibinfo {author} {\bibfnamefont {Jens}\ \bibnamefont {Stenger}},
  \bibinfo {author} {\bibfnamefont {Harsha~M.}\ \bibnamefont {Vaswani}},
  \bibinfo {author} {\bibfnamefont {Minhaeng}\ \bibnamefont {Cho}}, \bibinfo
  {author} {\bibfnamefont {Robert~E.}\ \bibnamefont {Blankenship}}, \ and\
  \bibinfo {author} {\bibfnamefont {Graham~R.}\ \bibnamefont {Fleming}},\
  }\bibfield  {title} {\enquote {\bibinfo {title} {Two-dimensional spectroscopy
  of electronic couplings in photosynthesis},}\ }\href {\doibase
  10.1038/nature03429} {\bibfield  {journal} {\bibinfo  {journal} {Nature}\
  }\textbf {\bibinfo {volume} {434}},\ \bibinfo {pages} {625} (\bibinfo {year}
  {2005})}\BibitemShut {NoStop}%
\bibitem [{\citenamefont {Thyrhaug}\ \emph {et~al.}(2018)\citenamefont
  {Thyrhaug}, \citenamefont {Tempelaar}, \citenamefont {Alcocer}, \citenamefont
  {{\v Z}{\'i}dek}, \citenamefont {B{\'i}na}, \citenamefont {Knoester},
  \citenamefont {Jansen},\ and\ \citenamefont {Zigmantas}}]{thyrhaug2018}%
  \BibitemOpen
  \bibfield  {author} {\bibinfo {author} {\bibfnamefont {E.}~\bibnamefont
  {Thyrhaug}}, \bibinfo {author} {\bibfnamefont {R.}~\bibnamefont {Tempelaar}},
  \bibinfo {author} {\bibfnamefont {M.~J.~P.}\ \bibnamefont {Alcocer}},
  \bibinfo {author} {\bibfnamefont {K.}~\bibnamefont {{\v Z}{\'i}dek}},
  \bibinfo {author} {\bibfnamefont {D.}~\bibnamefont {B{\'i}na}}, \bibinfo
  {author} {\bibfnamefont {J.}~\bibnamefont {Knoester}}, \bibinfo {author}
  {\bibfnamefont {T.~L.~C.}\ \bibnamefont {Jansen}}, \ and\ \bibinfo {author}
  {\bibfnamefont {D.}~\bibnamefont {Zigmantas}},\ }\bibfield  {title} {\enquote
  {\bibinfo {title} {Identification and characterization of diverse coherences
  in the {{Fenna}}-{{Matthews}}-{{Olson}} complex.}}\ }\href {\doibase
  10.1038/s41557-018-0060-5} {\bibfield  {journal} {\bibinfo  {journal} {Nature
  chemistry}\ }\textbf {\bibinfo {volume} {10}},\ \bibinfo {pages} {780--786}
  (\bibinfo {year} {2018})}\BibitemShut {NoStop}%
\bibitem [{\citenamefont {Xiang}\ \emph {et~al.}(2017)\citenamefont {Xiang},
  \citenamefont {Li}, \citenamefont {Pham}, \citenamefont {Paesani},\ and\
  \citenamefont {Xiong}}]{xiang2017}%
  \BibitemOpen
  \bibfield  {author} {\bibinfo {author} {\bibfnamefont {Bo}~\bibnamefont
  {Xiang}}, \bibinfo {author} {\bibfnamefont {Yingmin}\ \bibnamefont {Li}},
  \bibinfo {author} {\bibfnamefont {C.~Huy}\ \bibnamefont {Pham}}, \bibinfo
  {author} {\bibfnamefont {Francesco}\ \bibnamefont {Paesani}}, \ and\ \bibinfo
  {author} {\bibfnamefont {Wei}\ \bibnamefont {Xiong}},\ }\bibfield  {title}
  {\enquote {\bibinfo {title} {Ultrafast direct electron transfer at organic
  semiconductor and metal interfaces},}\ }\href {\doibase
  10.1126/sciadv.1701508} {\bibfield  {journal} {\bibinfo  {journal} {Science
  Advances}\ }\textbf {\bibinfo {volume} {3}},\ \bibinfo {pages} {e1701508}
  (\bibinfo {year} {2017})}\BibitemShut {NoStop}%
\bibitem [{\citenamefont {Breen}\ \emph {et~al.}(2017)\citenamefont {Breen},
  \citenamefont {Tempelaar}, \citenamefont {Bizimana}, \citenamefont {Kloss},
  \citenamefont {Reichman},\ and\ \citenamefont {Turner}}]{breen2017}%
  \BibitemOpen
  \bibfield  {author} {\bibinfo {author} {\bibfnamefont {Ilana}\ \bibnamefont
  {Breen}}, \bibinfo {author} {\bibfnamefont {Roel}\ \bibnamefont {Tempelaar}},
  \bibinfo {author} {\bibfnamefont {Laurie~A.}\ \bibnamefont {Bizimana}},
  \bibinfo {author} {\bibfnamefont {Benedikt}\ \bibnamefont {Kloss}}, \bibinfo
  {author} {\bibfnamefont {David~R.}\ \bibnamefont {Reichman}}, \ and\ \bibinfo
  {author} {\bibfnamefont {Daniel~B.}\ \bibnamefont {Turner}},\ }\bibfield
  {title} {\enquote {\bibinfo {title} {Triplet {{Separation Drives Singlet
  Fission}} after {{Femtosecond Correlated Triplet Pair Production}} in
  {{Rubrene}}},}\ }\href {\doibase 10.1021/jacs.7b02621} {\bibfield  {journal}
  {\bibinfo  {journal} {Journal of the American Chemical Society}\ }\textbf
  {\bibinfo {volume} {139}},\ \bibinfo {pages} {11745--11751} (\bibinfo {year}
  {2017})}\BibitemShut {NoStop}%
\bibitem [{\citenamefont {Sanvitto}\ and\ \citenamefont
  {{K{\'e}na-Cohen}}(2016)}]{sanvitto2016}%
  \BibitemOpen
  \bibfield  {author} {\bibinfo {author} {\bibfnamefont {Daniele}\ \bibnamefont
  {Sanvitto}}\ and\ \bibinfo {author} {\bibfnamefont {St{\'e}phane}\
  \bibnamefont {{K{\'e}na-Cohen}}},\ }\bibfield  {title} {\enquote {\bibinfo
  {title} {The road towards polaritonic devices},}\ }\href {\doibase
  10.1038/nmat4668} {\bibfield  {journal} {\bibinfo  {journal} {Nature
  Materials}\ }\textbf {\bibinfo {volume} {15}},\ \bibinfo {pages} {1061--1073}
  (\bibinfo {year} {2016})}\BibitemShut {NoStop}%
\bibitem [{\citenamefont {Zasedatelev}\ \emph {et~al.}(2019)\citenamefont
  {Zasedatelev}, \citenamefont {Baranikov}, \citenamefont {Urbonas},
  \citenamefont {Scafirimuto}, \citenamefont {Scherf}, \citenamefont
  {St{\"o}ferle}, \citenamefont {Mahrt},\ and\ \citenamefont
  {Lagoudakis}}]{zasedatelev2019}%
  \BibitemOpen
  \bibfield  {author} {\bibinfo {author} {\bibfnamefont {Anton~V.}\
  \bibnamefont {Zasedatelev}}, \bibinfo {author} {\bibfnamefont {Anton~V.}\
  \bibnamefont {Baranikov}}, \bibinfo {author} {\bibfnamefont {Darius}\
  \bibnamefont {Urbonas}}, \bibinfo {author} {\bibfnamefont {Fabio}\
  \bibnamefont {Scafirimuto}}, \bibinfo {author} {\bibfnamefont {Ullrich}\
  \bibnamefont {Scherf}}, \bibinfo {author} {\bibfnamefont {Thilo}\
  \bibnamefont {St{\"o}ferle}}, \bibinfo {author} {\bibfnamefont {Rainer~F.}\
  \bibnamefont {Mahrt}}, \ and\ \bibinfo {author} {\bibfnamefont {Pavlos~G.}\
  \bibnamefont {Lagoudakis}},\ }\bibfield  {title} {\enquote {\bibinfo {title}
  {A room-temperature organic polariton transistor},}\ }\href {\doibase
  10.1038/s41566-019-0392-8} {\bibfield  {journal} {\bibinfo  {journal} {Nature
  Photonics}\ }\textbf {\bibinfo {volume} {13}},\ \bibinfo {pages} {378}
  (\bibinfo {year} {2019})}\BibitemShut {NoStop}%
\bibitem [{\citenamefont {Monroe}(2002)}]{monroe2002}%
  \BibitemOpen
  \bibfield  {author} {\bibinfo {author} {\bibfnamefont {C.}~\bibnamefont
  {Monroe}},\ }\bibfield  {title} {\enquote {\bibinfo {title} {Quantum
  information processing with atoms and photons},}\ }\href {\doibase
  10.1038/416238a} {\bibfield  {journal} {\bibinfo  {journal} {Nature}\
  }\textbf {\bibinfo {volume} {416}},\ \bibinfo {pages} {238} (\bibinfo {year}
  {2002})}\BibitemShut {NoStop}%
\bibitem [{\citenamefont {Braunstein}\ and\ \citenamefont {{van
  Loock}}(2005)}]{braunstein2005}%
  \BibitemOpen
  \bibfield  {author} {\bibinfo {author} {\bibfnamefont {Samuel~L.}\
  \bibnamefont {Braunstein}}\ and\ \bibinfo {author} {\bibfnamefont {Peter}\
  \bibnamefont {{van Loock}}},\ }\bibfield  {title} {\enquote {\bibinfo {title}
  {Quantum information with continuous variables},}\ }\href {\doibase
  10.1103/RevModPhys.77.513} {\bibfield  {journal} {\bibinfo  {journal}
  {Reviews of Modern Physics}\ }\textbf {\bibinfo {volume} {77}},\ \bibinfo
  {pages} {513--577} (\bibinfo {year} {2005})}\BibitemShut {NoStop}%
\bibitem [{\citenamefont {Chulhai}\ \emph {et~al.}(2016)\citenamefont
  {Chulhai}, \citenamefont {Hu}, \citenamefont {Moore}, \citenamefont {Chen},\
  and\ \citenamefont {Jensen}}]{chulhai2016}%
  \BibitemOpen
  \bibfield  {author} {\bibinfo {author} {\bibfnamefont {Dhabih~V.}\
  \bibnamefont {Chulhai}}, \bibinfo {author} {\bibfnamefont {Zhongwei}\
  \bibnamefont {Hu}}, \bibinfo {author} {\bibfnamefont {Justin~E.}\
  \bibnamefont {Moore}}, \bibinfo {author} {\bibfnamefont {Xing}\ \bibnamefont
  {Chen}}, \ and\ \bibinfo {author} {\bibfnamefont {Lasse}\ \bibnamefont
  {Jensen}},\ }\bibfield  {title} {\enquote {\bibinfo {title} {Theory of
  {{Linear}} and {{Nonlinear Surface}}-{{Enhanced Vibrational
  Spectroscopies}}},}\ }\href {\doibase 10.1146/annurev-physchem-040215-112347}
  {\bibfield  {journal} {\bibinfo  {journal} {Annual Review of Physical
  Chemistry}\ }\textbf {\bibinfo {volume} {67}},\ \bibinfo {pages} {541--564}
  (\bibinfo {year} {2016})},\ \bibinfo {note} {\_eprint:
  https://doi.org/10.1146/annurev-physchem-040215-112347}\BibitemShut {NoStop}%
\bibitem [{\citenamefont {Boyd}\ and\ \citenamefont
  {Prato}(2008)}]{boyd2008nonlinear}%
  \BibitemOpen
  \bibfield  {author} {\bibinfo {author} {\bibfnamefont {R.W.}\ \bibnamefont
  {Boyd}}\ and\ \bibinfo {author} {\bibfnamefont {D.}~\bibnamefont {Prato}},\
  }\href@noop {} {\emph {\bibinfo {title} {Nonlinear Optics}}}\ (\bibinfo
  {publisher} {{Elsevier Science}},\ \bibinfo {year} {2008})\BibitemShut
  {NoStop}%
\bibitem [{\citenamefont {Kavokin}\ \emph {et~al.}(2017)\citenamefont
  {Kavokin}, \citenamefont {Baumberg}, \citenamefont {Malpuech},\ and\
  \citenamefont {Laussy}}]{kavokin2017microcavities}%
  \BibitemOpen
  \bibfield  {author} {\bibinfo {author} {\bibfnamefont {Alexey~V}\
  \bibnamefont {Kavokin}}, \bibinfo {author} {\bibfnamefont {Jeremy~J}\
  \bibnamefont {Baumberg}}, \bibinfo {author} {\bibfnamefont {Guillaume}\
  \bibnamefont {Malpuech}}, \ and\ \bibinfo {author} {\bibfnamefont
  {Fabrice~P}\ \bibnamefont {Laussy}},\ }\href@noop {} {\emph {\bibinfo {title}
  {Microcavities}}},\ Vol.~\bibinfo {volume} {21}\ (\bibinfo  {publisher}
  {{Oxford University Press}},\ \bibinfo {year} {2017})\BibitemShut {NoStop}%
\bibitem [{\citenamefont {Ebbesen}(2016)}]{ebbesen_hybrid_2016}%
  \BibitemOpen
  \bibfield  {author} {\bibinfo {author} {\bibfnamefont {Thomas~W.}\
  \bibnamefont {Ebbesen}},\ }\bibfield  {title} {\enquote {\bibinfo {title}
  {Hybrid {{Light}}\textendash{{Matter States}} in a {{Molecular}} and
  {{Material Science Perspective}}},}\ }\href {\doibase
  10.1021/acs.accounts.6b00295} {\bibfield  {journal} {\bibinfo  {journal}
  {Accounts of Chemical Research}\ }\textbf {\bibinfo {volume} {49}},\ \bibinfo
  {pages} {2403--2412} (\bibinfo {year} {2016})}\BibitemShut {NoStop}%
\bibitem [{\citenamefont {Sukharev}\ and\ \citenamefont
  {Nitzan}(2017)}]{sukharev_optics_2017}%
  \BibitemOpen
  \bibfield  {author} {\bibinfo {author} {\bibfnamefont {Maxim}\ \bibnamefont
  {Sukharev}}\ and\ \bibinfo {author} {\bibfnamefont {Abraham}\ \bibnamefont
  {Nitzan}},\ }\bibfield  {title} {\enquote {\bibinfo {title} {Optics of
  exciton-plasmon nanomaterials},}\ }\href {\doibase 10.1088/1361-648X/aa85ef}
  {\bibfield  {journal} {\bibinfo  {journal} {Journal of Physics: Condensed
  Matter}\ }\textbf {\bibinfo {volume} {29}},\ \bibinfo {pages} {443003}
  (\bibinfo {year} {2017})}\BibitemShut {NoStop}%
\bibitem [{\citenamefont {Ribeiro}\ \emph
  {et~al.}(2018{\natexlab{a}})\citenamefont {Ribeiro}, \citenamefont {{A.
  Mart{\'i}nez-Mart{\'i}nez}}, \citenamefont {Du}, \citenamefont
  {{Campos-Gonzalez-Angulo}},\ and\ \citenamefont
  {{Yuen-Zhou}}}]{ribeiro2018d}%
  \BibitemOpen
  \bibfield  {author} {\bibinfo {author} {\bibfnamefont {Raphael~F.}\
  \bibnamefont {Ribeiro}}, \bibinfo {author} {\bibfnamefont {Luis}\
  \bibnamefont {{A. Mart{\'i}nez-Mart{\'i}nez}}}, \bibinfo {author}
  {\bibfnamefont {Matthew}\ \bibnamefont {Du}}, \bibinfo {author}
  {\bibfnamefont {Jorge}\ \bibnamefont {{Campos-Gonzalez-Angulo}}}, \ and\
  \bibinfo {author} {\bibfnamefont {Joel}\ \bibnamefont {{Yuen-Zhou}}},\
  }\bibfield  {title} {\enquote {\bibinfo {title} {Polariton chemistry:
  Controlling molecular dynamics with optical cavities},}\ }\href {\doibase
  10.1039/C8SC01043A} {\bibfield  {journal} {\bibinfo  {journal} {Chemical
  Science}\ }\textbf {\bibinfo {volume} {9}},\ \bibinfo {pages} {6325--6339}
  (\bibinfo {year} {2018}{\natexlab{a}})}\BibitemShut {NoStop}%
\bibitem [{\citenamefont {Sukharev}\ \emph {et~al.}(2014)\citenamefont
  {Sukharev}, \citenamefont {Seideman}, \citenamefont {Gordon}, \citenamefont
  {Salomon},\ and\ \citenamefont {Prior}}]{sukharev_ultrafast_2014}%
  \BibitemOpen
  \bibfield  {author} {\bibinfo {author} {\bibfnamefont {Maxim}\ \bibnamefont
  {Sukharev}}, \bibinfo {author} {\bibfnamefont {Tamar}\ \bibnamefont
  {Seideman}}, \bibinfo {author} {\bibfnamefont {Robert~J.}\ \bibnamefont
  {Gordon}}, \bibinfo {author} {\bibfnamefont {Adi}\ \bibnamefont {Salomon}}, \
  and\ \bibinfo {author} {\bibfnamefont {Yehiam}\ \bibnamefont {Prior}},\
  }\bibfield  {title} {\enquote {\bibinfo {title} {Ultrafast {{Energy
  Transfer}} between {{Molecular Assemblies}} and {{Surface Plasmons}} in the
  {{Strong Coupling Regime}}},}\ }\href {\doibase 10.1021/nn4054528} {\bibfield
   {journal} {\bibinfo  {journal} {ACS Nano}\ }\textbf {\bibinfo {volume}
  {8}},\ \bibinfo {pages} {807--817} (\bibinfo {year} {2014})}\BibitemShut
  {NoStop}%
\bibitem [{\citenamefont {Vasa}\ and\ \citenamefont {Lienau}(2018)}]{vasa2018}%
  \BibitemOpen
  \bibfield  {author} {\bibinfo {author} {\bibfnamefont {Parinda}\ \bibnamefont
  {Vasa}}\ and\ \bibinfo {author} {\bibfnamefont {Christoph}\ \bibnamefont
  {Lienau}},\ }\bibfield  {title} {\enquote {\bibinfo {title} {Strong
  {{Light}}\textendash{{Matter Interaction}} in {{Quantum Emitter}}/{{Metal
  Hybrid Nanostructures}}},}\ }\href {\doibase 10.1021/acsphotonics.7b00650}
  {\bibfield  {journal} {\bibinfo  {journal} {ACS Photonics}\ }\textbf
  {\bibinfo {volume} {5}},\ \bibinfo {pages} {2--23} (\bibinfo {year}
  {2018})}\BibitemShut {NoStop}%
\bibitem [{\citenamefont {S.~Dovzhenko}\ \emph {et~al.}(2018)\citenamefont
  {S.~Dovzhenko}, \citenamefont {V.~Ryabchuk}, \citenamefont {P.~Rakovich},\
  and\ \citenamefont {R.~Nabiev}}]{s.dovzhenko2018}%
  \BibitemOpen
  \bibfield  {author} {\bibinfo {author} {\bibfnamefont {D.}~\bibnamefont
  {S.~Dovzhenko}}, \bibinfo {author} {\bibfnamefont {S.}~\bibnamefont
  {V.~Ryabchuk}}, \bibinfo {author} {\bibfnamefont {Yu}~\bibnamefont
  {P.~Rakovich}}, \ and\ \bibinfo {author} {\bibfnamefont {I.}~\bibnamefont
  {R.~Nabiev}},\ }\bibfield  {title} {\enquote {\bibinfo {title}
  {Light\textendash{}matter interaction in the strong coupling regime:
  Configurations, conditions, and applications},}\ }\href {\doibase
  10.1039/C7NR06917K} {\ \textbf {\bibinfo {volume} {10}},\ \bibinfo {pages}
  {3589--3605} (\bibinfo {year} {2018})}\BibitemShut {NoStop}%
\bibitem [{\citenamefont {Herrera}\ and\ \citenamefont
  {Owrutsky}(2020)}]{herrera2020}%
  \BibitemOpen
  \bibfield  {author} {\bibinfo {author} {\bibfnamefont {Felipe}\ \bibnamefont
  {Herrera}}\ and\ \bibinfo {author} {\bibfnamefont {Jeffrey}\ \bibnamefont
  {Owrutsky}},\ }\bibfield  {title} {\enquote {\bibinfo {title} {Molecular
  polaritons for controlling chemistry with quantum optics},}\ }\href {\doibase
  10.1063/1.5136320} {\bibfield  {journal} {\bibinfo  {journal} {The Journal of
  Chemical Physics}\ }\textbf {\bibinfo {volume} {152}},\ \bibinfo {pages}
  {100902} (\bibinfo {year} {2020})}\BibitemShut {NoStop}%
\bibitem [{\citenamefont {Agranovich}(2009)}]{agranovich2009excitations}%
  \BibitemOpen
  \bibfield  {author} {\bibinfo {author} {\bibfnamefont {Vladimir~M}\
  \bibnamefont {Agranovich}},\ }\href@noop {} {\emph {\bibinfo {title}
  {Excitations in Organic Solids}}},\ Vol.\ \bibinfo {volume} {142}\ (\bibinfo
  {publisher} {{OUP Oxford}},\ \bibinfo {year} {2009})\BibitemShut {NoStop}%
\bibitem [{\citenamefont {Coles}\ \emph {et~al.}(2014)\citenamefont {Coles},
  \citenamefont {Somaschi}, \citenamefont {Michetti}, \citenamefont {Clark},
  \citenamefont {Lagoudakis}, \citenamefont {Savvidis},\ and\ \citenamefont
  {Lidzey}}]{coles2014b}%
  \BibitemOpen
  \bibfield  {author} {\bibinfo {author} {\bibfnamefont {David~M.}\
  \bibnamefont {Coles}}, \bibinfo {author} {\bibfnamefont {Niccolo}\
  \bibnamefont {Somaschi}}, \bibinfo {author} {\bibfnamefont {Paolo}\
  \bibnamefont {Michetti}}, \bibinfo {author} {\bibfnamefont {Caspar}\
  \bibnamefont {Clark}}, \bibinfo {author} {\bibfnamefont {Pavlos~G.}\
  \bibnamefont {Lagoudakis}}, \bibinfo {author} {\bibfnamefont {Pavlos~G.}\
  \bibnamefont {Savvidis}}, \ and\ \bibinfo {author} {\bibfnamefont {David~G.}\
  \bibnamefont {Lidzey}},\ }\bibfield  {title} {\enquote {\bibinfo {title}
  {Polariton-mediated energy transfer between organic dyes in a strongly
  coupled optical microcavity},}\ }\href {\doibase 10.1038/nmat3950} {\bibfield
   {journal} {\bibinfo  {journal} {Nature Materials}\ }\textbf {\bibinfo
  {volume} {13}},\ \bibinfo {pages} {712} (\bibinfo {year} {2014})}\BibitemShut
  {NoStop}%
\bibitem [{\citenamefont {Zhong}\ \emph {et~al.}(2017)\citenamefont {Zhong},
  \citenamefont {Chervy}, \citenamefont {Zhang}, \citenamefont {Thomas},
  \citenamefont {George}, \citenamefont {Genet}, \citenamefont {Hutchison},\
  and\ \citenamefont {Ebbesen}}]{zhong2017}%
  \BibitemOpen
  \bibfield  {author} {\bibinfo {author} {\bibfnamefont {Xiaolan}\ \bibnamefont
  {Zhong}}, \bibinfo {author} {\bibfnamefont {Thibault}\ \bibnamefont
  {Chervy}}, \bibinfo {author} {\bibfnamefont {Lei}\ \bibnamefont {Zhang}},
  \bibinfo {author} {\bibfnamefont {Anoop}\ \bibnamefont {Thomas}}, \bibinfo
  {author} {\bibfnamefont {Jino}\ \bibnamefont {George}}, \bibinfo {author}
  {\bibfnamefont {Cyriaque}\ \bibnamefont {Genet}}, \bibinfo {author}
  {\bibfnamefont {James~A.}\ \bibnamefont {Hutchison}}, \ and\ \bibinfo
  {author} {\bibfnamefont {Thomas~W.}\ \bibnamefont {Ebbesen}},\ }\bibfield
  {title} {\enquote {\bibinfo {title} {Energy {{Transfer}} between {{Spatially
  Separated Entangled Molecules}}},}\ }\href {\doibase 10.1002/anie.201703539}
  {\bibfield  {journal} {\bibinfo  {journal} {Angewandte Chemie International
  Edition}\ }\textbf {\bibinfo {volume} {56}},\ \bibinfo {pages} {9034--9038}
  (\bibinfo {year} {2017})}\BibitemShut {NoStop}%
\bibitem [{\citenamefont {Du}\ \emph {et~al.}(2018)\citenamefont {Du},
  \citenamefont {{A. Mart{\'i}nez-Mart{\'i}nez}}, \citenamefont {F.~Ribeiro},
  \citenamefont {Hu}, \citenamefont {M.~Menon},\ and\ \citenamefont
  {{Yuen-Zhou}}}]{du2018}%
  \BibitemOpen
  \bibfield  {author} {\bibinfo {author} {\bibfnamefont {Matthew}\ \bibnamefont
  {Du}}, \bibinfo {author} {\bibfnamefont {Luis}\ \bibnamefont {{A.
  Mart{\'i}nez-Mart{\'i}nez}}}, \bibinfo {author} {\bibfnamefont {Raphael}\
  \bibnamefont {F.~Ribeiro}}, \bibinfo {author} {\bibfnamefont {Zixuan}\
  \bibnamefont {Hu}}, \bibinfo {author} {\bibfnamefont {Vinod}\ \bibnamefont
  {M.~Menon}}, \ and\ \bibinfo {author} {\bibfnamefont {Joel}\ \bibnamefont
  {{Yuen-Zhou}}},\ }\bibfield  {title} {\enquote {\bibinfo {title} {Theory for
  polariton-assisted remote energy transfer},}\ }\href {\doibase
  10.1039/C8SC00171E} {\bibfield  {journal} {\bibinfo  {journal} {Chemical
  Science}\ }\textbf {\bibinfo {volume} {9}},\ \bibinfo {pages} {6659--6669}
  (\bibinfo {year} {2018})}\BibitemShut {NoStop}%
\bibitem [{\citenamefont {Orgiu}\ \emph {et~al.}(2015)\citenamefont {Orgiu},
  \citenamefont {George}, \citenamefont {Hutchison}, \citenamefont {Devaux},
  \citenamefont {Dayen}, \citenamefont {Doudin}, \citenamefont {Stellacci},
  \citenamefont {Genet}, \citenamefont {Schachenmayer}, \citenamefont {Genes},
  \citenamefont {Pupillo}, \citenamefont {Samor{\`i}},\ and\ \citenamefont
  {Ebbesen}}]{orgiu2015}%
  \BibitemOpen
  \bibfield  {author} {\bibinfo {author} {\bibfnamefont {E.}~\bibnamefont
  {Orgiu}}, \bibinfo {author} {\bibfnamefont {J.}~\bibnamefont {George}},
  \bibinfo {author} {\bibfnamefont {J.~A.}\ \bibnamefont {Hutchison}}, \bibinfo
  {author} {\bibfnamefont {E.}~\bibnamefont {Devaux}}, \bibinfo {author}
  {\bibfnamefont {J.~F.}\ \bibnamefont {Dayen}}, \bibinfo {author}
  {\bibfnamefont {B.}~\bibnamefont {Doudin}}, \bibinfo {author} {\bibfnamefont
  {F.}~\bibnamefont {Stellacci}}, \bibinfo {author} {\bibfnamefont
  {C.}~\bibnamefont {Genet}}, \bibinfo {author} {\bibfnamefont
  {J.}~\bibnamefont {Schachenmayer}}, \bibinfo {author} {\bibfnamefont
  {C.}~\bibnamefont {Genes}}, \bibinfo {author} {\bibfnamefont
  {G.}~\bibnamefont {Pupillo}}, \bibinfo {author} {\bibfnamefont
  {P.}~\bibnamefont {Samor{\`i}}}, \ and\ \bibinfo {author} {\bibfnamefont
  {T.~W.}\ \bibnamefont {Ebbesen}},\ }\bibfield  {title} {\enquote {\bibinfo
  {title} {Conductivity in organic semiconductors hybridized with the vacuum
  field},}\ }\href {\doibase 10.1038/nmat4392} {\bibfield  {journal} {\bibinfo
  {journal} {Nature Materials}\ }\textbf {\bibinfo {volume} {14}},\ \bibinfo
  {pages} {1123--1129} (\bibinfo {year} {2015})}\BibitemShut {NoStop}%
\bibitem [{\citenamefont {Feist}\ and\ \citenamefont
  {{Garcia-Vidal}}(2015)}]{feist_extraordinary_2015}%
  \BibitemOpen
  \bibfield  {author} {\bibinfo {author} {\bibfnamefont {Johannes}\
  \bibnamefont {Feist}}\ and\ \bibinfo {author} {\bibfnamefont {Francisco~J.}\
  \bibnamefont {{Garcia-Vidal}}},\ }\bibfield  {title} {\enquote {\bibinfo
  {title} {Extraordinary {{Exciton Conductance Induced}} by {{Strong
  Coupling}}},}\ }\href {\doibase 10.1103/PhysRevLett.114.196402} {\bibfield
  {journal} {\bibinfo  {journal} {Physical Review Letters}\ }\textbf {\bibinfo
  {volume} {114}},\ \bibinfo {pages} {196402} (\bibinfo {year}
  {2015})}\BibitemShut {NoStop}%
\bibitem [{\citenamefont {Hagenm{\"u}ller}\ \emph {et~al.}(2017)\citenamefont
  {Hagenm{\"u}ller}, \citenamefont {Schachenmayer}, \citenamefont {Sch{\"u}tz},
  \citenamefont {Genes},\ and\ \citenamefont {Pupillo}}]{hagenmuller2017a}%
  \BibitemOpen
  \bibfield  {author} {\bibinfo {author} {\bibfnamefont {David}\ \bibnamefont
  {Hagenm{\"u}ller}}, \bibinfo {author} {\bibfnamefont {Johannes}\ \bibnamefont
  {Schachenmayer}}, \bibinfo {author} {\bibfnamefont {Stefan}\ \bibnamefont
  {Sch{\"u}tz}}, \bibinfo {author} {\bibfnamefont {Claudiu}\ \bibnamefont
  {Genes}}, \ and\ \bibinfo {author} {\bibfnamefont {Guido}\ \bibnamefont
  {Pupillo}},\ }\bibfield  {title} {\enquote {\bibinfo {title}
  {Cavity-{{Enhanced Transport}} of {{Charge}}},}\ }\href {\doibase
  10.1103/PhysRevLett.119.223601} {\bibfield  {journal} {\bibinfo  {journal}
  {Physical Review Letters}\ }\textbf {\bibinfo {volume} {119}},\ \bibinfo
  {pages} {223601} (\bibinfo {year} {2017})}\BibitemShut {NoStop}%
\bibitem [{\citenamefont {Thomas}\ \emph {et~al.}(2016)\citenamefont {Thomas},
  \citenamefont {George}, \citenamefont {Shalabney}, \citenamefont {Dryzhakov},
  \citenamefont {Varma}, \citenamefont {Moran}, \citenamefont {Chervy},
  \citenamefont {Zhong}, \citenamefont {Devaux}, \citenamefont {Genet},
  \citenamefont {Hutchison},\ and\ \citenamefont {Ebbesen}}]{thomas2016}%
  \BibitemOpen
  \bibfield  {author} {\bibinfo {author} {\bibfnamefont {Anoop}\ \bibnamefont
  {Thomas}}, \bibinfo {author} {\bibfnamefont {Jino}\ \bibnamefont {George}},
  \bibinfo {author} {\bibfnamefont {Atef}\ \bibnamefont {Shalabney}}, \bibinfo
  {author} {\bibfnamefont {Marian}\ \bibnamefont {Dryzhakov}}, \bibinfo
  {author} {\bibfnamefont {Sreejith~J.}\ \bibnamefont {Varma}}, \bibinfo
  {author} {\bibfnamefont {Joseph}\ \bibnamefont {Moran}}, \bibinfo {author}
  {\bibfnamefont {Thibault}\ \bibnamefont {Chervy}}, \bibinfo {author}
  {\bibfnamefont {Xiaolan}\ \bibnamefont {Zhong}}, \bibinfo {author}
  {\bibfnamefont {Elo{\"i}se}\ \bibnamefont {Devaux}}, \bibinfo {author}
  {\bibfnamefont {Cyriaque}\ \bibnamefont {Genet}}, \bibinfo {author}
  {\bibfnamefont {James~A.}\ \bibnamefont {Hutchison}}, \ and\ \bibinfo
  {author} {\bibfnamefont {Thomas~W.}\ \bibnamefont {Ebbesen}},\ }\bibfield
  {title} {\enquote {\bibinfo {title} {Ground-{{State Chemical Reactivity}}
  under {{Vibrational Coupling}} to the {{Vacuum Electromagnetic Field}}},}\
  }\href {\doibase 10.1002/anie.201605504} {\bibfield  {journal} {\bibinfo
  {journal} {Angewandte Chemie International Edition}\ }\textbf {\bibinfo
  {volume} {55}},\ \bibinfo {pages} {11462--11466} (\bibinfo {year}
  {2016})}\BibitemShut {NoStop}%
\bibitem [{\citenamefont {Thomas}\ \emph {et~al.}(2019)\citenamefont {Thomas},
  \citenamefont {{Lethuillier-Karl}}, \citenamefont {Nagarajan}, \citenamefont
  {Vergauwe}, \citenamefont {George}, \citenamefont {Chervy}, \citenamefont
  {Shalabney}, \citenamefont {Devaux}, \citenamefont {Genet}, \citenamefont
  {Moran},\ and\ \citenamefont {Ebbesen}}]{thomas2019}%
  \BibitemOpen
  \bibfield  {author} {\bibinfo {author} {\bibfnamefont {A.}~\bibnamefont
  {Thomas}}, \bibinfo {author} {\bibfnamefont {L.}~\bibnamefont
  {{Lethuillier-Karl}}}, \bibinfo {author} {\bibfnamefont {K.}~\bibnamefont
  {Nagarajan}}, \bibinfo {author} {\bibfnamefont {R.~M.~A.}\ \bibnamefont
  {Vergauwe}}, \bibinfo {author} {\bibfnamefont {J.}~\bibnamefont {George}},
  \bibinfo {author} {\bibfnamefont {T.}~\bibnamefont {Chervy}}, \bibinfo
  {author} {\bibfnamefont {A.}~\bibnamefont {Shalabney}}, \bibinfo {author}
  {\bibfnamefont {E.}~\bibnamefont {Devaux}}, \bibinfo {author} {\bibfnamefont
  {C.}~\bibnamefont {Genet}}, \bibinfo {author} {\bibfnamefont
  {J.}~\bibnamefont {Moran}}, \ and\ \bibinfo {author} {\bibfnamefont {T.~W.}\
  \bibnamefont {Ebbesen}},\ }\bibfield  {title} {\enquote {\bibinfo {title}
  {Tilting a ground-state reactivity landscape by vibrational strong
  coupling},}\ }\href {\doibase 10.1126/science.aau7742} {\bibfield  {journal}
  {\bibinfo  {journal} {Science}\ }\textbf {\bibinfo {volume} {363}},\ \bibinfo
  {pages} {615--619} (\bibinfo {year} {2019})}\BibitemShut {NoStop}%
\bibitem [{\citenamefont {{Campos-Gonzalez-Angulo}}\ \emph
  {et~al.}(2019)\citenamefont {{Campos-Gonzalez-Angulo}}, \citenamefont
  {Ribeiro},\ and\ \citenamefont {{Yuen-Zhou}}}]{campos-gonzalez-angulo2019a}%
  \BibitemOpen
  \bibfield  {author} {\bibinfo {author} {\bibfnamefont {Jorge~A.}\
  \bibnamefont {{Campos-Gonzalez-Angulo}}}, \bibinfo {author} {\bibfnamefont
  {Raphael~F.}\ \bibnamefont {Ribeiro}}, \ and\ \bibinfo {author}
  {\bibfnamefont {Joel}\ \bibnamefont {{Yuen-Zhou}}},\ }\bibfield  {title}
  {\enquote {\bibinfo {title} {Resonant catalysis of thermally activated
  chemical reactions with vibrational polaritons},}\ }\href {\doibase
  10.1038/s41467-019-12636-1} {\bibfield  {journal} {\bibinfo  {journal}
  {Nature Communications}\ }\textbf {\bibinfo {volume} {10}},\ \bibinfo {pages}
  {1--8} (\bibinfo {year} {2019})}\BibitemShut {NoStop}%
\bibitem [{\citenamefont {Li}\ \emph {et~al.}(2020)\citenamefont {Li},
  \citenamefont {Nitzan},\ and\ \citenamefont {Subotnik}}]{li2020}%
  \BibitemOpen
  \bibfield  {author} {\bibinfo {author} {\bibfnamefont {Tao~E.}\ \bibnamefont
  {Li}}, \bibinfo {author} {\bibfnamefont {Abraham}\ \bibnamefont {Nitzan}}, \
  and\ \bibinfo {author} {\bibfnamefont {Joseph~E.}\ \bibnamefont {Subotnik}},\
  }\bibfield  {title} {\enquote {\bibinfo {title} {On the {{Origin}} of
  {{Ground}}-{{State Vacuum}}-{{Field Catalysis}}: {{Equilibrium
  Consideration}}},}\ }\href@noop {} {\bibfield  {journal} {\bibinfo  {journal}
  {arXiv:2002.09977 [physics]}\ } (\bibinfo {year} {2020})},\ \Eprint
  {http://arxiv.org/abs/2002.09977} {arXiv:2002.09977 [physics]} \BibitemShut
  {NoStop}%
\bibitem [{\citenamefont {Fofang}\ \emph {et~al.}(2011)\citenamefont {Fofang},
  \citenamefont {Grady}, \citenamefont {Fan}, \citenamefont {Govorov},\ and\
  \citenamefont {Halas}}]{fofang2011}%
  \BibitemOpen
  \bibfield  {author} {\bibinfo {author} {\bibfnamefont {Nche~T.}\ \bibnamefont
  {Fofang}}, \bibinfo {author} {\bibfnamefont {Nathaniel~K.}\ \bibnamefont
  {Grady}}, \bibinfo {author} {\bibfnamefont {Zhiyuan}\ \bibnamefont {Fan}},
  \bibinfo {author} {\bibfnamefont {Alexander~O.}\ \bibnamefont {Govorov}}, \
  and\ \bibinfo {author} {\bibfnamefont {Naomi~J.}\ \bibnamefont {Halas}},\
  }\bibfield  {title} {\enquote {\bibinfo {title} {Plexciton {{Dynamics}}:
  {{Exciton}}-{{Plasmon Coupling}} in a {{J}}-{{Aggregate}}-{{Au Nanoshell
  Complex Provides}} a {{Mechanism}} for {{Nonlinearity}}},}\ }\href {\doibase
  10.1021/nl104352j} {\bibfield  {journal} {\bibinfo  {journal} {Nano Letters}\
  }\textbf {\bibinfo {volume} {11}},\ \bibinfo {pages} {1556--1560} (\bibinfo
  {year} {2011})}\BibitemShut {NoStop}%
\bibitem [{\citenamefont {Virgili}\ \emph {et~al.}(2011)\citenamefont
  {Virgili}, \citenamefont {Coles}, \citenamefont {Adawi}, \citenamefont
  {Clark}, \citenamefont {Michetti}, \citenamefont {Rajendran}, \citenamefont
  {Brida}, \citenamefont {Polli}, \citenamefont {Cerullo},\ and\ \citenamefont
  {Lidzey}}]{virgili2011}%
  \BibitemOpen
  \bibfield  {author} {\bibinfo {author} {\bibfnamefont {T.}~\bibnamefont
  {Virgili}}, \bibinfo {author} {\bibfnamefont {D.}~\bibnamefont {Coles}},
  \bibinfo {author} {\bibfnamefont {A.~M.}\ \bibnamefont {Adawi}}, \bibinfo
  {author} {\bibfnamefont {C.}~\bibnamefont {Clark}}, \bibinfo {author}
  {\bibfnamefont {P.}~\bibnamefont {Michetti}}, \bibinfo {author}
  {\bibfnamefont {S.~K.}\ \bibnamefont {Rajendran}}, \bibinfo {author}
  {\bibfnamefont {D.}~\bibnamefont {Brida}}, \bibinfo {author} {\bibfnamefont
  {D.}~\bibnamefont {Polli}}, \bibinfo {author} {\bibfnamefont
  {G.}~\bibnamefont {Cerullo}}, \ and\ \bibinfo {author} {\bibfnamefont
  {D.~G.}\ \bibnamefont {Lidzey}},\ }\bibfield  {title} {\enquote {\bibinfo
  {title} {Ultrafast polariton relaxation dynamics in an organic semiconductor
  microcavity},}\ }\href {\doibase 10.1103/PhysRevB.83.245309} {\bibfield
  {journal} {\bibinfo  {journal} {Physical Review B}\ }\textbf {\bibinfo
  {volume} {83}},\ \bibinfo {pages} {245309} (\bibinfo {year}
  {2011})}\BibitemShut {NoStop}%
\bibitem [{\citenamefont {Vasa}\ \emph {et~al.}(2013)\citenamefont {Vasa},
  \citenamefont {Wang}, \citenamefont {Pomraenke}, \citenamefont {Lammers},
  \citenamefont {Maiuri}, \citenamefont {Manzoni}, \citenamefont {Cerullo},\
  and\ \citenamefont {Lienau}}]{vasa2013}%
  \BibitemOpen
  \bibfield  {author} {\bibinfo {author} {\bibfnamefont {Parinda}\ \bibnamefont
  {Vasa}}, \bibinfo {author} {\bibfnamefont {Wei}\ \bibnamefont {Wang}},
  \bibinfo {author} {\bibfnamefont {Robert}\ \bibnamefont {Pomraenke}},
  \bibinfo {author} {\bibfnamefont {Melanie}\ \bibnamefont {Lammers}}, \bibinfo
  {author} {\bibfnamefont {Margherita}\ \bibnamefont {Maiuri}}, \bibinfo
  {author} {\bibfnamefont {Cristian}\ \bibnamefont {Manzoni}}, \bibinfo
  {author} {\bibfnamefont {Giulio}\ \bibnamefont {Cerullo}}, \ and\ \bibinfo
  {author} {\bibfnamefont {Christoph}\ \bibnamefont {Lienau}},\ }\bibfield
  {title} {\enquote {\bibinfo {title} {Real-time observation of ultrafast
  {{Rabi}} oscillations between excitons and plasmons in metal nanostructures
  with {{J}}-aggregates},}\ }\href {\doibase 10.1038/nphoton.2012.340}
  {\bibfield  {journal} {\bibinfo  {journal} {Nature Photonics}\ }\textbf
  {\bibinfo {volume} {7}},\ \bibinfo {pages} {128--132} (\bibinfo {year}
  {2013})}\BibitemShut {NoStop}%
\bibitem [{\citenamefont {Chervy}\ \emph {et~al.}(2016)\citenamefont {Chervy},
  \citenamefont {Xu}, \citenamefont {Duan}, \citenamefont {Wang}, \citenamefont
  {Mager}, \citenamefont {Frerejean}, \citenamefont {M{\"u}nninghoff},
  \citenamefont {Tinnemans}, \citenamefont {Hutchison}, \citenamefont {Genet},
  \citenamefont {Rowan}, \citenamefont {Rasing},\ and\ \citenamefont
  {Ebbesen}}]{chervy2016}%
  \BibitemOpen
  \bibfield  {author} {\bibinfo {author} {\bibfnamefont {Thibault}\
  \bibnamefont {Chervy}}, \bibinfo {author} {\bibfnamefont {Jialiang}\
  \bibnamefont {Xu}}, \bibinfo {author} {\bibfnamefont {Yulong}\ \bibnamefont
  {Duan}}, \bibinfo {author} {\bibfnamefont {Chunliang}\ \bibnamefont {Wang}},
  \bibinfo {author} {\bibfnamefont {Lo{\"i}c}\ \bibnamefont {Mager}}, \bibinfo
  {author} {\bibfnamefont {Maurice}\ \bibnamefont {Frerejean}}, \bibinfo
  {author} {\bibfnamefont {Joris A.~W.}\ \bibnamefont {M{\"u}nninghoff}},
  \bibinfo {author} {\bibfnamefont {Paul}\ \bibnamefont {Tinnemans}}, \bibinfo
  {author} {\bibfnamefont {James~A.}\ \bibnamefont {Hutchison}}, \bibinfo
  {author} {\bibfnamefont {Cyriaque}\ \bibnamefont {Genet}}, \bibinfo {author}
  {\bibfnamefont {Alan~E.}\ \bibnamefont {Rowan}}, \bibinfo {author}
  {\bibfnamefont {Theo}\ \bibnamefont {Rasing}}, \ and\ \bibinfo {author}
  {\bibfnamefont {Thomas~W.}\ \bibnamefont {Ebbesen}},\ }\bibfield  {title}
  {\enquote {\bibinfo {title} {High-{{Efficiency Second}}-{{Harmonic
  Generation}} from {{Hybrid Light}}-{{Matter States}}},}\ }\href {\doibase
  10.1021/acs.nanolett.6b02567} {\bibfield  {journal} {\bibinfo  {journal}
  {Nano Letters}\ }\textbf {\bibinfo {volume} {16}},\ \bibinfo {pages}
  {7352--7356} (\bibinfo {year} {2016})}\BibitemShut {NoStop}%
\bibitem [{\citenamefont {Dunkelberger}\ \emph {et~al.}(2016)\citenamefont
  {Dunkelberger}, \citenamefont {Spann}, \citenamefont {Fears}, \citenamefont
  {Simpkins},\ and\ \citenamefont {Owrutsky}}]{dunkelberger_modified_2016}%
  \BibitemOpen
  \bibfield  {author} {\bibinfo {author} {\bibfnamefont {A.~D.}\ \bibnamefont
  {Dunkelberger}}, \bibinfo {author} {\bibfnamefont {B.~T.}\ \bibnamefont
  {Spann}}, \bibinfo {author} {\bibfnamefont {K.~P.}\ \bibnamefont {Fears}},
  \bibinfo {author} {\bibfnamefont {B.~S.}\ \bibnamefont {Simpkins}}, \ and\
  \bibinfo {author} {\bibfnamefont {J.~C.}\ \bibnamefont {Owrutsky}},\
  }\bibfield  {title} {\enquote {\bibinfo {title} {Modified relaxation dynamics
  and coherent energy exchange in coupled vibration-cavity polaritons},}\
  }\href {\doibase 10.1038/ncomms13504} {\bibfield  {journal} {\bibinfo
  {journal} {Nature Communications}\ }\textbf {\bibinfo {volume} {7}} (\bibinfo
  {year} {2016}),\ 10.1038/ncomms13504}\BibitemShut {NoStop}%
\bibitem [{\citenamefont {Xiang}\ \emph {et~al.}(2018)\citenamefont {Xiang},
  \citenamefont {Ribeiro}, \citenamefont {Dunkelberger}, \citenamefont {Wang},
  \citenamefont {Li}, \citenamefont {Simpkins}, \citenamefont {Owrutsky},
  \citenamefont {{Yuen-Zhou}},\ and\ \citenamefont {Xiong}}]{xiang2018}%
  \BibitemOpen
  \bibfield  {author} {\bibinfo {author} {\bibfnamefont {Bo}~\bibnamefont
  {Xiang}}, \bibinfo {author} {\bibfnamefont {Raphael~F.}\ \bibnamefont
  {Ribeiro}}, \bibinfo {author} {\bibfnamefont {Adam~D.}\ \bibnamefont
  {Dunkelberger}}, \bibinfo {author} {\bibfnamefont {Jiaxi}\ \bibnamefont
  {Wang}}, \bibinfo {author} {\bibfnamefont {Yingmin}\ \bibnamefont {Li}},
  \bibinfo {author} {\bibfnamefont {Blake~S.}\ \bibnamefont {Simpkins}},
  \bibinfo {author} {\bibfnamefont {Jeffrey~C.}\ \bibnamefont {Owrutsky}},
  \bibinfo {author} {\bibfnamefont {Joel}\ \bibnamefont {{Yuen-Zhou}}}, \ and\
  \bibinfo {author} {\bibfnamefont {Wei}\ \bibnamefont {Xiong}},\ }\bibfield
  {title} {\enquote {\bibinfo {title} {Two-dimensional infrared spectroscopy of
  vibrational polaritons},}\ }\href {\doibase 10.1073/pnas.1722063115}
  {\bibfield  {journal} {\bibinfo  {journal} {Proceedings of the National
  Academy of Sciences}\ ,\ \bibinfo {pages} {201722063}} (\bibinfo {year}
  {2018})}\BibitemShut {NoStop}%
\bibitem [{\citenamefont {Dunkelberger}\ \emph {et~al.}(2018)\citenamefont
  {Dunkelberger}, \citenamefont {Davidson}, \citenamefont {Ahn}, \citenamefont
  {Simpkins},\ and\ \citenamefont {Owrutsky}}]{dunkelberger2018a}%
  \BibitemOpen
  \bibfield  {author} {\bibinfo {author} {\bibfnamefont {Adam~D.}\ \bibnamefont
  {Dunkelberger}}, \bibinfo {author} {\bibfnamefont {Roderick~B.}\ \bibnamefont
  {Davidson}}, \bibinfo {author} {\bibfnamefont {Wonmi}\ \bibnamefont {Ahn}},
  \bibinfo {author} {\bibfnamefont {Blake~S.}\ \bibnamefont {Simpkins}}, \ and\
  \bibinfo {author} {\bibfnamefont {Jeffrey~C.}\ \bibnamefont {Owrutsky}},\
  }\bibfield  {title} {\enquote {\bibinfo {title} {Ultrafast {{Transmission
  Modulation}} and {{Recovery}} via {{Vibrational Strong Coupling}}},}\ }\href
  {\doibase 10.1021/acs.jpca.7b10299} {\bibfield  {journal} {\bibinfo
  {journal} {The Journal of Physical Chemistry A}\ }\textbf {\bibinfo {volume}
  {122}},\ \bibinfo {pages} {965--971} (\bibinfo {year} {2018})}\BibitemShut
  {NoStop}%
\bibitem [{\citenamefont {Avramenko}\ and\ \citenamefont
  {Rury}(2020)}]{avramenko2020}%
  \BibitemOpen
  \bibfield  {author} {\bibinfo {author} {\bibfnamefont {Aleksandr~G.}\
  \bibnamefont {Avramenko}}\ and\ \bibinfo {author} {\bibfnamefont {Aaron~S.}\
  \bibnamefont {Rury}},\ }\bibfield  {title} {\enquote {\bibinfo {title}
  {Quantum {{Control}} of {{Ultrafast Internal Conversion Using Nanoconfined
  Virtual Photons}}},}\ }\href {\doibase 10.1021/acs.jpclett.9b03447}
  {\bibfield  {journal} {\bibinfo  {journal} {The Journal of Physical Chemistry
  Letters}\ }\textbf {\bibinfo {volume} {11}},\ \bibinfo {pages} {1013--1021}
  (\bibinfo {year} {2020})}\BibitemShut {NoStop}%
\bibitem [{\citenamefont {DelPo}\ \emph {et~al.}(2020)\citenamefont {DelPo},
  \citenamefont {Kudisch}, \citenamefont {Park}, \citenamefont {Khan},
  \citenamefont {Fassioli}, \citenamefont {Fausti}, \citenamefont {Rand},\ and\
  \citenamefont {Scholes}}]{delpo2020}%
  \BibitemOpen
  \bibfield  {author} {\bibinfo {author} {\bibfnamefont {Courtney~A.}\
  \bibnamefont {DelPo}}, \bibinfo {author} {\bibfnamefont {Bryan}\ \bibnamefont
  {Kudisch}}, \bibinfo {author} {\bibfnamefont {Kyu~Hyung}\ \bibnamefont
  {Park}}, \bibinfo {author} {\bibfnamefont {Saeed-Uz-Zaman}\ \bibnamefont
  {Khan}}, \bibinfo {author} {\bibfnamefont {Francesca}\ \bibnamefont
  {Fassioli}}, \bibinfo {author} {\bibfnamefont {Daniele}\ \bibnamefont
  {Fausti}}, \bibinfo {author} {\bibfnamefont {Barry~P.}\ \bibnamefont {Rand}},
  \ and\ \bibinfo {author} {\bibfnamefont {Gregory~D.}\ \bibnamefont
  {Scholes}},\ }\bibfield  {title} {\enquote {\bibinfo {title} {Polariton
  {{Transitions}} in {{Femtosecond Transient Absorption Studies}} of
  {{Ultrastrong Light}}\textendash{{Molecule Coupling}}},}\ }\href {\doibase
  10.1021/acs.jpclett.0c00247} {\bibfield  {journal} {\bibinfo  {journal} {The
  Journal of Physical Chemistry Letters}\ }\textbf {\bibinfo {volume} {11}},\
  \bibinfo {pages} {2667--2674} (\bibinfo {year} {2020})}\BibitemShut {NoStop}%
\bibitem [{\citenamefont {{Finkelstein-Shapiro}}\ \emph
  {et~al.}(2020)\citenamefont {{Finkelstein-Shapiro}}, \citenamefont {Mante},
  \citenamefont {Sarisozen}, \citenamefont {Wittenbecher}, \citenamefont
  {Minda}, \citenamefont {Balci}, \citenamefont {Pullerits},\ and\
  \citenamefont {Zigmantas}}]{finkelstein-shapiro2020}%
  \BibitemOpen
  \bibfield  {author} {\bibinfo {author} {\bibfnamefont {Daniel}\ \bibnamefont
  {{Finkelstein-Shapiro}}}, \bibinfo {author} {\bibfnamefont {Pierre-Adrien}\
  \bibnamefont {Mante}}, \bibinfo {author} {\bibfnamefont {Sema}\ \bibnamefont
  {Sarisozen}}, \bibinfo {author} {\bibfnamefont {Lukas}\ \bibnamefont
  {Wittenbecher}}, \bibinfo {author} {\bibfnamefont {Iulia}\ \bibnamefont
  {Minda}}, \bibinfo {author} {\bibfnamefont {Sinan}\ \bibnamefont {Balci}},
  \bibinfo {author} {\bibfnamefont {Tonu}\ \bibnamefont {Pullerits}}, \ and\
  \bibinfo {author} {\bibfnamefont {Donatas}\ \bibnamefont {Zigmantas}},\
  }\bibfield  {title} {\enquote {\bibinfo {title} {Radiative {{Transitions}}
  and {{Relaxation Pathways}} in {{Plasmon}}-{{Based Cavity Quantum
  Electrodynamics Systems}}},}\ }\href@noop {} {\bibfield  {journal} {\bibinfo
  {journal} {arXiv:2002.05642 [physics, physics:quant-ph]}\ } (\bibinfo {year}
  {2020})},\ \Eprint {http://arxiv.org/abs/2002.05642} {arXiv:2002.05642
  [physics, physics:quant-ph]} \BibitemShut {NoStop}%
\bibitem [{\citenamefont {Xiang}\ \emph
  {et~al.}(2019{\natexlab{a}})\citenamefont {Xiang}, \citenamefont {Ribeiro},
  \citenamefont {Chen}, \citenamefont {Wang}, \citenamefont {Du}, \citenamefont
  {{Yuen-Zhou}},\ and\ \citenamefont {Xiong}}]{xiang2019a}%
  \BibitemOpen
  \bibfield  {author} {\bibinfo {author} {\bibfnamefont {Bo}~\bibnamefont
  {Xiang}}, \bibinfo {author} {\bibfnamefont {Raphael~F.}\ \bibnamefont
  {Ribeiro}}, \bibinfo {author} {\bibfnamefont {Liying}\ \bibnamefont {Chen}},
  \bibinfo {author} {\bibfnamefont {Jiaxi}\ \bibnamefont {Wang}}, \bibinfo
  {author} {\bibfnamefont {Matthew}\ \bibnamefont {Du}}, \bibinfo {author}
  {\bibfnamefont {Joel}\ \bibnamefont {{Yuen-Zhou}}}, \ and\ \bibinfo {author}
  {\bibfnamefont {Wei}\ \bibnamefont {Xiong}},\ }\bibfield  {title} {\enquote
  {\bibinfo {title} {State-{{Selective Polariton}} to {{Dark State Relaxation
  Dynamics}}},}\ }\href {\doibase 10.1021/acs.jpca.9b04601} {\bibfield
  {journal} {\bibinfo  {journal} {The Journal of Physical Chemistry A}\
  }\textbf {\bibinfo {volume} {123}},\ \bibinfo {pages} {5918--5927} (\bibinfo
  {year} {2019}{\natexlab{a}})}\BibitemShut {NoStop}%
\bibitem [{\citenamefont {Ribeiro}\ \emph
  {et~al.}(2018{\natexlab{b}})\citenamefont {Ribeiro}, \citenamefont
  {Dunkelberger}, \citenamefont {Xiang}, \citenamefont {Xiong}, \citenamefont
  {Simpkins}, \citenamefont {Owrutsky},\ and\ \citenamefont
  {{Yuen-Zhou}}}]{ribeiro2018c}%
  \BibitemOpen
  \bibfield  {author} {\bibinfo {author} {\bibfnamefont {Raphael~F.}\
  \bibnamefont {Ribeiro}}, \bibinfo {author} {\bibfnamefont {Adam~D.}\
  \bibnamefont {Dunkelberger}}, \bibinfo {author} {\bibfnamefont
  {Bo}~\bibnamefont {Xiang}}, \bibinfo {author} {\bibfnamefont {Wei}\
  \bibnamefont {Xiong}}, \bibinfo {author} {\bibfnamefont {Blake~S.}\
  \bibnamefont {Simpkins}}, \bibinfo {author} {\bibfnamefont {Jeffrey~C.}\
  \bibnamefont {Owrutsky}}, \ and\ \bibinfo {author} {\bibfnamefont {Joel}\
  \bibnamefont {{Yuen-Zhou}}},\ }\bibfield  {title} {\enquote {\bibinfo {title}
  {Theory for {{Nonlinear Spectroscopy}} of {{Vibrational Polaritons}}},}\
  }\href {\doibase 10.1021/acs.jpclett.8b01176} {\bibfield  {journal} {\bibinfo
   {journal} {The Journal of Physical Chemistry Letters}\ }\textbf {\bibinfo
  {volume} {9}},\ \bibinfo {pages} {3766--3771} (\bibinfo {year}
  {2018}{\natexlab{b}})}\BibitemShut {NoStop}%
\bibitem [{\citenamefont {{P{\'e}rez-S{\'a}nchez}}\ and\ \citenamefont
  {{Yuen-Zhou}}(2020)}]{perez-sanchez2020}%
  \BibitemOpen
  \bibfield  {author} {\bibinfo {author} {\bibfnamefont {Juan~B.}\ \bibnamefont
  {{P{\'e}rez-S{\'a}nchez}}}\ and\ \bibinfo {author} {\bibfnamefont {Joel}\
  \bibnamefont {{Yuen-Zhou}}},\ }\bibfield  {title} {\enquote {\bibinfo {title}
  {Polariton {{Assisted Down}}-{{Conversion}} of {{Photons}} via {{Nonadiabatic
  Molecular Dynamics}}: {{A Molecular Dynamical Casimir Effect}}},}\ }\href
  {\doibase 10.1021/acs.jpclett.9b02870} {\bibfield  {journal} {\bibinfo
  {journal} {The Journal of Physical Chemistry Letters}\ }\textbf {\bibinfo
  {volume} {11}},\ \bibinfo {pages} {152--159} (\bibinfo {year}
  {2020})}\BibitemShut {NoStop}%
\bibitem [{\citenamefont {Hern{\'a}ndez}\ and\ \citenamefont
  {Herrera}(2019)}]{hernandez2019}%
  \BibitemOpen
  \bibfield  {author} {\bibinfo {author} {\bibfnamefont {Federico~J.}\
  \bibnamefont {Hern{\'a}ndez}}\ and\ \bibinfo {author} {\bibfnamefont
  {Felipe}\ \bibnamefont {Herrera}},\ }\bibfield  {title} {\enquote {\bibinfo
  {title} {Multi-level quantum {{Rabi}} model for anharmonic vibrational
  polaritons},}\ }\href {\doibase 10.1063/1.5121426} {\bibfield  {journal}
  {\bibinfo  {journal} {The Journal of Chemical Physics}\ }\textbf {\bibinfo
  {volume} {151}},\ \bibinfo {pages} {144116} (\bibinfo {year}
  {2019})}\BibitemShut {NoStop}%
\bibitem [{\citenamefont {Steck}(2017{\natexlab{a}})}]{steck2017}%
  \BibitemOpen
  \bibfield  {author} {\bibinfo {author} {\bibfnamefont {Daniel~A}\
  \bibnamefont {Steck}},\ }\href@noop {} {\emph {\bibinfo {title} {Classical
  and {{Modern Optics}}}}},\ \bibinfo {edition} {revision 1.7.4}\ ed.\
  (\bibinfo  {publisher} {{available online at http://steck.us/teaching}},\
  \bibinfo {year} {2017})\BibitemShut {NoStop}%
\bibitem [{\citenamefont {Houdr{\'e}}\ \emph {et~al.}(1996)\citenamefont
  {Houdr{\'e}}, \citenamefont {Stanley},\ and\ \citenamefont
  {Ilegems}}]{houdre_vacuum-field_1996}%
  \BibitemOpen
  \bibfield  {author} {\bibinfo {author} {\bibfnamefont {R.}~\bibnamefont
  {Houdr{\'e}}}, \bibinfo {author} {\bibfnamefont {R.~P.}\ \bibnamefont
  {Stanley}}, \ and\ \bibinfo {author} {\bibfnamefont {M.}~\bibnamefont
  {Ilegems}},\ }\bibfield  {title} {\enquote {\bibinfo {title} {Vacuum-field
  {{Rabi}} splitting in the presence of inhomogeneous broadening:
  {{Resolution}} of a homogeneous linewidth in an inhomogeneously broadened
  system},}\ }\href {\doibase 10.1103/PhysRevA.53.2711} {\bibfield  {journal}
  {\bibinfo  {journal} {Physical Review A}\ }\textbf {\bibinfo {volume} {53}},\
  \bibinfo {pages} {2711--2715} (\bibinfo {year} {1996})}\BibitemShut {NoStop}%
\bibitem [{\citenamefont {George}\ \emph {et~al.}(2015)\citenamefont {George},
  \citenamefont {Shalabney}, \citenamefont {Hutchison}, \citenamefont {Genet},\
  and\ \citenamefont {Ebbesen}}]{george_liquid-phase_2015}%
  \BibitemOpen
  \bibfield  {author} {\bibinfo {author} {\bibfnamefont {Jino}\ \bibnamefont
  {George}}, \bibinfo {author} {\bibfnamefont {Atef}\ \bibnamefont
  {Shalabney}}, \bibinfo {author} {\bibfnamefont {James~A.}\ \bibnamefont
  {Hutchison}}, \bibinfo {author} {\bibfnamefont {Cyriaque}\ \bibnamefont
  {Genet}}, \ and\ \bibinfo {author} {\bibfnamefont {Thomas~W.}\ \bibnamefont
  {Ebbesen}},\ }\bibfield  {title} {\enquote {\bibinfo {title} {Liquid-{{Phase
  Vibrational Strong Coupling}}},}\ }\href {\doibase
  10.1021/acs.jpclett.5b00204} {\bibfield  {journal} {\bibinfo  {journal} {The
  Journal of Physical Chemistry Letters}\ }\textbf {\bibinfo {volume} {6}},\
  \bibinfo {pages} {1027--1031} (\bibinfo {year} {2015})}\BibitemShut {NoStop}%
\bibitem [{\citenamefont {Long}\ and\ \citenamefont
  {Simpkins}(2015)}]{long_coherent_2015}%
  \BibitemOpen
  \bibfield  {author} {\bibinfo {author} {\bibfnamefont {J.~P.}\ \bibnamefont
  {Long}}\ and\ \bibinfo {author} {\bibfnamefont {B.~S.}\ \bibnamefont
  {Simpkins}},\ }\bibfield  {title} {\enquote {\bibinfo {title} {Coherent
  {{Coupling}} between a {{Molecular Vibration}} and
  {{Fabry}}\textendash{{Perot Optical Cavity}} to {{Give Hybridized States}} in
  the {{Strong Coupling Limit}}},}\ }\href {\doibase 10.1021/ph5003347}
  {\bibfield  {journal} {\bibinfo  {journal} {ACS Photonics}\ }\textbf
  {\bibinfo {volume} {2}},\ \bibinfo {pages} {130--136} (\bibinfo {year}
  {2015})}\BibitemShut {NoStop}%
\bibitem [{\citenamefont {Casey}\ and\ \citenamefont
  {Sparks}(2016)}]{casey_vibrational_2016}%
  \BibitemOpen
  \bibfield  {author} {\bibinfo {author} {\bibfnamefont {Shaelyn~R.}\
  \bibnamefont {Casey}}\ and\ \bibinfo {author} {\bibfnamefont {Justin~R.}\
  \bibnamefont {Sparks}},\ }\bibfield  {title} {\enquote {\bibinfo {title}
  {Vibrational {{Strong Coupling}} of {{Organometallic Complexes}}},}\ }\href
  {\doibase 10.1021/acs.jpcc.6b10493} {\bibfield  {journal} {\bibinfo
  {journal} {The Journal of Physical Chemistry C}\ }\textbf {\bibinfo {volume}
  {120}},\ \bibinfo {pages} {28138--28143} (\bibinfo {year}
  {2016})}\BibitemShut {NoStop}%
\bibitem [{\citenamefont {Ciuti}\ \emph {et~al.}(2005)\citenamefont {Ciuti},
  \citenamefont {Bastard},\ and\ \citenamefont
  {Carusotto}}]{ciuti_quantum_2005}%
  \BibitemOpen
  \bibfield  {author} {\bibinfo {author} {\bibfnamefont {Cristiano}\
  \bibnamefont {Ciuti}}, \bibinfo {author} {\bibfnamefont {G{\'e}rald}\
  \bibnamefont {Bastard}}, \ and\ \bibinfo {author} {\bibfnamefont {Iacopo}\
  \bibnamefont {Carusotto}},\ }\bibfield  {title} {\enquote {\bibinfo {title}
  {Quantum vacuum properties of the intersubband cavity polariton field},}\
  }\href {\doibase 10.1103/PhysRevB.72.115303} {\bibfield  {journal} {\bibinfo
  {journal} {Physical Review B}\ }\textbf {\bibinfo {volume} {72}},\ \bibinfo
  {pages} {115303} (\bibinfo {year} {2005})}\BibitemShut {NoStop}%
\bibitem [{\citenamefont {Todorov}\ \emph {et~al.}(2010)\citenamefont
  {Todorov}, \citenamefont {Andrews}, \citenamefont {Colombelli}, \citenamefont
  {De~Liberato}, \citenamefont {Ciuti}, \citenamefont {Klang}, \citenamefont
  {Strasser},\ and\ \citenamefont {Sirtori}}]{todorov_ultrastrong_2010}%
  \BibitemOpen
  \bibfield  {author} {\bibinfo {author} {\bibfnamefont {Y.}~\bibnamefont
  {Todorov}}, \bibinfo {author} {\bibfnamefont {A.~M.}\ \bibnamefont
  {Andrews}}, \bibinfo {author} {\bibfnamefont {R.}~\bibnamefont {Colombelli}},
  \bibinfo {author} {\bibfnamefont {S.}~\bibnamefont {De~Liberato}}, \bibinfo
  {author} {\bibfnamefont {C.}~\bibnamefont {Ciuti}}, \bibinfo {author}
  {\bibfnamefont {P.}~\bibnamefont {Klang}}, \bibinfo {author} {\bibfnamefont
  {G.}~\bibnamefont {Strasser}}, \ and\ \bibinfo {author} {\bibfnamefont
  {C.}~\bibnamefont {Sirtori}},\ }\bibfield  {title} {\enquote {\bibinfo
  {title} {Ultrastrong {{Light}}-{{Matter Coupling Regime}} with {{Polariton
  Dots}}},}\ }\href {\doibase 10.1103/PhysRevLett.105.196402} {\bibfield
  {journal} {\bibinfo  {journal} {Physical Review Letters}\ }\textbf {\bibinfo
  {volume} {105}},\ \bibinfo {pages} {196402} (\bibinfo {year}
  {2010})}\BibitemShut {NoStop}%
\bibitem [{\citenamefont {Strashko}\ and\ \citenamefont
  {Keeling}(2016)}]{strashko2016}%
  \BibitemOpen
  \bibfield  {author} {\bibinfo {author} {\bibfnamefont {Artem}\ \bibnamefont
  {Strashko}}\ and\ \bibinfo {author} {\bibfnamefont {Jonathan}\ \bibnamefont
  {Keeling}},\ }\bibfield  {title} {\enquote {\bibinfo {title} {Raman
  scattering with strongly coupled vibron-polaritons},}\ }\href {\doibase
  10.1103/PhysRevA.94.023843} {\bibfield  {journal} {\bibinfo  {journal}
  {Physical Review A}\ }\textbf {\bibinfo {volume} {94}},\ \bibinfo {pages}
  {023843} (\bibinfo {year} {2016})}\BibitemShut {NoStop}%
\bibitem [{\citenamefont {Gardiner}\ and\ \citenamefont
  {Collett}(1985)}]{gardiner_input_1985}%
  \BibitemOpen
  \bibfield  {author} {\bibinfo {author} {\bibfnamefont {C.~W.}\ \bibnamefont
  {Gardiner}}\ and\ \bibinfo {author} {\bibfnamefont {M.~J.}\ \bibnamefont
  {Collett}},\ }\bibfield  {title} {\enquote {\bibinfo {title} {Input and
  output in damped quantum systems: {{Quantum}} stochastic differential
  equations and the master equation},}\ }\href {\doibase
  10.1103/PhysRevA.31.3761} {\bibfield  {journal} {\bibinfo  {journal}
  {Physical Review A}\ }\textbf {\bibinfo {volume} {31}},\ \bibinfo {pages}
  {3761--3774} (\bibinfo {year} {1985})}\BibitemShut {NoStop}%
\bibitem [{\citenamefont {Steck}(2017{\natexlab{b}})}]{steck2007quantum}%
  \BibitemOpen
  \bibfield  {author} {\bibinfo {author} {\bibfnamefont {Daniel~A}\
  \bibnamefont {Steck}},\ }\href@noop {} {\emph {\bibinfo {title} {Quantum and
  Atom Optics}}},\ \bibinfo {edition} {revision 0.12.0}\ ed.\ (\bibinfo
  {publisher} {{available online at
  http://atomoptics-nas.uoregon.edu/\textasciitilde{}dsteck/teaching/quantum-optics/quantum-optics-notes.pdf}},\
  \bibinfo {year} {2017})\BibitemShut {NoStop}%
\bibitem [{\citenamefont {Li}\ \emph {et~al.}(2018)\citenamefont {Li},
  \citenamefont {Piryatinski}, \citenamefont {Jerke}, \citenamefont {Kandada},
  \citenamefont {Silva},\ and\ \citenamefont {Bittner}}]{li2018}%
  \BibitemOpen
  \bibfield  {author} {\bibinfo {author} {\bibfnamefont {Hao}\ \bibnamefont
  {Li}}, \bibinfo {author} {\bibfnamefont {Andrei}\ \bibnamefont
  {Piryatinski}}, \bibinfo {author} {\bibfnamefont {Jonathan}\ \bibnamefont
  {Jerke}}, \bibinfo {author} {\bibfnamefont {Ajay Ram~Srimath}\ \bibnamefont
  {Kandada}}, \bibinfo {author} {\bibfnamefont {Carlos}\ \bibnamefont {Silva}},
  \ and\ \bibinfo {author} {\bibfnamefont {Eric~R.}\ \bibnamefont {Bittner}},\
  }\bibfield  {title} {\enquote {\bibinfo {title} {Probing dynamical symmetry
  breaking using quantum-entangled photons},}\ }\href {\doibase
  10.1088/2058-9565/aa93b6} {\bibfield  {journal} {\bibinfo  {journal} {Quantum
  Science and Technology}\ }\textbf {\bibinfo {volume} {3}},\ \bibinfo {pages}
  {015003} (\bibinfo {year} {2018})}\BibitemShut {NoStop}%
\bibitem [{\citenamefont {Litinskaya}\ and\ \citenamefont
  {Reineker}(2006)}]{litinskaya_loss_2006}%
  \BibitemOpen
  \bibfield  {author} {\bibinfo {author} {\bibfnamefont {Marina}\ \bibnamefont
  {Litinskaya}}\ and\ \bibinfo {author} {\bibfnamefont {Peter}\ \bibnamefont
  {Reineker}},\ }\bibfield  {title} {\enquote {\bibinfo {title} {Loss of
  coherence of exciton polaritons in inhomogeneous organic microcavities},}\
  }\href {\doibase 10.1103/PhysRevB.74.165320} {\bibfield  {journal} {\bibinfo
  {journal} {Physical Review B}\ }\textbf {\bibinfo {volume} {74}},\ \bibinfo
  {pages} {165320} (\bibinfo {year} {2006})}\BibitemShut {NoStop}%
\bibitem [{\citenamefont {Kraack}\ \emph {et~al.}(2017)\citenamefont {Kraack},
  \citenamefont {Frei}, \citenamefont {Alberto},\ and\ \citenamefont
  {Hamm}}]{kraack2017}%
  \BibitemOpen
  \bibfield  {author} {\bibinfo {author} {\bibfnamefont {Jan~Philip}\
  \bibnamefont {Kraack}}, \bibinfo {author} {\bibfnamefont {Angelo}\
  \bibnamefont {Frei}}, \bibinfo {author} {\bibfnamefont {Roger}\ \bibnamefont
  {Alberto}}, \ and\ \bibinfo {author} {\bibfnamefont {Peter}\ \bibnamefont
  {Hamm}},\ }\bibfield  {title} {\enquote {\bibinfo {title} {Ultrafast
  {{Vibrational Energy Transfer}} in {{Catalytic Monolayers}} at
  {{Solid}}\textendash{{Liquid Interfaces}}},}\ }\href {\doibase
  10.1021/acs.jpclett.7b01034} {\bibfield  {journal} {\bibinfo  {journal} {The
  Journal of Physical Chemistry Letters}\ }\textbf {\bibinfo {volume} {8}},\
  \bibinfo {pages} {2489--2495} (\bibinfo {year} {2017})}\BibitemShut {NoStop}%
\bibitem [{\citenamefont {Craig}\ and\ \citenamefont
  {Thirunamachandran}(1998)}]{craig1998molecular}%
  \BibitemOpen
  \bibfield  {author} {\bibinfo {author} {\bibfnamefont {D.P.}\ \bibnamefont
  {Craig}}\ and\ \bibinfo {author} {\bibfnamefont {T.}~\bibnamefont
  {Thirunamachandran}},\ }\href@noop {} {\emph {\bibinfo {title} {Molecular
  {{Quantum Electrodynamics}}: {{An Introduction}} to {{Radiation}}-Molecule
  {{Interactions}}}}},\ Dover {{Books}} on {{Chemistry Series}}\ (\bibinfo
  {publisher} {{Dover Publications}},\ \bibinfo {year} {1998})\BibitemShut
  {NoStop}%
\bibitem [{\citenamefont {Rokaj}\ \emph {et~al.}(2018)\citenamefont {Rokaj},
  \citenamefont {Welakuh}, \citenamefont {Ruggenthaler},\ and\ \citenamefont
  {Rubio}}]{rokaj2018}%
  \BibitemOpen
  \bibfield  {author} {\bibinfo {author} {\bibfnamefont {Vasil}\ \bibnamefont
  {Rokaj}}, \bibinfo {author} {\bibfnamefont {Davis~M.}\ \bibnamefont
  {Welakuh}}, \bibinfo {author} {\bibfnamefont {Michael}\ \bibnamefont
  {Ruggenthaler}}, \ and\ \bibinfo {author} {\bibfnamefont {Angel}\
  \bibnamefont {Rubio}},\ }\bibfield  {title} {\enquote {\bibinfo {title}
  {Light\textendash{}matter interaction in the long-wavelength limit: No
  ground-state without dipole self-energy},}\ }\href {\doibase
  10.1088/1361-6455/aa9c99} {\bibfield  {journal} {\bibinfo  {journal} {Journal
  of Physics B: Atomic, Molecular and Optical Physics}\ }\textbf {\bibinfo
  {volume} {51}},\ \bibinfo {pages} {034005} (\bibinfo {year}
  {2018})}\BibitemShut {NoStop}%
\bibitem [{\citenamefont {De~Liberato}(2014)}]{deliberato2014}%
  \BibitemOpen
  \bibfield  {author} {\bibinfo {author} {\bibfnamefont {Simone}\ \bibnamefont
  {De~Liberato}},\ }\bibfield  {title} {\enquote {\bibinfo {title}
  {Light-{{Matter Decoupling}} in the {{Deep Strong Coupling Regime}}: {{The
  Breakdown}} of the {{Purcell Effect}}},}\ }\href {\doibase
  10.1103/PhysRevLett.112.016401} {\bibfield  {journal} {\bibinfo  {journal}
  {Physical Review Letters}\ }\textbf {\bibinfo {volume} {112}},\ \bibinfo
  {pages} {016401} (\bibinfo {year} {2014})}\BibitemShut {NoStop}%
\bibitem [{\citenamefont {del Pino}\ \emph {et~al.}(2015)\citenamefont {del
  Pino}, \citenamefont {Feist},\ and\ \citenamefont
  {{Garcia-Vidal}}}]{pino2015}%
  \BibitemOpen
  \bibfield  {author} {\bibinfo {author} {\bibfnamefont {Javier}\ \bibnamefont
  {del Pino}}, \bibinfo {author} {\bibfnamefont {Johannes}\ \bibnamefont
  {Feist}}, \ and\ \bibinfo {author} {\bibfnamefont {Francisco~J.}\
  \bibnamefont {{Garcia-Vidal}}},\ }\bibfield  {title} {\enquote {\bibinfo
  {title} {Quantum theory of collective strong coupling of molecular vibrations
  with a microcavity mode},}\ }\href {\doibase 10.1088/1367-2630/17/5/053040}
  {\bibfield  {journal} {\bibinfo  {journal} {New Journal of Physics}\ }\textbf
  {\bibinfo {volume} {17}},\ \bibinfo {pages} {053040} (\bibinfo {year}
  {2015})}\BibitemShut {NoStop}%
\bibitem [{\citenamefont {Zhang}\ \emph {et~al.}(2019)\citenamefont {Zhang},
  \citenamefont {Wang}, \citenamefont {Yi}, \citenamefont {Zubairy},
  \citenamefont {Scully},\ and\ \citenamefont {Mukamel}}]{zhang2019}%
  \BibitemOpen
  \bibfield  {author} {\bibinfo {author} {\bibfnamefont {Zhedong}\ \bibnamefont
  {Zhang}}, \bibinfo {author} {\bibfnamefont {Kai}\ \bibnamefont {Wang}},
  \bibinfo {author} {\bibfnamefont {Zhenhuan}\ \bibnamefont {Yi}}, \bibinfo
  {author} {\bibfnamefont {M.~Suhail}\ \bibnamefont {Zubairy}}, \bibinfo
  {author} {\bibfnamefont {Marlan~O.}\ \bibnamefont {Scully}}, \ and\ \bibinfo
  {author} {\bibfnamefont {Shaul}\ \bibnamefont {Mukamel}},\ }\bibfield
  {title} {\enquote {\bibinfo {title} {Polariton-{{Assisted Cooperativity}} of
  {{Molecules}} in {{Microcavities Monitored}} by {{Two}}-{{Dimensional
  Infrared Spectroscopy}}},}\ }\href {\doibase 10.1021/acs.jpclett.9b00979}
  {\bibfield  {journal} {\bibinfo  {journal} {The Journal of Physical Chemistry
  Letters}\ } (\bibinfo {year} {2019}),\
  10.1021/acs.jpclett.9b00979}\BibitemShut {NoStop}%
\bibitem [{\citenamefont {Agranovich}\ \emph {et~al.}(2003)\citenamefont
  {Agranovich}, \citenamefont {Litinskaia},\ and\ \citenamefont
  {Lidzey}}]{agranovich2003}%
  \BibitemOpen
  \bibfield  {author} {\bibinfo {author} {\bibfnamefont {V.~M.}\ \bibnamefont
  {Agranovich}}, \bibinfo {author} {\bibfnamefont {M.}~\bibnamefont
  {Litinskaia}}, \ and\ \bibinfo {author} {\bibfnamefont {D.~G.}\ \bibnamefont
  {Lidzey}},\ }\bibfield  {title} {\enquote {\bibinfo {title} {Cavity
  polaritons in microcavities containing disordered organic semiconductors},}\
  }\href {\doibase 10.1103/PhysRevB.67.085311} {\bibfield  {journal} {\bibinfo
  {journal} {Physical Review B}\ }\textbf {\bibinfo {volume} {67}},\ \bibinfo
  {pages} {085311} (\bibinfo {year} {2003})}\BibitemShut {NoStop}%
\bibitem [{\citenamefont {Agranovich}\ and\ \citenamefont
  {Gartstein}(2007)}]{agranovich_nature_2007}%
  \BibitemOpen
  \bibfield  {author} {\bibinfo {author} {\bibfnamefont {V.~M.}\ \bibnamefont
  {Agranovich}}\ and\ \bibinfo {author} {\bibfnamefont {Yu.~N.}\ \bibnamefont
  {Gartstein}},\ }\bibfield  {title} {\enquote {\bibinfo {title} {Nature and
  dynamics of low-energy exciton polaritons in semiconductor microcavities},}\
  }\href {\doibase 10.1103/PhysRevB.75.075302} {\bibfield  {journal} {\bibinfo
  {journal} {Physical Review B}\ }\textbf {\bibinfo {volume} {75}},\ \bibinfo
  {pages} {075302} (\bibinfo {year} {2007})}\BibitemShut {NoStop}%
\bibitem [{\citenamefont {Botzung}\ \emph {et~al.}(2020)\citenamefont
  {Botzung}, \citenamefont {Hagenm{\"u}ller}, \citenamefont {Sch{\"u}tz},
  \citenamefont {Dubail}, \citenamefont {Pupillo},\ and\ \citenamefont
  {Schachenmayer}}]{botzung2020}%
  \BibitemOpen
  \bibfield  {author} {\bibinfo {author} {\bibfnamefont {Thomas}\ \bibnamefont
  {Botzung}}, \bibinfo {author} {\bibfnamefont {David}\ \bibnamefont
  {Hagenm{\"u}ller}}, \bibinfo {author} {\bibfnamefont {Stefan}\ \bibnamefont
  {Sch{\"u}tz}}, \bibinfo {author} {\bibfnamefont {J{\'e}r{\^o}me}\
  \bibnamefont {Dubail}}, \bibinfo {author} {\bibfnamefont {Guido}\
  \bibnamefont {Pupillo}}, \ and\ \bibinfo {author} {\bibfnamefont {Johannes}\
  \bibnamefont {Schachenmayer}},\ }\bibfield  {title} {\enquote {\bibinfo
  {title} {Dark state localization of quantum emitters in a cavity},}\
  }\href@noop {} {\bibfield  {journal} {\bibinfo  {journal} {arXiv:2003.07179
  [cond-mat, physics:quant-ph]}\ } (\bibinfo {year} {2020})},\ \Eprint
  {http://arxiv.org/abs/2003.07179} {arXiv:2003.07179 [cond-mat,
  physics:quant-ph]} \BibitemShut {NoStop}%
\bibitem [{\citenamefont {Gardiner}\ and\ \citenamefont
  {Haken}(1991)}]{gardiner1991quantum}%
  \BibitemOpen
  \bibfield  {author} {\bibinfo {author} {\bibfnamefont {Crispin~W}\
  \bibnamefont {Gardiner}}\ and\ \bibinfo {author} {\bibfnamefont {Hermann}\
  \bibnamefont {Haken}},\ }\href@noop {} {\emph {\bibinfo {title} {Quantum
  Noise}}},\ Vol.~\bibinfo {volume} {26}\ (\bibinfo  {publisher} {{Springer
  Berlin}},\ \bibinfo {year} {1991})\BibitemShut {NoStop}%
\bibitem [{\citenamefont {Leegwater}\ and\ \citenamefont
  {Mukamel}(1992)}]{leegwater1992}%
  \BibitemOpen
  \bibfield  {author} {\bibinfo {author} {\bibfnamefont {Jan~A.}\ \bibnamefont
  {Leegwater}}\ and\ \bibinfo {author} {\bibfnamefont {Shaul}\ \bibnamefont
  {Mukamel}},\ }\bibfield  {title} {\enquote {\bibinfo {title}
  {Exciton-scattering mechanism for enhanced nonlinear response of molecular
  nanostructures},}\ }\href {\doibase 10.1103/PhysRevA.46.452} {\bibfield
  {journal} {\bibinfo  {journal} {Physical Review A}\ }\textbf {\bibinfo
  {volume} {46}},\ \bibinfo {pages} {452--464} (\bibinfo {year}
  {1992})}\BibitemShut {NoStop}%
\bibitem [{\citenamefont {Wang}\ \emph {et~al.}(1994)\citenamefont {Wang},
  \citenamefont {Chernyak},\ and\ \citenamefont {Mukamel}}]{wang1994a}%
  \BibitemOpen
  \bibfield  {author} {\bibinfo {author} {\bibfnamefont {Ningjun}\ \bibnamefont
  {Wang}}, \bibinfo {author} {\bibfnamefont {Vladimir}\ \bibnamefont
  {Chernyak}}, \ and\ \bibinfo {author} {\bibfnamefont {Shaul}\ \bibnamefont
  {Mukamel}},\ }\bibfield  {title} {\enquote {\bibinfo {title} {Cooperative
  ultrafast nonlinear optical response of molecular nanostructures},}\ }\href
  {\doibase 10.1063/1.466495} {\bibfield  {journal} {\bibinfo  {journal} {The
  Journal of Chemical Physics}\ }\textbf {\bibinfo {volume} {100}},\ \bibinfo
  {pages} {2465--2480} (\bibinfo {year} {1994})}\BibitemShut {NoStop}%
\bibitem [{\citenamefont {Chernyak}\ \emph {et~al.}(1995)\citenamefont
  {Chernyak}, \citenamefont {Wang},\ and\ \citenamefont
  {Mukamel}}]{chernyak1995}%
  \BibitemOpen
  \bibfield  {author} {\bibinfo {author} {\bibfnamefont {Vladimir}\
  \bibnamefont {Chernyak}}, \bibinfo {author} {\bibfnamefont {Ningjun}\
  \bibnamefont {Wang}}, \ and\ \bibinfo {author} {\bibfnamefont {Shaul}\
  \bibnamefont {Mukamel}},\ }\bibfield  {title} {\enquote {\bibinfo {title}
  {Four-wave mixing and luminescence of confined excitons in molecular
  aggregates and nanostructures. many-body green function approach},}\ }\href
  {\doibase 10.1016/0370-1573(95)00015-1} {\bibfield  {journal} {\bibinfo
  {journal} {Physics Reports}\ }\textbf {\bibinfo {volume} {263}},\ \bibinfo
  {pages} {213--309} (\bibinfo {year} {1995})}\BibitemShut {NoStop}%
\bibitem [{\citenamefont {Muallem}\ \emph {et~al.}(2016)\citenamefont
  {Muallem}, \citenamefont {Palatnik}, \citenamefont {Nessim},\ and\
  \citenamefont {Tischler}}]{muallem2016}%
  \BibitemOpen
  \bibfield  {author} {\bibinfo {author} {\bibfnamefont {Merav}\ \bibnamefont
  {Muallem}}, \bibinfo {author} {\bibfnamefont {Alexander}\ \bibnamefont
  {Palatnik}}, \bibinfo {author} {\bibfnamefont {Gilbert~D.}\ \bibnamefont
  {Nessim}}, \ and\ \bibinfo {author} {\bibfnamefont {Yaakov~R.}\ \bibnamefont
  {Tischler}},\ }\bibfield  {title} {\enquote {\bibinfo {title} {Strong
  {{Light}}-{{Matter Coupling}} and {{Hybridization}} of {{Molecular
  Vibrations}} in a {{Low}}-{{Loss Infrared Microcavity}}},}\ }\href {\doibase
  10.1021/acs.jpclett.6b00617} {\bibfield  {journal} {\bibinfo  {journal} {The
  Journal of Physical Chemistry Letters}\ }\textbf {\bibinfo {volume} {7}},\
  \bibinfo {pages} {2002--2008} (\bibinfo {year} {2016})}\BibitemShut {NoStop}%
\bibitem [{\citenamefont {Menghrajani}\ \emph {et~al.}(2019)\citenamefont
  {Menghrajani}, \citenamefont {Nash},\ and\ \citenamefont
  {Barnes}}]{menghrajani2019}%
  \BibitemOpen
  \bibfield  {author} {\bibinfo {author} {\bibfnamefont {Kishan~S.}\
  \bibnamefont {Menghrajani}}, \bibinfo {author} {\bibfnamefont {Geoffrey~R.}\
  \bibnamefont {Nash}}, \ and\ \bibinfo {author} {\bibfnamefont {William~L.}\
  \bibnamefont {Barnes}},\ }\bibfield  {title} {\enquote {\bibinfo {title}
  {Vibrational {{Strong Coupling}} with {{Surface Plasmons}} and the
  {{Presence}} of {{Surface Plasmon Stop Bands}}},}\ }\href {\doibase
  10.1021/acsphotonics.9b00662} {\bibfield  {journal} {\bibinfo  {journal} {ACS
  Photonics}\ }\textbf {\bibinfo {volume} {6}},\ \bibinfo {pages} {2110--2116}
  (\bibinfo {year} {2019})}\BibitemShut {NoStop}%
\bibitem [{\citenamefont {Shalabney}\ \emph {et~al.}(2015)\citenamefont
  {Shalabney}, \citenamefont {George}, \citenamefont {Hutchison}, \citenamefont
  {Pupillo}, \citenamefont {Genet},\ and\ \citenamefont
  {Ebbesen}}]{shalabney_coherent_2015}%
  \BibitemOpen
  \bibfield  {author} {\bibinfo {author} {\bibfnamefont {A.}~\bibnamefont
  {Shalabney}}, \bibinfo {author} {\bibfnamefont {J.}~\bibnamefont {George}},
  \bibinfo {author} {\bibfnamefont {J.}~\bibnamefont {Hutchison}}, \bibinfo
  {author} {\bibfnamefont {G.}~\bibnamefont {Pupillo}}, \bibinfo {author}
  {\bibfnamefont {C.}~\bibnamefont {Genet}}, \ and\ \bibinfo {author}
  {\bibfnamefont {T.~W.}\ \bibnamefont {Ebbesen}},\ }\bibfield  {title}
  {\enquote {\bibinfo {title} {Coherent coupling of molecular resonators with a
  microcavity mode},}\ }\href {\doibase 10.1038/ncomms6981} {\bibfield
  {journal} {\bibinfo  {journal} {Nature Communications}\ }\textbf {\bibinfo
  {volume} {6}} (\bibinfo {year} {2015}),\ 10.1038/ncomms6981}\BibitemShut
  {NoStop}%
\bibitem [{\citenamefont {Born}\ and\ \citenamefont
  {Wolf}(2013)}]{born2013principles}%
  \BibitemOpen
  \bibfield  {author} {\bibinfo {author} {\bibfnamefont {Max}\ \bibnamefont
  {Born}}\ and\ \bibinfo {author} {\bibfnamefont {Emil}\ \bibnamefont {Wolf}},\
  }\href@noop {} {\emph {\bibinfo {title} {Principles of Optics:
  Electromagnetic Theory of Propagation, Interference and Diffraction of
  Light}}}\ (\bibinfo {year} {2013})\BibitemShut {NoStop}%
\bibitem [{\citenamefont {Tavis}\ and\ \citenamefont
  {Cummings}(1968)}]{tavis_exact_1968}%
  \BibitemOpen
  \bibfield  {author} {\bibinfo {author} {\bibfnamefont {Michael}\ \bibnamefont
  {Tavis}}\ and\ \bibinfo {author} {\bibfnamefont {Frederick~W.}\ \bibnamefont
  {Cummings}},\ }\bibfield  {title} {\enquote {\bibinfo {title} {Exact
  {{Solution}} for an {{N}}-{{Molecule}}-{{Radiation}}-{{Field
  Hamiltonian}}},}\ }\href {\doibase 10.1103/PhysRev.170.379} {\bibfield
  {journal} {\bibinfo  {journal} {Physical Review}\ }\textbf {\bibinfo {volume}
  {170}},\ \bibinfo {pages} {379--384} (\bibinfo {year} {1968})}\BibitemShut
  {NoStop}%
\bibitem [{\citenamefont {Varada}\ \emph {et~al.}(1987)\citenamefont {Varada},
  \citenamefont {Kumar},\ and\ \citenamefont {Agarwal}}]{varada1987}%
  \BibitemOpen
  \bibfield  {author} {\bibinfo {author} {\bibfnamefont {G.~V.}\ \bibnamefont
  {Varada}}, \bibinfo {author} {\bibfnamefont {M.~Sanjay}\ \bibnamefont
  {Kumar}}, \ and\ \bibinfo {author} {\bibfnamefont {G.~S.}\ \bibnamefont
  {Agarwal}},\ }\bibfield  {title} {\enquote {\bibinfo {title} {Quantum effects
  of the atom-cavity interaction on four-wave mixing},}\ }\href {\doibase
  10.1016/0030-4018(87)90299-9} {\bibfield  {journal} {\bibinfo  {journal}
  {Optics Communications}\ }\textbf {\bibinfo {volume} {62}},\ \bibinfo {pages}
  {328--332} (\bibinfo {year} {1987})}\BibitemShut {NoStop}%
\bibitem [{\citenamefont {Xiang}\ \emph
  {et~al.}(2019{\natexlab{b}})\citenamefont {Xiang}, \citenamefont {Ribeiro},
  \citenamefont {Li}, \citenamefont {Dunkelberger}, \citenamefont {Simpkins},
  \citenamefont {{Yuen-Zhou}},\ and\ \citenamefont {Xiong}}]{xiang2019b}%
  \BibitemOpen
  \bibfield  {author} {\bibinfo {author} {\bibfnamefont {Bo}~\bibnamefont
  {Xiang}}, \bibinfo {author} {\bibfnamefont {Raphael~F.}\ \bibnamefont
  {Ribeiro}}, \bibinfo {author} {\bibfnamefont {Yingmin}\ \bibnamefont {Li}},
  \bibinfo {author} {\bibfnamefont {Adam~D.}\ \bibnamefont {Dunkelberger}},
  \bibinfo {author} {\bibfnamefont {Blake~B.}\ \bibnamefont {Simpkins}},
  \bibinfo {author} {\bibfnamefont {Joel}\ \bibnamefont {{Yuen-Zhou}}}, \ and\
  \bibinfo {author} {\bibfnamefont {Wei}\ \bibnamefont {Xiong}},\ }\bibfield
  {title} {\enquote {\bibinfo {title} {Manipulating optical nonlinearities of
  molecular polaritons by delocalization},}\ }\href {\doibase
  10.1126/sciadv.aax5196} {\bibfield  {journal} {\bibinfo  {journal} {Science
  Advances}\ }\textbf {\bibinfo {volume} {5}},\ \bibinfo {pages} {eaax5196}
  (\bibinfo {year} {2019}{\natexlab{b}})}\BibitemShut {NoStop}%
\bibitem [{\citenamefont {Cheng}\ \emph {et~al.}(2018)\citenamefont {Cheng},
  \citenamefont {Dhanker}, \citenamefont {Gray}, \citenamefont {Mukhopadhyay},
  \citenamefont {Kennehan}, \citenamefont {Asbury}, \citenamefont {Sokolov},\
  and\ \citenamefont {Giebink}}]{cheng2018}%
  \BibitemOpen
  \bibfield  {author} {\bibinfo {author} {\bibfnamefont {Chiao-Yu}\
  \bibnamefont {Cheng}}, \bibinfo {author} {\bibfnamefont {Rijul}\ \bibnamefont
  {Dhanker}}, \bibinfo {author} {\bibfnamefont {Christopher~L.}\ \bibnamefont
  {Gray}}, \bibinfo {author} {\bibfnamefont {Sukrit}\ \bibnamefont
  {Mukhopadhyay}}, \bibinfo {author} {\bibfnamefont {Eric~R.}\ \bibnamefont
  {Kennehan}}, \bibinfo {author} {\bibfnamefont {John~B.}\ \bibnamefont
  {Asbury}}, \bibinfo {author} {\bibfnamefont {Anatoliy}\ \bibnamefont
  {Sokolov}}, \ and\ \bibinfo {author} {\bibfnamefont {Noel~C.}\ \bibnamefont
  {Giebink}},\ }\bibfield  {title} {\enquote {\bibinfo {title} {Charged
  {{Polaron Polaritons}} in an {{Organic Semiconductor Microcavity}}},}\ }\href
  {\doibase 10.1103/PhysRevLett.120.017402} {\bibfield  {journal} {\bibinfo
  {journal} {Physical Review Letters}\ }\textbf {\bibinfo {volume} {120}},\
  \bibinfo {pages} {017402} (\bibinfo {year} {2018})}\BibitemShut {NoStop}%
\bibitem [{\citenamefont {Barachati}\ \emph {et~al.}(2018)\citenamefont
  {Barachati}, \citenamefont {Simon}, \citenamefont {Getmanenko}, \citenamefont
  {Barlow}, \citenamefont {Marder},\ and\ \citenamefont
  {{K{\'e}na-Cohen}}}]{barachati2018}%
  \BibitemOpen
  \bibfield  {author} {\bibinfo {author} {\bibfnamefont {F{\'a}bio}\
  \bibnamefont {Barachati}}, \bibinfo {author} {\bibfnamefont {Janos}\
  \bibnamefont {Simon}}, \bibinfo {author} {\bibfnamefont {Yulia~A.}\
  \bibnamefont {Getmanenko}}, \bibinfo {author} {\bibfnamefont {Stephen}\
  \bibnamefont {Barlow}}, \bibinfo {author} {\bibfnamefont {Seth~R.}\
  \bibnamefont {Marder}}, \ and\ \bibinfo {author} {\bibfnamefont
  {St{\'e}phane}\ \bibnamefont {{K{\'e}na-Cohen}}},\ }\bibfield  {title}
  {\enquote {\bibinfo {title} {Tunable {{Third}}-{{Harmonic Generation}} from
  {{Polaritons}} in the {{Ultrastrong Coupling Regime}}},}\ }\href {\doibase
  10.1021/acsphotonics.7b00305} {\bibfield  {journal} {\bibinfo  {journal} {ACS
  Photonics}\ }\textbf {\bibinfo {volume} {5}},\ \bibinfo {pages} {119--125}
  (\bibinfo {year} {2018})}\BibitemShut {NoStop}%
\bibitem [{\citenamefont {Wang}\ \emph {et~al.}(2020)\citenamefont {Wang},
  \citenamefont {Seidel}, \citenamefont {Nagarajan}, \citenamefont {Chervy},
  \citenamefont {Genet},\ and\ \citenamefont {Ebbesen}}]{wang2020}%
  \BibitemOpen
  \bibfield  {author} {\bibinfo {author} {\bibfnamefont {Kuidong}\ \bibnamefont
  {Wang}}, \bibinfo {author} {\bibfnamefont {Marcus}\ \bibnamefont {Seidel}},
  \bibinfo {author} {\bibfnamefont {Kalaivanan}\ \bibnamefont {Nagarajan}},
  \bibinfo {author} {\bibfnamefont {Thibault}\ \bibnamefont {Chervy}}, \bibinfo
  {author} {\bibfnamefont {Cyriaque}\ \bibnamefont {Genet}}, \ and\ \bibinfo
  {author} {\bibfnamefont {Thomas~W.}\ \bibnamefont {Ebbesen}},\ }\bibfield
  {title} {\enquote {\bibinfo {title} {Large optical nonlinearity enhancement
  under electronic strong coupling},}\ }\href@noop {} {\bibfield  {journal}
  {\bibinfo  {journal} {arXiv:2005.13325 [physics]}\ } (\bibinfo {year}
  {2020})},\ \Eprint {http://arxiv.org/abs/2005.13325} {arXiv:2005.13325
  [physics]} \BibitemShut {NoStop}%
\bibitem [{\citenamefont {So}\ \emph {et~al.}(2000)\citenamefont {So},
  \citenamefont {Dong}, \citenamefont {Masters},\ and\ \citenamefont
  {Berland}}]{so2000}%
  \BibitemOpen
  \bibfield  {author} {\bibinfo {author} {\bibfnamefont {Peter T.~C.}\
  \bibnamefont {So}}, \bibinfo {author} {\bibfnamefont {Chen~Y.}\ \bibnamefont
  {Dong}}, \bibinfo {author} {\bibfnamefont {Barry~R.}\ \bibnamefont
  {Masters}}, \ and\ \bibinfo {author} {\bibfnamefont {Keith~M.}\ \bibnamefont
  {Berland}},\ }\bibfield  {title} {\enquote {\bibinfo {title} {Two-{{Photon
  Excitation Fluorescence Microscopy}}},}\ }\href {\doibase
  10.1146/annurev.bioeng.2.1.399} {\bibfield  {journal} {\bibinfo  {journal}
  {Annual Review of Biomedical Engineering}\ }\textbf {\bibinfo {volume} {2}},\
  \bibinfo {pages} {399--429} (\bibinfo {year} {2000})},\ \bibinfo {note}
  {\_eprint: https://doi.org/10.1146/annurev.bioeng.2.1.399}\BibitemShut
  {NoStop}%
\bibitem [{\citenamefont {Kraack}\ and\ \citenamefont
  {Hamm}(2016)}]{kraack2016a}%
  \BibitemOpen
  \bibfield  {author} {\bibinfo {author} {\bibfnamefont {Jan~Philip}\
  \bibnamefont {Kraack}}\ and\ \bibinfo {author} {\bibfnamefont {Peter}\
  \bibnamefont {Hamm}},\ }\bibfield  {title} {\enquote {\bibinfo {title}
  {Vibrational ladder-climbing in surface-enhanced, ultrafast infrared
  spectroscopy},}\ }\href {\doibase 10.1039/C6CP02589G} {\bibfield  {journal}
  {\bibinfo  {journal} {Physical Chemistry Chemical Physics}\ }\textbf
  {\bibinfo {volume} {18}},\ \bibinfo {pages} {16088--16093} (\bibinfo {year}
  {2016})}\BibitemShut {NoStop}%
\bibitem [{\citenamefont {Morichika}\ \emph {et~al.}(2019)\citenamefont
  {Morichika}, \citenamefont {Murata}, \citenamefont {Sakurai}, \citenamefont
  {Ishii},\ and\ \citenamefont {Ashihara}}]{morichika2019}%
  \BibitemOpen
  \bibfield  {author} {\bibinfo {author} {\bibfnamefont {Ikki}\ \bibnamefont
  {Morichika}}, \bibinfo {author} {\bibfnamefont {Kei}\ \bibnamefont {Murata}},
  \bibinfo {author} {\bibfnamefont {Atsunori}\ \bibnamefont {Sakurai}},
  \bibinfo {author} {\bibfnamefont {Kazuyuki}\ \bibnamefont {Ishii}}, \ and\
  \bibinfo {author} {\bibfnamefont {Satoshi}\ \bibnamefont {Ashihara}},\
  }\bibfield  {title} {\enquote {\bibinfo {title} {Molecular ground-state
  dissociation in the condensed phase employing plasmonic field enhancement of
  chirped mid-infrared pulses},}\ }\href {\doibase 10.1038/s41467-019-11902-6}
  {\bibfield  {journal} {\bibinfo  {journal} {Nature Communications}\ }\textbf
  {\bibinfo {volume} {10}},\ \bibinfo {pages} {3893} (\bibinfo {year}
  {2019})}\BibitemShut {NoStop}%
\bibitem [{\citenamefont {Juraschek}\ \emph {et~al.}(2019)\citenamefont
  {Juraschek}, \citenamefont {Neuman}, \citenamefont {Flick},\ and\
  \citenamefont {Narang}}]{juraschek2019}%
  \BibitemOpen
  \bibfield  {author} {\bibinfo {author} {\bibfnamefont {Dominik~M.}\
  \bibnamefont {Juraschek}}, \bibinfo {author} {\bibfnamefont {Tom{\'a}{\v s}}\
  \bibnamefont {Neuman}}, \bibinfo {author} {\bibfnamefont {Johannes}\
  \bibnamefont {Flick}}, \ and\ \bibinfo {author} {\bibfnamefont {Prineha}\
  \bibnamefont {Narang}},\ }\bibfield  {title} {\enquote {\bibinfo {title}
  {Cavity control of nonlinear phononics},}\ }\href@noop {} {\bibfield
  {journal} {\bibinfo  {journal} {arXiv:1912.00122 [cond-mat,
  physics:physics]}\ } (\bibinfo {year} {2019})},\ \Eprint
  {http://arxiv.org/abs/1912.00122} {arXiv:1912.00122 [cond-mat,
  physics:physics]} \BibitemShut {NoStop}%
\end{thebibliography}%


\begin{thebibliography}{10}%
\makeatletter
\providecommand \@ifxundefined [1]{%
 \@ifx{#1\undefined}
}%
\providecommand \@ifnum [1]{%
 \ifnum #1\expandafter \@firstoftwo
 \else \expandafter \@secondoftwo
 \fi
}%
\providecommand \@ifx [1]{%
 \ifx #1\expandafter \@firstoftwo
 \else \expandafter \@secondoftwo
 \fi
}%
\providecommand \natexlab [1]{#1}%
\providecommand \enquote  [1]{``#1''}%
\providecommand \bibnamefont  [1]{#1}%
\providecommand \bibfnamefont [1]{#1}%
\providecommand \citenamefont [1]{#1}%
\providecommand \href@noop [0]{\@secondoftwo}%
\providecommand \href [0]{\begingroup \@sanitize@url \@href}%
\providecommand \@href[1]{\@@startlink{#1}\@@href}%
\providecommand \@@href[1]{\endgroup#1\@@endlink}%
\providecommand \@sanitize@url [0]{\catcode `\\12\catcode `\$12\catcode
  `\&12\catcode `\#12\catcode `\^12\catcode `\_12\catcode `\%12\relax}%
\providecommand \@@startlink[1]{}%
\providecommand \@@endlink[0]{}%
\providecommand \url  [0]{\begingroup\@sanitize@url \@url }%
\providecommand \@url [1]{\endgroup\@href {#1}{\urlprefix }}%
\providecommand \urlprefix  [0]{URL }%
\providecommand \Eprint [0]{\href }%
\providecommand \doibase [0]{http://dx.doi.org/}%
\providecommand \selectlanguage [0]{\@gobble}%
\providecommand \bibinfo  [0]{\@secondoftwo}%
\providecommand \bibfield  [0]{\@secondoftwo}%
\providecommand \translation [1]{[#1]}%
\providecommand \BibitemOpen [0]{}%
\providecommand \bibitemStop [0]{}%
\providecommand \bibitemNoStop [0]{.\EOS\space}%
\providecommand \EOS [0]{\spacefactor3000\relax}%
\providecommand \BibitemShut  [1]{\csname bibitem#1\endcsname}%
\let\auto@bib@innerbib\@empty
\bibitem [{\citenamefont {Gardiner}\ and\ \citenamefont
  {Collett}(1985)}]{gardiner_input_1985}%
  \BibitemOpen
  \bibfield  {author} {\bibinfo {author} {\bibfnamefont {C.~W.}\ \bibnamefont
  {Gardiner}}\ and\ \bibinfo {author} {\bibfnamefont {M.~J.}\ \bibnamefont
  {Collett}},\ }\bibfield  {title} {\enquote {\bibinfo {title} {Input and
  output in damped quantum systems: {{Quantum}} stochastic differential
  equations and the master equation},}\ }\href {\doibase
  10.1103/PhysRevA.31.3761} {\bibfield  {journal} {\bibinfo  {journal}
  {Physical Review A}\ }\textbf {\bibinfo {volume} {31}},\ \bibinfo {pages}
  {3761--3774} (\bibinfo {year} {1985})}\BibitemShut {NoStop}%
\bibitem [{\citenamefont {Steck}(2017{\natexlab{a}})}]{steck2007quantum}%
  \BibitemOpen
  \bibfield  {author} {\bibinfo {author} {\bibfnamefont {Daniel~A}\
  \bibnamefont {Steck}},\ }\href@noop {} {\emph {\bibinfo {title} {Quantum and
  Atom Optics}}},\ \bibinfo {edition} {revision 0.12.0}\ ed.\ (\bibinfo
  {publisher} {{available online at
  http://atomoptics-nas.uoregon.edu/\textasciitilde{}dsteck/teaching/quantum-optics/quantum-optics-notes.pdf}},\
  \bibinfo {year} {2017})\BibitemShut {NoStop}%
\bibitem [{\citenamefont {Steck}(2017{\natexlab{b}})}]{steck2017}%
  \BibitemOpen
  \bibfield  {author} {\bibinfo {author} {\bibfnamefont {Daniel~A}\
  \bibnamefont {Steck}},\ }\href@noop {} {\emph {\bibinfo {title} {Classical
  and {{Modern Optics}}}}},\ \bibinfo {edition} {revision 1.7.4}\ ed.\
  (\bibinfo  {publisher} {{available online at http://steck.us/teaching}},\
  \bibinfo {year} {2017})\BibitemShut {NoStop}%
\bibitem [{\citenamefont {Kavokin}\ \emph {et~al.}(2017)\citenamefont
  {Kavokin}, \citenamefont {Baumberg}, \citenamefont {Malpuech},\ and\
  \citenamefont {Laussy}}]{kavokin2017microcavities}%
  \BibitemOpen
  \bibfield  {author} {\bibinfo {author} {\bibfnamefont {Alexey~V}\
  \bibnamefont {Kavokin}}, \bibinfo {author} {\bibfnamefont {Jeremy~J}\
  \bibnamefont {Baumberg}}, \bibinfo {author} {\bibfnamefont {Guillaume}\
  \bibnamefont {Malpuech}}, \ and\ \bibinfo {author} {\bibfnamefont
  {Fabrice~P}\ \bibnamefont {Laussy}},\ }\href@noop {} {\emph {\bibinfo {title}
  {Microcavities}}},\ Vol.~\bibinfo {volume} {21}\ (\bibinfo  {publisher}
  {{Oxford University Press}},\ \bibinfo {year} {2017})\BibitemShut {NoStop}%
\bibitem [{\citenamefont {Mukamel}(1999)}]{mukamel1999principles}%
  \BibitemOpen
  \bibfield  {author} {\bibinfo {author} {\bibfnamefont {Shaul}\ \bibnamefont
  {Mukamel}},\ }\href@noop {} {\emph {\bibinfo {title} {Principles of Nonlinear
  Optical Spectroscopy}}}\ (\bibinfo  {publisher} {{Oxford University Press on
  Demand}},\ \bibinfo {year} {1999})\BibitemShut {NoStop}%
\bibitem [{\citenamefont {Boyd}\ and\ \citenamefont
  {Prato}(2008)}]{boyd2008nonlinear}%
  \BibitemOpen
  \bibfield  {author} {\bibinfo {author} {\bibfnamefont {R.W.}\ \bibnamefont
  {Boyd}}\ and\ \bibinfo {author} {\bibfnamefont {D.}~\bibnamefont {Prato}},\
  }\href@noop {} {\emph {\bibinfo {title} {Nonlinear Optics}}}\ (\bibinfo
  {publisher} {{Elsevier Science}},\ \bibinfo {year} {2008})\BibitemShut
  {NoStop}%
\bibitem [{\citenamefont {Xiang}\ \emph {et~al.}(2018)\citenamefont {Xiang},
  \citenamefont {Ribeiro}, \citenamefont {Dunkelberger}, \citenamefont {Wang},
  \citenamefont {Li}, \citenamefont {Simpkins}, \citenamefont {Owrutsky},
  \citenamefont {{Yuen-Zhou}},\ and\ \citenamefont {Xiong}}]{xiang2018}%
  \BibitemOpen
  \bibfield  {author} {\bibinfo {author} {\bibfnamefont {Bo}~\bibnamefont
  {Xiang}}, \bibinfo {author} {\bibfnamefont {Raphael~F.}\ \bibnamefont
  {Ribeiro}}, \bibinfo {author} {\bibfnamefont {Adam~D.}\ \bibnamefont
  {Dunkelberger}}, \bibinfo {author} {\bibfnamefont {Jiaxi}\ \bibnamefont
  {Wang}}, \bibinfo {author} {\bibfnamefont {Yingmin}\ \bibnamefont {Li}},
  \bibinfo {author} {\bibfnamefont {Blake~S.}\ \bibnamefont {Simpkins}},
  \bibinfo {author} {\bibfnamefont {Jeffrey~C.}\ \bibnamefont {Owrutsky}},
  \bibinfo {author} {\bibfnamefont {Joel}\ \bibnamefont {{Yuen-Zhou}}}, \ and\
  \bibinfo {author} {\bibfnamefont {Wei}\ \bibnamefont {Xiong}},\ }\bibfield
  {title} {\enquote {\bibinfo {title} {Two-dimensional infrared spectroscopy of
  vibrational polaritons},}\ }\href {\doibase 10.1073/pnas.1722063115}
  {\bibfield  {journal} {\bibinfo  {journal} {Proceedings of the National
  Academy of Sciences}\ ,\ \bibinfo {pages} {201722063}} (\bibinfo {year}
  {2018})}\BibitemShut {NoStop}%
\bibitem [{\citenamefont {Ribeiro}\ \emph {et~al.}(2018)\citenamefont
  {Ribeiro}, \citenamefont {Dunkelberger}, \citenamefont {Xiang}, \citenamefont
  {Xiong}, \citenamefont {Simpkins}, \citenamefont {Owrutsky},\ and\
  \citenamefont {{Yuen-Zhou}}}]{ribeiro2018c}%
  \BibitemOpen
  \bibfield  {author} {\bibinfo {author} {\bibfnamefont {Raphael~F.}\
  \bibnamefont {Ribeiro}}, \bibinfo {author} {\bibfnamefont {Adam~D.}\
  \bibnamefont {Dunkelberger}}, \bibinfo {author} {\bibfnamefont
  {Bo}~\bibnamefont {Xiang}}, \bibinfo {author} {\bibfnamefont {Wei}\
  \bibnamefont {Xiong}}, \bibinfo {author} {\bibfnamefont {Blake~S.}\
  \bibnamefont {Simpkins}}, \bibinfo {author} {\bibfnamefont {Jeffrey~C.}\
  \bibnamefont {Owrutsky}}, \ and\ \bibinfo {author} {\bibfnamefont {Joel}\
  \bibnamefont {{Yuen-Zhou}}},\ }\bibfield  {title} {\enquote {\bibinfo {title}
  {Theory for {{Nonlinear Spectroscopy}} of {{Vibrational Polaritons}}},}\
  }\href {\doibase 10.1021/acs.jpclett.8b01176} {\bibfield  {journal} {\bibinfo
   {journal} {The Journal of Physical Chemistry Letters}\ }\textbf {\bibinfo
  {volume} {9}},\ \bibinfo {pages} {3766--3771} (\bibinfo {year}
  {2018})}\BibitemShut {NoStop}%
\bibitem [{\citenamefont {Xiang}\ \emph
  {et~al.}(2019{\natexlab{a}})\citenamefont {Xiang}, \citenamefont {Ribeiro},
  \citenamefont {Chen}, \citenamefont {Wang}, \citenamefont {Du}, \citenamefont
  {{Yuen-Zhou}},\ and\ \citenamefont {Xiong}}]{xiang2019a}%
  \BibitemOpen
  \bibfield  {author} {\bibinfo {author} {\bibfnamefont {Bo}~\bibnamefont
  {Xiang}}, \bibinfo {author} {\bibfnamefont {Raphael~F.}\ \bibnamefont
  {Ribeiro}}, \bibinfo {author} {\bibfnamefont {Liying}\ \bibnamefont {Chen}},
  \bibinfo {author} {\bibfnamefont {Jiaxi}\ \bibnamefont {Wang}}, \bibinfo
  {author} {\bibfnamefont {Matthew}\ \bibnamefont {Du}}, \bibinfo {author}
  {\bibfnamefont {Joel}\ \bibnamefont {{Yuen-Zhou}}}, \ and\ \bibinfo {author}
  {\bibfnamefont {Wei}\ \bibnamefont {Xiong}},\ }\bibfield  {title} {\enquote
  {\bibinfo {title} {State-{{Selective Polariton}} to {{Dark State Relaxation
  Dynamics}}},}\ }\href {\doibase 10.1021/acs.jpca.9b04601} {\bibfield
  {journal} {\bibinfo  {journal} {The Journal of Physical Chemistry A}\
  }\textbf {\bibinfo {volume} {123}},\ \bibinfo {pages} {5918--5927} (\bibinfo
  {year} {2019}{\natexlab{a}})}\BibitemShut {NoStop}%
\bibitem [{\citenamefont {Xiang}\ \emph
  {et~al.}(2019{\natexlab{b}})\citenamefont {Xiang}, \citenamefont {Ribeiro},
  \citenamefont {Li}, \citenamefont {Dunkelberger}, \citenamefont {Simpkins},
  \citenamefont {{Yuen-Zhou}},\ and\ \citenamefont {Xiong}}]{xiang2019b}%
  \BibitemOpen
  \bibfield  {author} {\bibinfo {author} {\bibfnamefont {Bo}~\bibnamefont
  {Xiang}}, \bibinfo {author} {\bibfnamefont {Raphael~F.}\ \bibnamefont
  {Ribeiro}}, \bibinfo {author} {\bibfnamefont {Yingmin}\ \bibnamefont {Li}},
  \bibinfo {author} {\bibfnamefont {Adam~D.}\ \bibnamefont {Dunkelberger}},
  \bibinfo {author} {\bibfnamefont {Blake~B.}\ \bibnamefont {Simpkins}},
  \bibinfo {author} {\bibfnamefont {Joel}\ \bibnamefont {{Yuen-Zhou}}}, \ and\
  \bibinfo {author} {\bibfnamefont {Wei}\ \bibnamefont {Xiong}},\ }\bibfield
  {title} {\enquote {\bibinfo {title} {Manipulating optical nonlinearities of
  molecular polaritons by delocalization},}\ }\href {\doibase
  10.1126/sciadv.aax5196} {\bibfield  {journal} {\bibinfo  {journal} {Science
  Advances}\ }\textbf {\bibinfo {volume} {5}},\ \bibinfo {pages} {eaax5196}
  (\bibinfo {year} {2019}{\natexlab{b}})}\BibitemShut {NoStop}%
\end{thebibliography}%
\end{document}


\title{Supporting Information for Enhanced optical nonlinearities under strong light-matter coupling}
\author{Raphael F. Ribeiro} 
\affiliation{Department of Chemistry and Biochemistry, University of California San Diego, La Jolla, CA 92093}
\author{Jorge A. Campos-Gonzalez-Angulo}
\affiliation{Department of Chemistry and Biochemistry, University of California San Diego, La Jolla, CA 92093}
\author{Noel C. Giebink}
\affiliation{Department of Electrical Engineering, The Pennsylvania State University, University Park, PA,
16802}
\author{Wei Xiong}
\affiliation{Department of Chemistry and Biochemistry, University of California San Diego, La Jolla, CA 92093}
\affiliation{Materials Science and Engineering Program, University of California San Diego, La Jolla, CA 92093}
\author{Joel Yuen-Zhou}
\affiliation{Department of Chemistry and Biochemistry, University of California San Diego, La Jolla, CA 92093}
\date{\today}

\maketitle
\tableofcontents

\section{Basic definitions for empty microcavity\cite{gardiner_input_1985,steck2007quantum, steck2017}}
We employ input-output theory\cite{gardiner_input_1985,steck2007quantum} to describe the open quantum system dynamics of a planar optical microcavity consisting of two highly-reflective symmetric mirrors \cite{kavokin2017microcavities} separated by a distance $L_c$. The input radiation is taken to have zero momentum along the transverse direction to the cavity longitudinal axis. Since we work in the limiting case where a single cavity mode interacts with the material system, we only consider the free space electromagnetic modes to the right and left of living in a 1D space with length $L$ sufficiently large for the corresponding field operators to satisfy periodic boundary conditions. 

We suppose the system is probed in transmission geometry, where the incident light irradiates the ``left" mirror and the optical signal is generated by the photon flux traversing the ``right" mirror. The output photon flux is given by
\begin{align}
 \braket{\left[b_{\t{out}}^{\t{R}}(t)\right]^\dagger b_{\t{out}}(t)}, 
\end{align} where the output annihilation operator $b_{\t{out}}(t)$ is written in terms of the ``right" free space modes at a future time $t_1 > t$ 
\cite{gardiner_input_1985}:
\begin{align}
b_{\t{out}}^{\t{R}}(t) = \frac{i}{\sqrt{2\pi}} \int_{-\infty}^{\infty} \mrm{d}\omega' b_{1}^{\t{R}}(\omega') e^{-i\omega'(t-t_1)}, ~~ t_1 > t	
\end{align}
where $b_{1}^{\t{R}}(\omega')$ is the Heisenberg annihilation operator for a photon with frequency $\omega'$ in the free space to the right of the optical cavity at $t_1$. 
In the absence of any input on the system from the right mirror, the input-output relations allow us to directly relate the (right) output EM power at time $t$ with the state of the cavity at the same moment. In particular\cite{gardiner_input_1985,steck2007quantum},
\begin{align}
b_{\t{out}}^{\t{R}}(t) = \sqrt{\frac{\kappa}	{2}}b(t), \label{eq:inout}
\end{align}
where $b$ is the cavity mode annihilation operator and $\kappa$ is the \textit{total} cavity leakage rate (including field decay through both mirrors). The latter is proportional to the mirrors transmission probability $|t|^2$, as well as inversely related to the cavity round-trip time $\tau_c = 2L_c/c$, ($L_c$ is the cavity longitudinal length) \cite{steck2017}, i.e.,
\begin{align}
\kappa = \frac{|t|^2}{\tau_{c}} =|t|^2\frac{c}{2L_c}. \label{eq:kappa}
\end{align}
 
In this work, we suppose the microcavity is driven by a superposition of coherent state fields which are nearly-resonant and weakly interact with the cavity ($\kappa$ is much smaller than the cavity photon frequency). The rotating-wave approximation is employed throughout, as is customary in an input-output treatment \cite{steck2007quantum}. We suppose the electric field of the external source which drives the system is expressed as: \begin{align}
   	E_{\t{in}}^{\t{L}}(t) & = \sum_{\omega >0} \left[E_{\t{in}}^{(+)}(\omega) e^{-i\omega t} + E_{\t{in}}^{(-)}(\omega) e^{i\omega t}\right] \nonum
   	&= \sum_{\omega>0}i\sqrt{\frac{\hbar \omega}{2\epsilon_0 V}} \left[\alpha_{\t{in}}(\omega)e^{-i\omega t}-\alpha_{\t{in}}^\dagger(\omega) e^{i\omega t}\right], \label{eq:ein}
   \end{align}
 where $\epsilon_0$ is the free space permittivity, $V = SL$ is the quantization volume of the l.h.s (or r.h.s.) free space, and $\alpha_{\t{in}} \in \mathbb{C}$ is a coherent state amplitude characterizing the phase and intensity of the input external field mode with frequency $\omega$. The photon flux corresponding to each frequency in \ref{eq:ein} is given by $|\alpha_{\text{in}}(\omega)|^2 c/L$\footnote{Here, we used the following expression for the mean photon flux 
\begin{align} 
\phi_{\t{in}} & = \frac{S\epsilon_0 c}{\hbar \omega T}\int_{-T/2}^{T/2} \braket{\left[E_{\t{in}}^{\t{L}}(t)\right]^\dagger E_{\t{in}}^{\t{L}}(t)} \nonum
& = \sum_{\omega > 0}|\alpha_{\text{in}}(\omega)|^2 c/L
\end{align} which is thus given in units of photon number per unit time.}. The photon input operator is thus \footnote{Note that our input fields are obtained from superpositions of coherent states of the electromagnetic field in the l.h.s. free space defined by 
\begin{align} b_{\t{in}}^{\t{L}}(t) = \frac{i}{\sqrt{2\pi}} \int_{-\infty}^\infty \mrm{d}\omega' b_0^L(\omega') e^{-i\omega'(t-t_0)} = 
 \sum_{\omega} b_{\t{in}}^{\t{L}}(\omega) e^{-i\omega t},\end{align} where $b_0^L(\omega')$ is the annihilation operator of the left mode with frequency $\omega'$ evaluated at a time $t_0<t$. \cite{gardiner_input_1985}}
\begin{align} b_{\t{in}}^{\t{L}}(\omega) = i\sqrt{\frac{\mathcal{P}(\omega)}{\hbar\omega}} e^{i\theta_{\t{in}}(\omega)}, 
\end{align} where $\mathcal{P}(\omega) = \hbar \omega |\alpha_{\t{in}}(\omega)|^2 c/L$ and $\theta_{\t{in}}(\omega)$ is determined by the relationship $\alpha_{\t{in}}(\omega) = |\alpha_{\t{in}}(\omega)| e^{i\theta_{\t{in}}(\omega)}$

\par We conclude this section by reviewing some relationships between the cavity electromagnetic field intensity in the presence of steady driving, and the corresponding free space intensity. This identification will be essential for the comparison of the nonlinear optical response of a hybrid cavity with that of the bare molecular material. 

\par First, we recall that the mirrors of a \textit{good cavity} have nearly vanishing photon transmission probability $|t|^2 \rightarrow 0$. The cavity is usually characterized by (a) the total photon leakage rate $\kappa$ (Eq. \ref{eq:kappa}) dependent on both geometric parameters (e.g., the cavity length) and the quality of the mirrors (via its dependence on $|t|^2$), and (b)  its finesse coefficient $\mathcal{F} = \pi \sqrt{|r|}/(1-|r|)$ (where $r$ is the field reflection probability amplitude) \cite{steck2017}, which depends only on the quality of the mirrors. As we demonstrate below, the finesse provides a  simple measure of the steady-state intracavity (resonant) electromagnetic field intensity $I_c$ enhancement compared to free space. In particular, at a cavity antinode, it follows that\cite{steck2017},
\begin{align}
I_c \approx \frac{2\mathcal{F}}{\pi} I_0,
\end{align}
 where $I_0$ is the free space electromagnetic field intensity \cite{steck2017}. In terms of the finesse, the cavity leakage rate $\kappa$ can be written as 
 \begin{align} \kappa = \frac{\pi c}{L_c\mathcal{F}}.\end{align}
Alternatively, $\mathcal{F}$ is a given as a simple function of the cavity quality factor $Q = \omega_c/\kappa$ \cite{steck2017}:
\begin{align}
\mathcal{F} & = \frac{\pi c}{L_c \omega_c}\frac{\omega_c}{\kappa} \nonum
& = \frac{Q}{m},
\end{align}
where $m \in \mathbb{Z}$ is the longitudinal quantum number of the cavity mode, and we used that the symmetric planar cavity mode frequency corresponding to $m$ is given by $\omega_c = c m \pi/L_c$. 

We conclude this section by deriving the cavity electric field enhancement factor from input-output theory. Consider an \textit{empty} cavity driven by an external field with power $\mathcal{P}(\omega) = \hbar \omega |\alpha_{\t{in}}(\omega)|^2 c/L$, where $\alpha_{\t{in}}(\omega) \in \mathbb{C}$. Using the previously defined parametrization $b_{\t{in}}^{\t{L}}(\omega)=i\sqrt{\mathcal{P}(\omega)/\hbar\omega} e^{i\theta_{\t{in}}(\omega)}$ for the input field, it follows from the input-output treatment of an empty driven cavity that the steady-state positive-frequency component of the empty cavity electric field (in the rotating-wave approximation) $E_c^{(+)}(\omega)$ is given by\footnote{This equation can be simply derived by using the Heisenberg equations of motion for the driven cavity mode operator \begin{align}
 & (i\hbar\partial_t -\hbar \tilde{\omega}_c)b(t) = -i\hbar\sqrt{\frac{\kappa}{2}} b_{\t{in}}^{\t{L}}(t) \nonum
&  \implies b(\omega) = -i \hbar\sqrt{\frac{\kappa}{2}}\frac{b_{\t{in}}^{\t{L}}(\omega)}{\hbar\omega-\hbar\omega_c+i\kappa/2}, \label{eq:bempty}\\
& \implies E_c^{(+)}(\omega) = i \sqrt{\frac{\hbar\omega_c}{2\epsilon_0 SL_c}}b(\omega) = \sqrt{\frac{\hbar\omega_c}{2\epsilon_0 SL_c}}\sqrt{\frac{\kappa}{2}}\frac{b_{\t{in}}^{\t{L}}(\omega)}{\omega-\omega_c+i\kappa/2},
\end{align} where we used $\tilde{\omega}_c = \omega_c - i\kappa/2$, and $b(t) = \sum_{\omega} b(\omega) e^{-i\omega t}$.}:
\begin{align}
E_c^{(+)}(\omega) & = E_0^c \sqrt{\frac{\kappa}{2}}	\frac{-i b_{\t{in}}^{\t{L}}(\omega)}{\omega-\omega_c +i \kappa/2} \nonum
&  = i\sqrt{\frac{\hbar \omega_c}{2\epsilon_0 S L_c}} |\alpha_{\t{in}}(\omega)|\sqrt{\frac{2c}{\kappa L}} \frac{\kappa/2}{\omega-\omega_c+i\kappa/2} e^{i\theta_{\t{in}}(\omega)}& \nonum
 & \approx \sqrt{\frac{2\mathcal{F}}{\pi}}\frac{\kappa/2}{\omega-\omega_c+i\kappa/2}E_{\t{in}}^{(+)}(\omega).
 \label{eq:e0ec}
\end{align}
where we used $E_0^c = i\sqrt{\hbar \omega_c/2\epsilon_0 S L_c}$. In the last line we employed $\omega = \omega_c + \delta$ and the limit where $\delta \omega/\omega_c \rightarrow 0$, i.e., $\omega\approx \omega_c$. This approximation is consistent with the weak-coupling and near-resonant assumptions of input-output theory\cite{gardiner_input_1985,steck2007quantum}, and is usually satisfied when $\omega$ is resonant with polaritons in the strong coupling limit (with Rabi splitting significantly weaker than the relevant bare molecule and cavity frequencies).\\

Equation \ref{eq:e0ec} demonstrates the well-known results that under resonant driving ($\omega=\omega_c$) (a) the cavity electromagnetic field intensity is enhanced by a factor of $2\mathcal{F}/\pi$ (at cavity antinodes) compared to free space, and (b) the cavity field is phase-shifted by $-\pi/2$ relative to the phase of the external field.

\section{Nonlinear susceptibility of strongly coupled molecular system}\label{sec:scav}
In this section, we derive the steady-state third-order polarization induced by continuous-wave input fields acting on a molecular system strongly-coupled to an optical cavity as described in the main text. This  polarization is the source of the nonlinear optical signal discussed above. The results obtained here are essential for the computation of the nonlinear molecular absorption under strong light-matter coupling which is described in the next section.

To obtain the material nonlinear polarization we solve perturbatively the equations of motion (EOM) for the expectation value of the molecular polarization operator in terms of the driving input fields.
 We perform this procedure in this subsection assuming the system remains in a pure state at all times. 

From the effective Hamiltonian introduced in the main text (Eq. 4), we obtain the Heisenberg-Langevin EOM for the expectation value of the cavity-photon annihilation operator using $i\hbar \partial_t b(t) = [b(t),H] -i\hbar\kappa b(t)/2$:
\begin{align}
\left(i\hbar \frac{\mathrm{d}}{\mathrm{d}t} -\hbar\tilde{\omega}_c\right)\braket{b(t)} = -i\hbar\sqrt{\frac{\kappa}{2}}b_{\t{in}}^{\t{L}}(t) - \mu \bar{E_0^c} \sum_{i=1}^N \braket{a_i(t)},\label{eq:beom}
\end{align}
where the cavity leakage rate $\kappa$ (derived within input-output theory) was introduced by the replacement $\omega_c \rightarrow \tilde{\omega}_c = \omega_c -i\kappa/2$. Eq. \ref{eq:beom} describes the time-dependent response of the cavity to a collective molecular polarization and to the driving by the input field. Similarly, the  molecular response to the cavity electromagnetic field is expressed by the analogous Heisenberg-Langevin EOMs satisfied by the time-dependent single-molecule and collective material polarizations:
\begin{align}
\left(i\hbar\frac{\mathrm{d}}{\mathrm{d}t} -\hbar\tilde{\omega}_0\right)\braket{\mu a_i(t)} & = -\mu^2 E_0^c \braket{b(t)} -2\hbar\Delta \mu \braket{a_i^\dagger(t) a_i(t)a_i(t)} \nonum
& = -\mu^2 E_0^c \braket{b(t)} -2\hbar\Delta \mu \braket{a_i^\dagger(t)} \braket{a_i(t)a_i(t)}, \label{eq:aiseom}\\
\left(i\hbar\frac{\mathrm{d}}{\mathrm{d}t} -\hbar\tilde{\omega}_0\right)\braket{\mu \sum_{i=1}^N a_i(t)} & = -N\mu^2 E_0^c \braket{b(t)} -2\hbar\Delta \mu \braket{\sum_{i=1}^N a_i^\dagger(t) a_i(t)a_i(t)} \nonum & = -N \mu^2 E_0^c \braket{b(t)} -2\hbar\Delta \mu \sum_{i=1}^N \braket{a_i^\dagger(t)} \braket{a_i(t)a_i(t)}, \label{eq:aieom}
\end{align}
where $\tilde{\omega}_0 = \omega_0 -i\gamma_m/2$, and we obtained the final equations in each case using
the factorization property of normal-ordered (all annihilation operators are to the right of the creation) pure-state correlation functions which, for $\braket{a^\dagger(t) a(t) a(t)}$ is valid to $O\left(\left|E_{\t{in}}\right|^3\right)$ (see e.g., Ref. \cite{mukamel1999principles}). Hence, Eqs. \ref{eq:aiseom} and \ref{eq:aieom} are valid to $O\left(\left|E_{\t{in}}\right|^3\right)$ \footnote{
More explicitly, we used,

\begin{align*}
\langle a_{i}^{\dagger}(t)a_{i}(t)a_{i}(t)\rangle & =\langle\psi|a_{i}^{\dagger}(t)a_{i}(t)a_{i}(t)|\psi\rangle\\
 & =\langle\psi(t)|a_{i}^{\dagger}a_{i}a_{i}|\psi(t)\rangle\\
 & =\sum_{n_{i}}\langle\psi(t)|a_{i}^{\dagger}|n_{i}\rangle\langle n_{i}|a_{i}a_{i}|\psi(t)\rangle\\
 & =\sum_{n_{i}}c_{n_{i}+1}^{*}(t)\sqrt{(n_{i}+1)(n_{i}+1)(n_{i}+2)}c_{n+2}(t)\\
 & =\underbrace{c_{1}^{*}(t)\sqrt{2}c_{2}(t)}_{=O(|E_{\text{in}}|^{3})}+O\left(|E_{\text{in}}|^{5}\right)\\
 & = \langle\psi(t)|a_{i}^{\dagger}|0\rangle\langle0|a_{i}a_{i}|\psi(t)\rangle +O\left(|E_{\text{in}}|^{5}\right) \\
 & \approx\underbrace{\langle\psi(t)|a_{i}^{\dagger}|\psi(t)\rangle}_{=O(|E_{\text{in}}|)}\underbrace{\langle\psi(t)|a_{i}a_{i}|\psi(t)\rangle}_{=O(|E_{\text{in}}|^{2})} \\
 & \approx\langle a_{i}^{\dagger}(t)\rangle\langle a_{i}(t)a_{i}(t)\rangle,
\end{align*}
where we used the resolution of the identity for the $i$th oscillator
$I_{i}=\sum|n_{i}\rangle\langle n_{i}|$ in terms of Fock states $\{|n_{i}\rangle\}$,
and kept contributions to the wavefunction $|\psi(t)\rangle$ to $O(|E_{\text{in}}|^{3})$.}.

Each of the time-dependent expectation values appearing in the Eqs. \ref{eq:beom} and \ref{eq:aieom} admits an expansion in powers of the input field $\braket{b_{\t{in}}^{\t{L}}}$ (since the cavity is only weakly-coupled to the external modes). For instance, we can write $\braket{a_i(t)} = \sum_{p} \braket{a_i(t)}^{(p)}$, where $\braket{a_i(t)}^p = \t{O}\left[\left(b_{\t{in}}^{\t{L}}\right)^p\right]$. Hereafter, we will employ the following frequency-domain expansion of the expectation value of time-dependent operators, e.g.,
\begin{align} \braket{O}(t) = \sum_{\omega} \braket{O}(\omega) e^{-i \omega t} = \sum_{\omega>0}\left(\braket{O}^{(+)}(\omega)e^{-i\omega t} +\braket{O}^{(-)}(\omega) e^{i \omega t}\right).\end{align}
 Performing an expansion of both sides of Eqs. \ref{eq:beom} and \ref{eq:aieom} in powers of the input electric field amplitude  we find the third-order contribution to the cavity and molecular annihilation operator expectation values satisfy the following coupled equations in the frequency domain:
\begin{align}
& (\hbar\omega-\hbar\tilde{\omega}_c)\braket{b}^{(3)}(\omega) = - \mu\bar{E_0^c} \sum_{i=1}^N \braket{a_i}^{(3)}(\omega), \label{eq:b3_int}\\
&\left(\hbar\omega -\hbar\tilde{\omega}_0\right)\sum_{i=1}^N\braket{\mu a_i}^{(3)}(\omega) = -N \mu^2 E_0^c \braket{b}^{(3)}(\omega) -2\hbar\Delta \mu \sum_{\omega_a}\sum_{\omega_b} \sum_{i=1}^N \braket{a_i^\dagger}^{(1)}(-\omega_a) \braket{a_ia_i}^{(2)}(\omega_b)\delta_{\omega,-\omega_a+\omega_b}.\label{eq:p3_int}
\end{align}
The positive frequency material third-order polarization component with frequency $\omega$ is given by $\braket{P}^{(3)}(\omega) = \mu\sum_{i=1}^N \braket{a_i}^{(3)}(\omega)$. As shown above, it can be expressed in terms of the photonic variable $\braket{b}^{(3)}(\omega)$ and lower-order molecular correlators. Inserting the formal solution  of Eq. \ref{eq:p3_int} into Eq.\ref{eq:b3_int}, we find:
\begin{align}
&\braket{b}^{(3)}(\omega) = 2\hbar\Delta \mu \bar{E}_0^c \sum_{\omega_a\omega_b}\sum_{i=1}^N\frac{\braket{a_i^\dagger}^{(1)}(-\omega_a)\braket{a_i a_i}^{(2)}(\omega_b)}{(\hbar\omega-\hbar\tilde{\omega}_c)(\hbar\omega-\hbar\tilde{\omega}_0)-N|\mu E_0^c|^2} \delta_{\omega,-\omega_a+\omega_b},\label{eq:bf3} \\
&\sum_{i=1}^4\mu\braket{a_i}^{(3)}(\omega) = -2\hbar\Delta \mu \sum_{\omega_a\omega_b}\sum_{i=1}^N\frac{\hbar\omega-\hbar\tilde{\omega}_c}{(\hbar\omega-\hbar\tilde{\omega}_c)(\hbar\omega-\hbar\tilde{\omega}_0)-N|\mu E_0^c|^2} \braket{a_i^\dagger}^{(1)}(-\omega_a)\braket{a_i a_i}^{(2)}(\omega_b)\delta_{\omega,-\omega_a+\omega_b}. \label{eq:af3}
\end{align}
The first-order molecular expectation values $\braket{\mu a_i(\omega_a)}^{(1)}$ describe the linear polarization induced on each molecule. By solving the coupled  cavity-matter equations (Eqs. \ref{eq:beom} and \ref{eq:aieom}) to first-order in the input field, we can obtain the  linear molecular polarization in the strongly coupled device. In the frequency domain the equations to be solved are:
\begin{align} 
& \left(\hbar\omega-\hbar\tilde{\omega}_c\right)\braket{b}^{(1)}(\omega) = -i\hbar\sqrt{\frac{\kappa}{2}}b_{\t{in}}^{\t{L}}(\omega) - \mu\bar{E_0^c} \sum_{i=1}^N \braket{a_i}^{(1)}(\omega),  \\
&\left(\hbar\omega-\hbar\tilde{\omega}_0\right)\sum_{i=1}^N \braket{a_i}^{(1)}(\omega) = -N\mu E_0^c \braket{b}^{(1)}(\omega).
\end{align}
The explicit solution for the linear polarization $\braket{P}^{(1)}(\omega) \equiv \sum_{i=1}^N \braket{\mu a_i}^{(1)}(\omega)$ induced by the input field is given by:
\begin{align}
	\braket{P}^{(1)}(\omega) & = i\hbar\sqrt{\frac{\kappa}{2}}\frac{N \mu^2 E_0^c b_{\t{in}}^{\t{L}}(\omega)}{(\hbar\omega-\hbar\tilde{\omega}_c)(\hbar\omega-\hbar\tilde{\omega}_0)-\left|\mu E_0^c\right|^2 N} \nonum
& =  N \mu G_{mm}(\omega) \mu \left[E_{0}^c  G_{pp}^{(0)}(\omega) i\hbar\sqrt{\frac{\kappa}{2}}   b_{\t{in}}^{\t{L}}(\omega)\right], \label{eq:aisol}
\end{align}
where $G_{pp}^{(0)}(\omega) = 1/(\hbar \omega-\hbar \tilde{\omega}_c)$ is the bare cavity photon frequency-domain propagator, and $G_{mm}(\omega)$ is the single-molecule response function renormalized due to the material strong interaction with the optical cavity 
\begin{align}
G_{mm}(\omega) = \frac{1}{\hbar\omega -\hbar\tilde{\omega}_0- \frac{\left|\mu E_0^c\right|^2 N}{\hbar\omega-\hbar\tilde{\omega}_c}}.
\end{align}
Note the light-matter weak-coupling limit for the molecular response function $G_{mm}^{(0)} =1/(hbar\omega-\hbar \tilde{\omega}_0)$ can be straightforwardly obtained from the above expression by performing a power series expansion in terms of $|\mu E_0^c|$. The linear response induced by the external field on the cavity photon is similarly given by:
\begin{align}
	\braket{b}^{(1)}(\omega) & = -i\hbar\sqrt{\frac{\kappa}{2}}\frac{(\hbar\omega-\hbar\tilde{\omega}_0)b_{\t{in}}^{\t{L}}(\omega)}{(\hbar\omega-\hbar\tilde{\omega}_c)(\hbar\omega-\hbar\tilde{\omega}_0)-\left|\mu E_0^c\right|^2N} \nonum & =G_{pp}(\omega)\left[-i\hbar\sqrt{\frac{\kappa}{2}}b_{\t{in}}^{\t{L}}(\omega)\right], \label{eq:bsol}
\end{align}
where $G_{pp}(\omega)$ is the frequency-domain representation of the cavity photon retarded propagator under strong coupling conditions
\begin{align}
G_{pp}(\omega) = \frac{1}{\hbar \omega-\hbar\tilde{\omega}_c - \frac{\left|\mu E_0^c\right|^2 N}{\hbar\omega-\hbar \tilde{\omega}_0}}.
\end{align} 
Note that the hybrid cavity linear response field amplitude given by Eq. \ref{eq:bsol} has the same form as that for an empty cavity (\ref{eq:bempty}). The bare cavity result is obtained trivially by simply taking $\mu \rightarrow 0$ in Eq. \ref{eq:bsol}. The following relationship between the cavity and molecular polarization retarded Green functions will be useful later:
\begin{align}
G_{pp}(\omega) =  G_{pp}^{(0)}(\omega) \frac{G_{mm}(\omega)}{G_{mm}^{(0)}(\omega)} \label{eq:gppgmm}\end{align}

The last expectation value which we need to compute in order to obtain the hybrid cavity third-order response is $\braket{a_i(t) a_i(t)}^{(2)}$ (see Eq. \ref{eq:bf3}). The time-dependence of this function is coupled to the other totally-symmetric (with respect to permutation of the molecular indices) 2-particle variables of the system, namely, $\braket{b(t)b(t)}^{(2)}$ which describes the evolution of the two-cavity photon state, $\braket{a_i(t) b(t)}^{(2)}$ which probes the correlated propagation of a photon and the $i$th molecule phonon, and $\braket{a_i(t) a_j(t)}^{(2)}, i \neq j$, that describes propagation of vibrational excited-states in distinct molecules. 

The system of Heisenberg-Langevin equations for the bright two-particle variables mentioned above can be derived using the operator equations of motion generated Hamiltonian in Eq. (4) of the main text, together with the same replacements effected above $\omega_0 \rightarrow \omega_0 - i\gamma_m/2$, and $\omega_c \rightarrow \omega_c - i\kappa/2$. It follows from the input-output treatment \cite{steck2007quantum} that under the assumptions of Markovian molecular bath, and in the absence of an input molecular polarization, the resulting two-particle EOMs are given by: 
\small
\begin{align}
&\left[i\hbar\frac{\mathrm{d}}{\mathrm{d}t} -2\left(\hbar\tilde{\omega}_0-\hbar\Delta\right)\right]\braket{a_i(t)a_i(t)}^{(2)} = -2\mu E_0^c \braket{a_i(t)b(t)}^{(2)}, \label{eq:aiai}\\
&\left[i\hbar\frac{\mathrm{d}}{\mathrm{d}t} -(\hbar\tilde{\omega}_c+\hbar\tilde{\omega}_0)\right]\braket{a_i(t)b(t)}^{(2)}= -\mu E_0^c\braket{b(t) b(t)}^{(2)}- \mu \bar{E_0^c}\sum_{j=1}^N\braket{a_i(t)a_j(t)}^{(2)}  -i\hbar\sqrt{\frac{\kappa}{2}}b_{\t{in}}^{\t{L}}(t)\braket{a_i(t)}^{(1)}, \label{eq:aib}\\
&\left[i\hbar\frac{\mathrm{d}}{\mathrm{d}t} -2\hbar\tilde{\omega}_c\right]\braket{b(t)b(t)}^{(2)} = -\mu \bar{E_0^c}\sum_i \braket{a_i(t)b(t)}^{(2)}-2i\hbar\sqrt{\frac{\kappa}{2}}b_{\text{in}}(t)\braket{b(t)}^{(1)},\label{eq:bb}\\
&\left[i\hbar\frac{\mathrm{d}}{\mathrm{d}t} -2\hbar\tilde{\omega}_0\right]\braket{a_i(t)a_j(t)}^{(2)} = -\mu E_0^c\left[\braket{a_i(t)b(t)}^{(2)}+\braket{a_j(t)b(t)}^{(2)}\right], ~~j \neq i \label{eq:aiaj}.
\end{align}
\normalsize
These equations show, as expected, that two-particle states are driven by the input field only in the presence of non-vanishing first-order photonic or molecular polarization (represented by $\braket{b(t)}^{(1)}$ and $\braket{a_i(t)}^{(1)}$). To solve this system in the frequency domain, we note that the electromagnetic field interacts equally with each molecule, and therefore, $\braket{a_i(t)b(t)} = \braket{a_j(t)b(t)}$, for all $i,j \in \{1,...,N\}$. From the same argument, it also follows that the correlators $\braket{a_i(t)a_j(t)}_{i\neq j}$, and $\braket{a_i(t) a_i(t)}$ are independent of the molecular indices. These considerations imply that, while the system of two-particle eqs. given above has $(N+1)^2$ unknowns, only four of those are independent. In order to proceed, we need $\braket{a_ia_i}^{(2)}(\omega)$ which can be written as:
 \begin{align}
&\braket{a_i a_i}^{(2)}(\omega) =  \frac{2(\hbar\omega-2\hbar\tilde{\omega}_0)}{D(\omega)} \left(\mu E_0^c\right)^2f_{\t{ext}}^{bb}(\omega) - \frac{2(\hbar\omega-2\hbar\tilde{\omega}_c)(\hbar\omega-2\hbar\tilde{\omega}_0)}{D(\omega)}\mu E_0^c f_{\t{ext}}^{mb}(\omega), \label{eq:aiaisol} 
\end{align}
where $f_{\t{ext}}^{bb}(\omega) = -2i\hbar\sqrt{\frac{\kappa}{2}}\braket{b_{\t{in}}^{\t{L}}b^{(1)}}^{(2)}(\omega)$ and $f_{\t{ext}}^{mb}(\omega)=-i\hbar\sqrt{\frac{\kappa}{2}}\braket{b_{\t{in}}^{\t{L}}a_i^{(1)}}^{(2)}(\omega)$, and $D(\omega)$ is a 4th-order polynomial, with its roots corresponding to the bright resonances of the doubly-excited manifold of the system. Denoting by $D^{(0)}(\omega)$ the bare noninteracting 2-particle resonances
\begin{align} D^{(0)}(\omega) = (\hbar\omega-\hbar\tilde{\omega}_c-\hbar\tilde{\omega}_0)(\hbar\omega-2\hbar\tilde{\omega}_0+2\hbar\Delta)(\hbar\omega-2\hbar\tilde{\omega}_c)(\hbar\omega-2\hbar\tilde{\omega}_0), \end{align} it follows that the interacting complex two-particle energy eigenvalues are given by the roots of 
\begin{align}  D(\omega) = & D^{(0)}(\omega)   -2g^2N(\hbar\omega-2\hbar\tilde{\omega}_0)(\hbar\omega-2\hbar\tilde{\omega}_0+2\hbar\Delta) -2g^2(N-1)(\hbar\omega-2\hbar\tilde{\omega}_0+2\hbar\Delta)(\hbar\omega-2\hbar\tilde{\omega}_c)  \nonum & -2g^2(\hbar\omega-2\hbar\tilde{\omega}_c)(\hbar\omega-2\hbar\tilde{\omega}_0), \end{align}
where $g^2 = |\mu E_0^c|^2$ as in the main text. We can also write Eq. \ref{eq:aiaisol} in terms of retarded single and two-particle Green functions in the frequency domain:
\begin{align}
\braket{a_i a_i}^{(2)}(\omega) & =  -2i\hbar \sqrt{\frac{\kappa}{2}}G_{mm,pp}(\omega) \left(\mu E_0^c\right)^2\braket{b_{\t{in}}^{\t{L}}b^{(1)}}(\omega) +i\hbar \sqrt{\frac{\kappa}{2}} G_{mm,mp}(\omega)\mu E_0^c \braket{b_{\t{in}}^{\t{L}} a_i^{(1)}}(\omega), \nonum
& = -\hbar^2\frac{\kappa}{2}\sum_{uv}\left[2G_{mm,pp}(\omega_u+\omega_v) G_{pp}(\omega_u)+ G_{mm,mp}(\omega_u+\omega_v)G_{mm}(\omega_u)G_{pp}^{(0)}(\omega_u)\right] \left(\mu E_0^c\right)^2b_{\t{in}}^{\t{L}}(\omega_v) b_{\t{in}}^{\t{L}}(\omega_u)\delta_{\omega,\omega_u+\omega_v}\end{align}
where $G_{mm,pp}(\omega)$ corresponds to the Fourier transform of the probability amplitude for a  two-cavity photon state to undergo a transition into a state where a given molecule is doubly excited, and $G_{mm,mp}(\omega)$ is the transition amplitude into the doubly-excited state of a given molecule from an initial state containing a photon and a single vibrational excitation of the same molecule. These propagators can be written explicitly as:
\begin{align}
	& G_{mm,pp}(\omega) =  \frac{2 (\hbar\omega-2\hbar\omega_0+i\hbar\gamma_m)}{D(\omega)},\\
	& G_{mm,mp}(\omega) = \frac{2(\hbar\omega-2\hbar\omega_c+i\hbar\kappa)(\hbar\omega-2\hbar\omega_0+i\hbar\gamma_m)}{D(\omega)}.
\end{align}
Using the relation introduced in Eq. \ref{eq:gppgmm}, we rewrite the two-particle molecular response as:
\begin{align}
\braket{a_i a_i}^{(2)}(\omega) = & \frac{1}{2}\sum_{uv}\left\{2G_{mm,pp}(\omega_u+\omega_v)\left[G_{mm}^{(0)}(\omega_u)\right]^{-1}\left[G_{pp}^{(0)}(\omega_v)\right]^{-1}+ G_{mm,mp}(\omega_u+\omega_v)\left[G_{pp}^{(0)}(\omega_v)\right]^{-1}\right\} \nonum
& \times G_{mm}(\omega_u) \left(\mu E_0^c\right)^2  \left[-i\hbar \sqrt{\frac{\kappa}{2}}G_{pp}^{(0)}(\omega_v) b_{\t{in}}^{\t{L}}(\omega_v)\right]\left[-i\hbar \sqrt{\frac{\kappa}{2}} G_{pp}^{(0)}(\omega_u)b_{\t{in}}^{\t{L}}(\omega_u)\right]\delta_{\omega,\omega_u+\omega_v}.
\end{align}
By symmetrizing the summand of the previous equation, we obtain:
\begin{align}
\braket{a_i a_i}^{(2)}(\omega) = & \sum_{uv}\Gamma_{mm,mm}(\omega_u+\omega_v) G_{mm}(\omega_u) G_{mm}(\omega_v) \left[-\mu E_0^c i\hbar \sqrt{\frac{\kappa}{2}} G_{pp}^{(0)}(\omega_v)b_{\t{in}}^{\t{L}}(\omega_v)\right]\left[-\mu E_0^c i\hbar \sqrt{\frac{\kappa}{2}} G_{pp}^{(0)}(\omega_u)b_{\t{in}}^{\t{L}}(\omega_u)\right]\nonum & \times \delta_{\omega,\omega_u+\omega_v},
\end{align}
where $\Gamma$ is the two-particle scattering matrix, and $\Gamma_{mm,mm}$ is the amplitude for the elastic scattering of two excitations in the same molecule. It may be written as:
\begin{align}
\Gamma_{mm,mm}(\omega) = \frac{(\hbar\omega-2\hbar\tilde{\omega}_0)(\hbar\omega-\hbar\tilde{\omega}_0-\hbar\tilde{\omega}_c)\left[(\hbar\omega-2\hbar\tilde{\omega}_0)(\hbar\omega-2\hbar\tilde{\omega}_c)-4g^2 N \right]}{D(\omega_u+\omega_v)}.
\end{align}
The nonlinear component of the molecular polarization $\braket{P(\omega_s)}^{(3)} = \mu \sum_{i=1}^N\braket{a_i(\omega_s)}^{(3)}$ can now be given the explicit form:
\begin{align}
\braket{P(\omega_s)}^{(3)} = &\sum_{\omega_u\omega_v\omega_w} 2N\hbar\Delta \mu^4 G_{mm}(\omega_s)\bar{G}_{mm}(\omega_w)\Gamma_{mm,mm}(\omega_u+ \omega_v)G_{mm}(\omega_v) G_{mm}(\omega_u)\times \nonum
&G_{pp}^{(0)}(\omega_v)\bar{G}_{pp}^{(0)}(\omega_w)G_{pp}^{(0)}(\omega_u)\left(\frac{\hbar \kappa}{2}\sqrt{\frac{2\mathcal{F}}{\pi}}\right)^3 E_{\t{in}}^{(+)}(\omega_v)E_{\t{in}}^{(-)}(\omega_w) E_{\t{in}}^{(+)}(\omega_u)\delta_{\omega_s,\omega_v-\omega_w+\omega_u},
\end{align}
where we used $-iE_0^c \sqrt\frac{\kappa}{2} b_{\t{in}}^{\t{L}}(\omega) \approx \frac{\kappa}{2} \sqrt{\frac{2\mathcal{F}}{\pi}} E_{\text{in}}^{(+)}(\omega)$ (from Eq. \ref{eq:e0ec}). From the above expression and the definition of the molecular nonlinear susceptibility \cite{boyd2008nonlinear}, we find
\begin{align}
\chi^{(3)}(-\omega_s;\omega_v,-\omega_w,\omega_u) = 2\hbar\Delta N \mu G_{mm}(\omega_s)\bar{G}_{mm}(\omega_w)\Gamma_{mm,mm}(\omega_u+\omega_v)G_{mm}(\omega_u) G_{mm}(\omega_v)\times
\nonum
& \left[\mu \sqrt{\frac{2\mathcal{F}}{\pi}}\frac{\hbar\kappa}{2} G_{pp}^{(0)}(\omega_v)\right]\left[\mu \sqrt{\frac{2\mathcal{F}}{\pi}} \frac{\hbar \kappa}{2} \bar{G}_{pp}^{(0)}(\omega_w)\right] \left[\mu \sqrt{\frac{2\mathcal{F}}{\pi}}\frac{\hbar \kappa}{2} G_{pp}^{(0)}(\omega_u)\right]\delta_{\omega_s,\omega_v-\omega_w+\omega_u}. \label{eq:chi3_ne}
\end{align}

\section{Nonlinear absorption spectrum under strong coupling}\label{sec:nlabs_strong}
In this section, we compute the nonlinear part of the absorption spectrum of an optical microcavity strongly coupled to the molecular polarization. In particular, we will calculate the nonlinear part (in the input electric field amplitude) of the external field power dissipated by the molecular system under steady-state conditions.

 Mathematically, the steady-state regime is characterized by a time-independent molecular excited-state population i.e., $\partial_t \sum_{i=1}^N\left[a_i^\dagger(t) a_i(t)\right] = 0$. Using the Heisenberg-Langevin equation for $\sum_{i=1}^N a_i^\dagger(t) a_i(t)$, we find that steady-state implies:
\begin{align}
& 0 = -i\hbar\gamma_m \sum_{i=1}^N a_i^\dagger(t) a_i(t)+ \mu \left[\bar{E}_0^c b^\dagger(t) \sum_{i=1}^N a_i(t) - E_0^c \sum_{i=1}^N a_i^\dagger(t) b(t)\right] , \nonum
& \implies \sum_{i=1}^N \gamma_m a_i^\dagger a_i = 2\t{Im}
\left[\sum_{i=1}^N \frac{\mu \bar{E}_0^c }{\hbar}b^\dagger a_i\right]
\end{align}
The last equality expresses the balance between the steady-state rate of molecular excited-state decay (l.h.s.) and driving by the external field mediated by the cavity (r.h.s). Hence, the photon absorption rate by the molecular system can be written as:
\begin{align}
W = \frac{2}{\hbar} \t{Im}
\braket{E_c^\dagger P}_{\t{ss}},
\end{align}
where $E_c^\dagger = \bar{E}_0^c b^\dagger$ and $P$ are the (complex conjugate) cavity electric field amplitude and collective molecular polarization in steady-state, respectively. Both $E_c$ and $P$ admit power series expansions in the external fields (see Sec. \ref{sec:scav}). The first non-vanishing nonlinear terms scales cubically with the input field $b_{\t{in}}^{\t{L}}$. Therefore, it follows that the nonlinear response contribution to the photon absorption rate $W$ scales as $|E_{\t{in}}|^4$. To obtain this quantity, we will solve the coupled Heisenberg-Langevin EOMs for population and coherence variables in the presence of driving by the external input fields. From now on, we will denote steady-state quantities by the usual expectation value notation without the subscript ``ss", as we will always work under steady-state conditions. Moreover, we will disregard the frequency dependence of all quantities until we obtain the final expression for the nonlinear absorption. In this section, we take the input field to be a monochromatic beam, i.e., $b_{in}(\omega') = 0$ for all $\omega' \neq \omega$. 

In steady-state, the cavity-molecular polarization coherence $\braket{E_c^\dagger P}$ satisfies
\begin{align}
\left(\hbar \tilde{\omega}_c^*-\hbar \tilde{\omega}_0\right)\braket{E_c^\dagger P} =  -\mu^2 |E_0^c|^2 \left(N \braket{b^\dagger b} - \sum_{ij=1}^N \braket{a_i^\dagger a_j} \right)-2\hbar\Delta \mu \sum_{i=1}^N \braket{E_c^\dagger a_i^\dagger a_i a_i}-i\hbar \sqrt{\frac{\kappa}{2}} \braket{\bar{E}_0^c \left(b_{\t{in}}^{\t{L}}\right)^\dagger P} .
\end{align}
Because we only care about the $O\left(\left|b_{\t{in}}^{\t{L}}\right|^4\right)$ absorption component, and we have assumed our system is in a pure-state, it follows by the same argument employed in Sec. \ref{sec:scav} that $\braket{b^\dagger a_i^\dagger a_i a_i} =  \braket{b^\dagger a_i^\dagger} \braket{a_i a_i}$ . Thus,
\begin{align}
\left(\hbar \tilde{\omega}_c^*-\hbar \tilde{\omega}_0\right)\braket{E_c^\dagger P}^{(4)} =  
-\mu^2 |E_0^c|^2 \left(N \braket{b^\dagger b}^{(4)} - \sum_{ij=1}^N \braket{a_i^\dagger a_j}^{(4)} \right)-2\hbar\Delta \mu \sum_{i=1}^N \braket{E_c^\dagger a_i^\dagger}^{(2)}\braket{a_i a_i}^{(2)}-i\hbar \sqrt{\frac{\kappa}{2}} \bar{E}_0^c \left(b_{\t{in}}^\t{L}\right)^{\dagger}\braket{P}^{(3)}, \label{eq:ecbarp4}
\end{align}
where we also used that the input fields are classical states uncorrelated with the cavity. Our task is now to express the steady-state cavity photon number $N_{\t{p}} = \braket{b^\dagger b}$, total molecular excited-state population $N_m = \sum_{i=1}^N \braket{a_i^\dagger a_i}$ and intermolecular coherences $\braket{a_i^\dagger a_j}_{i\neq j}$ in terms of the input field operators to the desired orders. The  steady-state cavity photon number satisfies
\begin{align}
N_p^{(4)} = -\frac{2}{\hbar \kappa}
\t{Im}\braket{E_c^\dagger P}^{(4)}
-\sqrt{\frac{2}{\kappa}} \t{Re}\braket{\left(b_{\t{in}}^L\right)^{\dagger} b}^{(4)},\end{align}
whereas the total molecular excited-stated population and intermolecular coherences are given by:
\begin{align}
& N_m^{(4)} = \frac{2}{\hbar \gamma_m} \t{Im}\braket{E_c^\dagger P}^{(4)}, \\
& \sum_{i>j} \sum_{j=1}^N \left(\braket{a_i^\dagger a_j}^{(4)}+\braket{a_j^\dagger a_i}^{(4)}\right)= \frac{2 (N-1)}{\hbar \gamma_m}\t{Im}\braket{E_c^\dagger P}^{(4)}+\frac{4\Delta}{\gamma_m} \sum_{ij=1}^N\t{Im}\left[\braket{a_j^\dagger a_j^\dagger}^{(2)}\braket{a_j a_i}^{(2)}\right],
\end{align}
where to obtain the last line, we used $\t{Im}\left[\braket{a_j^\dagger a_j^\dagger}^{(2)}\braket{a_j a_j}^{(2)}\right] = \t{Im}\left[\left|\braket{a_j^\dagger a_j^\dagger}^{(2)}\right|^2\right] = 0.$ Using the last three results, we find the intermediate result
\begin{align}
N N_p^{(4)} - \sum_{ij=1}^n \braket{a_i^\dagger a_j}^{(4)} = -\frac{2N}{\hbar \eta}\t{Im}\braket{E_c^\dagger P}^{(4)} -N \sqrt{\frac{2}{\kappa}}\t{Re}\braket{\left(b_{\t{in}}^{\t{L}}\right)^\dagger b}^{(4)}-\frac{4\Delta}{\gamma_m}\sum_{ij=1}^N \t{Im}\left[\braket{a_j^\dagger a_j^\dagger}^{(2)}\braket{a_j a_i}^{(2)}\right],
\end{align}
where $\eta^{-1} \equiv \kappa^{-1}+\eta^{-1}$. We now have all of the quantities required to obtain the rate of nonlinear absorption $W$. In particular, it follows from inserting our last result in Eq. \ref{eq:ecbarp4} that
\begin{align}
\left(\hbar\tilde{\omega}_c^*-\hbar\tilde{\omega}_0\right)\braket{E_c^\dagger P}^{(4)}  = &  \frac{\Omega_R^2}{2\hbar\eta}\t{Im}\left[\braket{E_c^\dagger P}^{(4)}\right]  + \frac{\Omega_R^2}{4} \sqrt{\frac{2}{\kappa}} \t{Re}\left[\left(b_{\t{in}}^{\t{L}}\right)^\dagger \braket{b}^{(3)}\right]+ \frac{\Omega_R^2 \Delta}{N\gamma_m}\sum_{ij=1}^N \t{Im}\left[\braket{a_j^\dagger a_j^\dagger}\braket{a_j a_i}\right] \nonum
& -2\hbar\Delta \mu \sum_{i=1}^N \braket{a_i a_i}^{(2)} \braket{a_i^\dagger E_c^\dagger}^{(2)} - i\hbar \sqrt{\frac{\kappa}{2}}\bar{E}_0^c \braket{b_{\t{in}}^{\t{L}}}^\dagger \braket{P}^{(3)}.\label{eq:intaib}
\end{align}
where we used $\Omega_R = 2\left|\mu E_0^c\right|\sqrt{N}$.  Note the l.h.s of the previous equation can be written as:
\begin{align}
\left(\hbar \tilde{\omega}_c^* -\hbar\tilde{\omega}_0\right)\left[\t{Re}\braket{E_c^\dagger P}^{(4)}
+i \t{Im}\braket{E_c^\dagger P}^{(4)}\right] = &(\hbar\omega_c-\hbar\omega_0) \t{Re}\left[\braket{E_c^\dagger P}^{(4)}\right] - \frac{\hbar\eta_s}{2}\t{Im}\left[\braket{E_c^\dagger P}^{(4)}\right] \nonum
&+i\left[\frac{\hbar\eta_s}{2}\t{Re}\left(\braket{E_c^\dagger P}^{(4)}\right)+(\hbar\omega_c-\hbar\omega_0)\t{Im}\left(\braket{E_c^\dagger P}^{(4)}\right)\right].\end{align}
where we introduced the notation $\eta_s = \kappa + \gamma_m$. Using the above to equate the imaginary part of the left and right-hand-side of Eq. \ref{eq:intaib}, we find that:
\begin{align}
& \t{Re}\left(\braket{E_c^\dagger P}^{(4)}\right)+\frac{\hbar(\omega_c-\omega_0)}{\eta_s}W^{(4)} =  -\frac{4\Delta \mu N}{\eta_s} \t{Im}\left[\braket{a_i a_i}^{(2)} \braket{a_i^\dagger E_c^\dagger}^{(2)}\right] -\frac{\sqrt{2\kappa}}{\eta_s} \t{Re}\left[\bar{E}_0^c \left(b_{\t{in}}^{\t{L}}\right)^\dagger \braket{P}^{(3)}\right],
\end{align}
\begin{align}
 (\omega_c-\omega_0) \t{Re}\left(\braket{E_c^\dagger P}^{(4)}\right) -\frac{\hbar\eta_s}{4} W^{\t{NL}} = & \frac{\Omega_R^2}{4\hbar\eta}W^{\t{NL}}  + \frac{\Omega_R^2}{4\hbar}\left\{\sqrt{\frac{2}{\kappa}} \t{Re}\left[\left(b_{\t{in}}^{\t{L}}\right)^\dagger \braket{b}^{(3)}\right]+ \frac{4\Delta}{N\gamma_m}\sum_{ij=1}^N \t{Im}\left[\braket{a_j^\dagger a_j^\dagger}^{(2)}\braket{a_j a_i}^{(2)}\right]\right\}\nonum
& -2\Delta \mu \sum_{i=1}^N \t{Re}\left[\braket{a_i a_i}^{(2)} \braket{a_i^\dagger E_c^\dagger}^{(2)}\right] +  \sqrt{\frac{\kappa}{2}}\t{Im}\left[\bar{E}_0^c \braket{b_{\t{in}}^{\t{L}}}^\dagger \braket{P}^{(3)}\right],
\end{align}
where we used $W^{\t{NL}} \equiv  W^{(4)} =  \frac{2}{\hbar}\t{Im}\left(\braket{E_c^\dagger P}^{(4)}\right)$. We can now eliminate $\t{Re}\left(\braket{E_c^\dagger P}^{(4)}\right)$ and solve for $W^{\t{NL}}$ in terms of the input field variables. This procedure gives
\begin{align}
W^{\t{NL}} =& -\frac{2\eta\hbar\eta_s}{2\eta \left[(\hbar\omega_c-\hbar\omega_0)^2+(\hbar\eta_s)^2/4\right]+\Omega_R^2 \eta_s/2}\frac{\Omega_R^2}{4\hbar}\left\{\sqrt{\frac{2}{\kappa}} \t{Re}\left[\left(b_{\t{in}}^{\t{L}}\right)^\dagger\braket{b}^{(3)}\right]+ \frac{4\Delta(N-1)}{\gamma_m}\t{Im}\left[\braket{a_j^\dagger a_j^\dagger}^{(2)}\braket{a_j a_i}_{j\neq i}^{(2)}(2\omega)\right]\right\}\nonum
& +\frac{2\eta\hbar\eta_s}{2\eta \left[(\hbar\omega_c-\hbar\omega_0)^2+(\hbar\eta_s)^2/4\right]+\Omega_R^2 \eta_s/2} \left\{2\Delta\mu N\t{Re}\left[\braket{a_i a_i}^{(2)} \braket{a_i^\dagger E_c^\dagger}^{(2)}\right] +\sqrt{\frac{\kappa}{2}}\t{Im}\left[\bar{E}_0^c \braket{b_{\t{in}}^{\t{L}}}^\dagger \braket{P}^{(3)}\right]\right\}  \nonum
& + \frac{2\eta\hbar\eta_s(\omega_c-\omega_0)}{2\eta \left[(\hbar\omega_c-\hbar\omega_0)^2+(\hbar\eta_s)^2/4\right]+\Omega_R^2 \eta_s/2}  \left\{\frac{4\Delta \mu N}{\eta_s} \t{Im}\left[\braket{a_i a_i}^{(2)} \braket{a_i^\dagger E_c^\dagger}\right] +\frac{\sqrt{2\kappa}}{\eta_s} \t{Re}\left[\bar{E}_0^c \left(b_{\t{in}}^{\t{L}}\right)^\dagger \braket{P}^{(3)}\right] \right\}, 
\end{align}
where we used that $\t{Im}\left[\braket{a_j^\dagger a_j^\dagger}\braket{a_j a_i}\right] = 0$ when $i=j$.
Thus, our final expression for the total nonlinear absorption is given by:
\begin{align}
W^{\t{NL}}(\omega) = & -\frac{\eta \Omega_R^2\eta_s}{2\eta \left[(\hbar\omega_c-\hbar\omega_0)^2+(\hbar\eta_s)^2/4\right]+\Omega_R^2 \eta_s/2}\t{Re}\left[\sqrt{\frac{1}{2\kappa}}\left[b_{\t{in}}^{\t{L}}(\omega)\right]^\dagger \braket{b}^{(3)}(\omega)\right]  \nonum
& -\frac{\eta\Omega_R^2\eta_s}{2\eta \left[(\hbar\omega_c-\hbar\omega_0)^2+(\hbar\eta_s)^2/4\right]+\Omega_R^2 \eta_s/2}\frac{2\Delta(N-1)}{\gamma_m}\t{Im}\left[\braket{a_j^\dagger a_j^\dagger}^{(2)}(-2\omega)\braket{a_j a_i}_{j\neq i}^{(2)}(2\omega)\right] \nonum
&+\frac{2\eta(2\hbar\Delta N)\eta_s}{2\eta \left[(\hbar\omega_c-\hbar\omega_0)^2+(\hbar\eta_s)^2/4\right]+\Omega_R^2 \eta_s/2} \t{Re}\left[\braket{a_i^\dagger a_i^\dagger}^{(2)}(-2\omega) \braket{\mu a_i E_c}^{(2)}(2\omega)\right] \nonum
&+  \frac{\eta\eta_s}{2\eta \left[(\hbar\omega_c-\hbar\omega_0)^2+(\hbar\eta_s)^2/4\right]+\Omega_R^2 \eta_s/2} \t{Im}\left[\hbar\sqrt{2\kappa}\bar{E}_0^c \left[b_{\t{in}}^{\t{L}}(\omega)\right]^\dagger \braket{P}^{(3)}(\omega)\right] \nonum
& +   \frac{2\eta(\hbar\omega_c-\hbar\omega_0)4\Delta  N}{2\eta \left[(\hbar\omega_c-\hbar\omega_0)^2+(\hbar\eta_s)^2/4\right]+\Omega_R^2 \eta_s/2}   \t{Im}\left[\braket{a_i a_i}^{(2)}(2\omega) \braket{\mu a_i^\dagger E_c^\dagger}^{(2)}(-2\omega)\right] \nonum
& + \frac{2\eta(\hbar\omega_c-\hbar\omega_0)\sqrt{2\kappa}}{2\eta \left[(\hbar\omega_c-\hbar\omega_0)^2+(\hbar\eta_s)^2/4\right]+\Omega_R^2 \eta_s/2}    \t{Re}\left[\bar{E}_0^c \left[b_{\t{in}}^{\t{L}}(\omega)\right]^\dagger \braket{P}^{(3)}(\omega)\right] \label{eq:wnlgen}.
\end{align}
Each of the above terms can be further simplified by using results obtained in Sec. \ref{sec:scav}. For instance, the identities $\braket{b}^{(3)}(\omega) = -\bar{E}_0^c G_{pp}^{(0)}(\omega) \braket{P}^{(3)}(\omega)$ and $i\bar{E}_0^c \left[b_{\t{in}}^{\t{L}}(\omega) \right]^{\dagger}= \sqrt{\frac{\kappa}{2}} \sqrt{\frac{2\mathcal{F}}{\pi}} E_{\t{in}}^{(-)}(\omega)$ (see Eq. \ref{eq:e0ec}) can be employed to simplify the first line of the last equation, while the 2nd, 3rd, and 5th lines can be simplified using the following results from Eqs. \ref{eq:aiai} and \ref{eq:aiaj}
\begin{align}
& \braket{a_i a_j}_{i\neq j}^{(2)}(\omega) = -\frac{\mu E_0^c}{\hbar\omega-2\hbar\tilde{\omega}_0} \left[\braket{a_j b}^{(2)}(\omega) +\braket{a_i b}^{(2)}(\omega)\right], ~\t{and} \nonum
& \braket{a_ib}^{(2)}(\omega) = -\frac{\hbar\omega-2\hbar\tilde{\omega}_0+2\hbar\Delta}{\mu E_0^c}\braket{a_i a_i}^{(2)}(\omega), ~\t{which imply that} \nonum
& \implies \braket{a_i a_j}_{i\neq j}^{(2)}(\omega) = \frac{(\hbar\omega-2\hbar\omega_0+2\hbar\Delta+i\hbar\gamma_m)(\hbar\omega-2\hbar\omega_0-i\hbar\gamma_m)}{(\hbar\omega-2\hbar\omega_0)^2+\hbar^2\gamma_m^2}\times 2\braket{a_i a_i}^{(2)}(\omega).
\end{align}

\subsection{Zero detuning}
The physical content of the terms in Eq. \ref{eq:wnlgen} becomes clearer in the zero-detuning case where $\omega_c \approx \omega_0$, in which case the last two lines of Eq. \ref{eq:wnlgen} vanish. Taking advantage also that when the strong coupling condition is satisfied $\Omega_R \gg \hbar \eta$ and $\Omega_R \gg \hbar \eta_s$, the nonlinear absorption can be written as a sum of four simple contributions
\begin{align} W^{\t{NL}}(\omega) = \sum_{\alpha=1}^4 W^{\t{NL}_\alpha}(\omega) \left|E_{\t{in}}^{(+)}(\omega)\right|^4,
\end{align}
where
\begin{align}
& W^{\t{NL}_1} (\omega) \approx -\frac{2\eta\kappa}{\hbar}\left[\frac{1}{4(\omega-\omega_0)^2+\kappa^2}+\frac{1}{(\Omega_R/\hbar)^2} \right]  \t{Re}\left[\sqrt{\frac{2\mathcal{F}}{\pi}}  \chi^{(3)}(\omega)\right], \label{eq:w1} \\
& W^{\t{NL}_2}(\omega) \approx \frac{4\eta}{\hbar}
\frac{\omega-\omega_0}{4(\omega-\omega_0)^2+\kappa^2}\t{Im}\left[\sqrt{\frac{2\mathcal{F}}{\pi}}  \chi^{(3)}(\omega)\right], \label{eq:w2}\\
& W^{\t{NL}_3} (\omega) \approx \eta \frac{4\Delta^2 N}{(\omega-\omega_0)^2+\gamma_m^2/4}\frac{\left|\braket{a_i a_i}^{(2)}(2\omega)\right|^2}{|E_{\t{in}}^{(+)}(\omega)|^4} \label{eq:w3}, \\
& W^{\t{NL}_4} (\omega) \approx -\eta \frac{(2\omega -\omega_{20})2\Delta N }{(\Omega_R/2\hbar)^2} \frac{\left|\braket{a_i a_i}^{(2)}(2\omega)\right|^2}{|E_{\t{in}}^{(+)}(\omega)|^4}\label{eq:w4}.
\end{align}

\section{Nonlinear response of bare molecular system}\label{sec:wout}
In order to describe the free space nonlinear polarization induced on a bare molecule ensemble driven by external continuous-wave fields, we employ an effective Hamiltonian that is similar to that used to model the molecular system in an optical cavity. The main difference is that each molecule now interacts with several EM modes (with vanishing momentum along the $x,y$ directions) quantized with periodic boundary conditions. The total Hamiltonian in the rotating-wave approximation is:
\begin{align}
H = \sum_{\omega} \hbar \omega b_{\omega}^{\dagger} b_{\omega} + \sum_{i=1}^N \left(\hbar \omega_0 a_i^\dagger a_i - \hbar \Delta a_i^\dagger a_i^\dagger a_i a_i\right) - \mu \sum_{i=1}^N\sum_{\omega > 0}  \left(E_{0\omega} a_i^\dagger b_{\omega} e^{i\omega z_i/c}+\bar{E}_{0\omega}a_i b_\omega^\dagger e^{-i\omega z_i/c}\right), \label{eq:bareham}
\end{align}
where $\omega = c k$, $k = 2\pi m/L, m \in \mathbb{Z}$, and $z_i$ is the projection of the position of molecule $i$ on the field direction of propagation, and $E_{0\omega}= i\sqrt{\frac{\hbar \omega}{2\epsilon_0 V}}$. The input field which drives the material polarization is introduced as a boundary condition to the electromagnetic mode operators in the Heisenberg picture, i.e., the input field satisfies the homogeneous part of the EM field equations. We assume the $E_{0\omega}$ are classical variables, as in the computation performed with the optical cavity in the previous sections.  \\

 The equation of motion for the expectation value of the molecular polarization is given by:
 \begin{align}
& \left(i\hbar\partial_t -\hbar\tilde{\omega}_0\right)\braket{\mu a_i(t)} = -2\hbar\Delta \mu \braket{a_i^\dagger(t) a_i(t) a_i(t)} -\mu^2 \sum_{\omega}E_{0\omega} \braket{b_{\omega}(t)} e^{i \omega z_i/c}.\label{eq:a_i0}
 \end{align}
Assuming the system is always in a pure state, the equation of motion for the third-order component of $\braket{a_i(t)}$ is given by:
\begin{align}
\left(i\hbar\partial_t-\hbar\tilde{\omega}_0\right)\braket{a_i(t)}^{(3)} = -2\hbar\Delta \braket{a_i^{\dagger}(t)}^{(1)}\braket{a_i(t) a_i(t)}^{(2)}.\label{eq:ai0td3}
\end{align}
The time evolution of the relevant first and second-order molecular expectation values are given by the solutions of the equations
\begin{align}
&\left(i\hbar\partial_t -\hbar\tilde{\omega}_0\right) \braket{a_i(t)}^{(1)} = - \mu \sum_{\omega'} E_{\omega'\t{in}}(t), \\
&\left(i\hbar\partial_t -\hbar\tilde{\omega}_{20} \right) \braket{a_i(t)a_i(t)}^{(2)}= -2\mu \sum_{\omega'} \braket{a_i(t)}^{(1)}E_{\omega'\t{in}}(t) e^{-i\omega' z_i/c},
\end{align}
where we made the replacement $E_{\omega\t{in}}(t) = E_{0\omega} \braket{b_{\omega}(t)}$. Using the long wavelength limit, and thus disregarding the spatial dispersion of the electromagnetic field (as in the computations performed for a molecular system in a cavity), the frequency-domain solutions of the prior equations are:
 \begin{align}
& \braket{a_i}^{(1)}(\omega) = -\mu \sum_{\omega_u}\frac{E_{\omega_u\t{in}}}{\hbar\omega-\hbar\tilde{\omega}_0}\delta_{\omega_u,\omega} , \label{eq:ai01}\\
& \braket{a_ia_i}^{(2)}(\omega)= \sum_{\omega_u}\sum_{\omega_v} \frac{2\mu^2E_{\omega_u\t{in}}E_{\omega_v\t{in}}}{(\hbar\omega_{u}+\hbar\omega_{v}-\hbar\tilde{\omega}_{20})(\hbar\omega_u-\hbar\tilde{\omega}_0)} \delta_{\omega,\omega_{u}+\omega_v}, \label{eq:aiai02}
\end{align}
where $\hbar\tilde{\omega}_{20} = \hbar\omega_{20}-i\hbar\gamma_m$, and  $\hbar\omega_{20} = 2\hbar\omega_0 -2\hbar\Delta$ is the energy difference between the doubly-excited vibrational state and the ground-state. These results can also be written in terms of bare molecule single-particle and two-particle retarded response functions in the frequency domain: 
\begin{align}
	& \braket{a_i}^{(1)}(\omega) = -\mu G_{mm}^{(0)}(\omega)E_{\omega\t{in}}, \nonum
	&\braket{a_i a_i }^{(2)}(\omega)= \mu^2\sum_{\omega_u,\omega_v} G_{mm,mm}^{(0)}(\omega_u+\omega_v)G_{mm}^{(0)}(\omega_u)E_{\omega_u\t{in}}E_{\omega_v\t{in}}\delta_{\omega,\omega_u+\omega_v},
\end{align}
where $G_{mm}^{(0)}(\omega) = 1/(\hbar\omega-\hbar\tilde{\omega}_0)$ and $G_{mm,mm}^{(0)}(\omega) = 2/(\hbar\omega-\hbar\tilde{\omega}_{20})$ are the Fourier transform of the single-particle and two-particle retarded molecular Green functions, respectively. In the time domain, they measure the probability amplitude that a single and a two-phonon state exist for a time $t$ after their creation. Note that the last equation may also be written in terms of a vibration-vibrational scattering matrix element $\Gamma_{mm,mm}^{(0)}(\omega) = (\hbar\omega -2\hbar\tilde{\omega}_0)/(\hbar\omega -2\hbar\tilde{\omega}_{20})$ as follows
\begin{align}
\braket{a_i a_i}^{(2)}(\omega) = \mu^2 \sum_{\omega_u,\omega_v} \Gamma_{mm,mm}^{(0)}(\omega_u+\omega_v) G_{mm}^{(0)}(\omega_u)G_{mm}^{(0)}(\omega_v)E_{\omega_u\t{in}} E_{\omega_v\t{in}}\delta_{\omega,\omega_u+\omega_v}.
\end{align}

Direct insertion of Eqs. \ref{eq:ai01} and \ref{eq:aiai02} into the frequency-domain representation of Eq. \ref{eq:ai0td3} gives the following solution:
\begin{align}
\braket{a_i}^{(3)}(\omega_s) =4\mu^3\hbar\Delta\sum_{\omega_u\omega_v\omega_w} \frac{E_{\omega_u\t{in}}\bar{E}_{\omega_{w}\t{in}}E_{\omega_v\t{in}}}{(\hbar\omega_s-\hbar\tilde{\omega}_0)(\hbar\omega_w-\hbar\tilde{\omega}_0^*)(\hbar\omega_u+\hbar\omega_v-\hbar\tilde{\omega}_{20})(\hbar\omega_u-\hbar\tilde{\omega}_0)}\delta_{\omega_s,\omega_u+\omega_v-\omega_w}.
\end{align}
In terms of the bare molecule Green functions and phonon-phonon scattering amplitudes, the bare third-order molecular nonlinear polarization $P^{(3)}_0(\omega_s) = \mu \sum_{i=1}^N \braket{a_i(\omega_s)}^{(3)}$ can be written as 
\begin{align}
\braket{P}^{(3)}_0(\omega_s)= \sum_{\omega_u\omega_v\omega_w} 2\hbar\Delta N \mu^4  G_{mm}^{(0)}(\omega_s) \bar{G}_{mm}^{(0)}(\omega_w) \Gamma_{mm,mm}^{(0)}(\omega_u+\omega_v) G_{mm}(\omega_v)G_{mm}(\omega_u) E_{\omega_u\t{in}}\bar{E}_{\omega_{w}\t{in}}E_{\omega_v\t{in}}\delta_{\omega_s,\omega_u+\omega_v-\omega_w},
\end{align}
which implies the bare nonlinear susceptibility
\begin{align}
\chi_0^{(3)}(-\omega_s; \omega_v, -\omega_w, \omega_u)=   2\hbar\Delta N\mu^4 G_{mm}^{(0)}(\omega_s) \bar{G}_{mm}^{(0)}(\omega_w) \Gamma_{mm,mm}^{(0)}(\omega_u+\omega_v) G_{mm}^{(0)}(\omega_v) G_{mm}^{(0)}(\omega_u)\delta_{\omega_s,\omega_v-\omega_w+\omega_u}. \label{eq:chi3_0} 
\end{align}

\section{Nonlinear absorption spectrum of bare molecular system}
The steady-state rate of photon absorption by the molecular system in free space can be computed from the Hamiltonian in Eq. \ref{eq:bareham}. In particular, the steady-state condition stipulates that in the presence of an external radiation field, the rate of excitation of the molecular system is equal to its rate of decay, and therefore $\partial_t \sum_{i=1}^N\braket{a_i^\dagger(t) a_i(t)} = 0$, where $t$ is an arbitrary time during which the system satisfies the condition given above.

 Using Heisenberg-Langevin equations of motion for the description of the response of the molecular system to the external electromagnetic field we find that:
\begin{align}
\gamma_m \sum_{i=1}^N \braket{a_i^\dagger(t) a_i(t)} = \sum_{i=1}^N\sum_{\omega} \t{Im}
\left[\frac{2\mu \bar{E}_{0\omega}}{\hbar}\braket{b_\omega^\dagger(t) a_i(t)}\right].
\end{align}
The l.h.s of the above equality corresponds to energy extracted from (or transferred to) the molecular system by the bath, whereas the r.h.s describes the pumping of the molecular system by the electromagnetic field. Assuming the usual weak coupling condition to be valid in free space, and taking the external field to be given by a macroscopic coherent state with negligible quantum fluctuations, it follows that the bare rate of photon absorption is given by:
\begin{align}
W_0 = \frac{2}{\hbar}\sum_\omega \t{Im}
\left[\bar{E}_{\omega\t{in}}(t)\braket{P(t)}_0\right],\end{align}
where $\braket{P(t)}_0$ refers to the free space (weakly coupled to the EM field) molecular polarization, i.e., $\braket{P(t)}_0 = \braket{\sum_{i=1}^N \mu a_i(t)}_0$. Thus, the nonlinear contribution to the molecular absorption spectrum is given by:
\begin{align}
W_0^{\t{NL}} = \frac{2}{\hbar}\t{Im}
\left[\bar{E}_{\t{in}}(t_{\t{ss}}) \braket{P(t_{\t{ss}})}_0^{(3)}\right],\end{align}
where $t_{\t{ss}}$ is sufficiently long that the system is in steady-state. Equivalently, we can write
\begin{align}
W_0^{\t{NL}}(\omega) & = \frac{2}{\hbar}\t{Im}
\sum_{\omega_s}\bar{E}_{\omega_s\t{in}}\braket{P(\omega_s)}_0^{(3)} \nonum
& = \frac{2}{\hbar}\sum_{\omega_s}\sum_{\omega_u\omega_v\omega_w}\t{Im}
\left[\chi_0^{(3)}(-\omega_s; \omega_v,-\omega_w,\omega_u)\bar{E}_{\omega_s\t{in}} E_{\omega_u\t{in}} \bar{E}_{\omega_w\t{in}} E_{\omega_v\t{in}}\right] \delta_{\omega_s,\omega_v-\omega_w+\omega_u} 
\end{align}

The nonlinear absorption spectrum for photons with frequency $\omega$ is given by:
\begin{align}
W_0^{\t{NL}}(\omega) & = \frac{2|E_{\omega\t{in}}|^4}{\hbar}\t{Im}
\left[\chi_0^{(3)}(-\omega;\omega,-\omega,\omega)
\right]. \label{eq:w0nl}
\end{align}
Using Eq. \ref{eq:chi3_0}, we find the nonlinear rate of absorption of photons by the molecular system is given by:
\begin{align}
W_0^{\t{NL}}(\omega) & =N \frac{2}{\hbar} \frac{4\hbar\Delta \mu^4}{\left[(\hbar\omega-\hbar\omega_0)^2+\hbar^2\gamma_m^2/4\right]^2}\frac{\hbar\gamma_m (\hbar\omega_0-\hbar\omega)}{(2\hbar\omega-\hbar\omega_{20})^2+\hbar^2\gamma_m^2}|E_{\omega\t{in}}|^4 \nonum
& +N \frac{2}{\hbar} \frac{4\hbar\Delta \mu^4}{\left[(\hbar\omega-\hbar\omega_0)^2+\hbar^2\gamma_m^2/4\right]^2} \frac{\hbar\gamma_m (\hbar\omega_{20}/2-\hbar\omega)}{(2\hbar\omega-\hbar\omega_{20})^2+\hbar^2\gamma_m^2}
|E_{\omega\t{in}}|^4.
\end{align}
Each of the two terms in the above rate of nonlinear absorption correspond to a distinct nonlinear absorption resonance. This can be seen by noting that the first term vanishes when $\omega = \omega_0$, whereas the second vanishes when $2\omega$  is resonant with the two-photon transition with frequency $\omega_{20} = 2 \omega_0-2\Delta$. When $\Delta/\gamma_m \gg 1$, the lineshapes corresponding to the two possible nonlinear absorption resonances are well separated, and we can isolate the contribution to $W_0^{\t{NL}}(\omega)$ corresponding to two-photon absorption:

\begin{align}
W_0^{\t{TPA}}(\omega) \equiv N \frac{2}{\hbar} \frac{4\hbar\Delta \mu^4}{\left[(\hbar\omega-\hbar\omega_0)^2+\hbar^2\gamma_m^2/4\right]^2}\frac{\hbar\gamma_m (\hbar\omega_0-\hbar\omega)}{(2\hbar\omega-\hbar\omega_{20})^2+\hbar^2\gamma_m^2}|E_{\omega\t{in}}|^4.
\end{align}
The textbook expression for the two-photon absorption rate \cite{boyd2008nonlinear} follows from the last result by taking the limit where $\Delta \gg \gamma$, and by assuming only probe frequencies $\omega$ around the TPA resonance at $\omega_0 -\Delta$ (so that no other quantum transitions interference with the absorption)  In this case, it follows that\footnote{Specifically, letting $\omega = \omega_0 -\Delta -\epsilon$ with $\epsilon \rightarrow 0$ and $\gamma_m/\Delta \rightarrow 0$, we have
\begin{align}
\frac{\hbar \Delta (\hbar\omega_0-\hbar\omega)}{\left[(\hbar\omega-\hbar\omega_0)^2+\hbar^2\gamma_m^2/4\right]^2} & \approx \frac{\hbar^2\Delta^2(1+\epsilon/\Delta)}{\left[(\hbar\omega-\hbar\omega_0)^2\right]^2} \nonum
& \approx \frac{\hbar^2\Delta^2}{(\hbar\omega-\hbar\omega_0)^2}\frac{(1+\epsilon/\Delta)}{\hbar^2\Delta^2(1+\epsilon/\Delta)^2}\nonum
& = \frac{1}{(\hbar\omega-\hbar\omega_0)^2}\left[1+O(\epsilon/\Delta)\right].
\end{align}
}:
 \begin{align}
W_0^{\t{TPA}}(\omega) \approx\frac{2\pi N}{\hbar}\frac{2\mu^4}{(\hbar\omega-\hbar\omega_0)^2}\rho_2(2\omega) |E_{\omega\t{in}}|^4,
\end{align}
where $\rho_2(2\omega) = -\frac{1}{\pi} \t{Im}\left[G_{mm,mm}^{(2)}(2\omega)\right]$: \begin{align} \rho_2(2\hbar\omega) = -\frac{1}{\pi}\t{Im}\left[\frac{2}{2\hbar\omega-\hbar\omega_{20}+i\hbar\gamma}\right] =\frac{1}{\pi} \frac{2\hbar\gamma}{(2\hbar\omega-\hbar\omega_{20})^2+\hbar^2\gamma_m^2}. \end{align}
\section{Quantitative comparison of molecular nonlinear susceptibility of bare and strongly coupled systems}
\begin{figure}[t]
\includegraphics[width=\columnwidth]{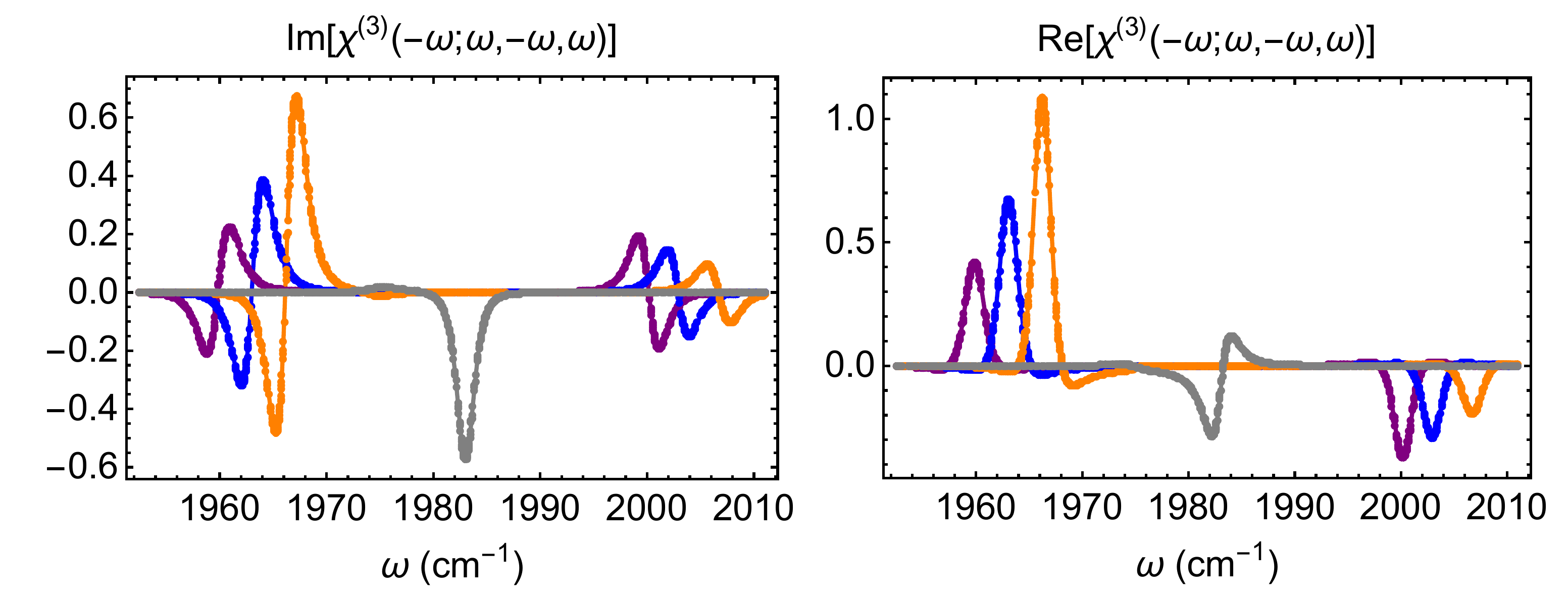}
\caption{Left (Right): Imaginary (Real) parts of $\chi^{(3)}(-\omega;\omega,-\omega,\omega)$ and $\chi_0^{(3)}(-\omega;\omega,-\omega,\omega)$ for a system with $\omega_0 = 1983~\t{cm}^{-1}, \gamma_m = 3~\t{cm}^{-1},  \kappa = 6 ~\t{cm}^{-1}, \Omega_R = 40~\t{cm}^{-1},$ and $\Delta = 8~\t{cm}^{-1}$. The grey curve corresponds to results obtained for the bare molecular system. Purple (orange) corresponds to $\omega_c - \omega_ 0 = 7 ~\t{cm}^{-1}$ $(\omega_c - \omega_ 0 = -7 ~\t{cm}^{-1})$, and blue describes the results obtained for $\omega_c = \omega_0$.}\label{fig:chi3ri}
\end{figure}

In this section, we provide a quantitative discussion of the molecular nonlinear susceptibility under strong coupling with a cavity. In Fig. \ref{fig:chi3ri}, we employ a prototypical system of $\t{W(CO)}_6$ molecules in hexane \cite{xiang2018} with $\omega_0 = 1983 ~\t{cm}^{-1}$, $\gamma_m = 3~\t{cm}^{-1}, \Delta =8 ~\t{cm}^{-1}$ to illustrate and compare the real and imaginary parts of the bare nonlinear susceptibility (Eq. \ref{eq:chi3_0}) to that obtained for the same system under strong coupling with an optical cavity (Eq. \ref{eq:chi3_ne}) with $\kappa = 6~ \t{cm}^{-1}$, $\Omega_R = 40~\t{cm}^{-1}$, and the following cavity frequencies: $\omega_c = 1977~\t{cm}^{-1}, 1983~\t{cm}^{-1}, 1990~\t{cm}^{-1}$ \cite{ribeiro2018c, xiang2019a,xiang2019b}. For the sake of simplicity, we show results for a monochromatic input field with frequency $\omega$ (thus, $\omega_u = \omega_w = \omega_w = \omega)$.

Fig. \ref{fig:chi3ri} shows that the bare and strongly coupled molecular system display strikingly contrasting nonlinear polarization. The imaginary part of the bare nonlinear susceptibility shows absorptive lineshapes, whereas dispersive behavior can be observed for the polaritonic. The opposite is true for the corresponding real parts. The absorptive lineshapes for $\t{Im} \left[\chi_0^{(3)}(-\omega; \omega,-\omega,\omega)\right]$ centered at $\omega_0$ and $\omega_0 - \Delta$ (see small bump of grey curve around $\omega = 1975~\t{cm}^{-1}$) are expected since this function is directly proportional to the nonlinear absorption rate (Eq. \ref{eq:w0nl}) by the bare molecules. The weak resonance at $\omega_0-\Delta$ corresponds to two-photon absorption by the molecular subsystem which absorb two input photons with $\omega = \omega_0 - \Delta$ to generate a population of molecules with energy $\hbar\omega = 2\hbar\omega_0 -2\hbar\Delta$ in the doubly-excited state, whereas the resonance at $\omega_0$ results from stimulated emission by excited-state population and ground-state bleach which contribute to the reduced nonlinear photon absorption probability at the fundamental frequency $\omega_0$ (thus giving rise to the observed negative amplitude).

 It is harder to interpret $\t{Im}\left[\chi^{(3)}(-\omega; \omega,-\omega,\omega)\right]$. As discussed in Sec. \ref{sec:nlabs_strong}, by virtue of the cavity-matter strong coupling, the  nonlinear polarization contribution to the energy absorbed by the molecular subsystem is \textit{not} directly proportional to the imaginary part of  $\chi^{(3)}(-\omega;\omega,-\omega,\omega)$. Nevertheless, the most obvious features of the molecular nonlinear susceptibility under strong coupling are visible from Fig. \ref{fig:chi3ri}. For instance, the absorptive lineshapes displayed by $\t{Re}\left[\chi^{(3)}(-\omega;\omega,-\omega,\omega)\right]$ are all centered at the LP and the UP frequencies for each of the studied systems. Stronger nonlinear polarization always happens at $\omega = \omega_{\t{LP}}$ in comparison to $\omega=\omega_{\t{UP}}$. This happens because, while for $N \gg 1$, the nonlinear response mediated by LP and UP arises mainly from their interaction with molecular doubly excited-states (see Sec. \ref{sec:opt_spec}), larger spectral overlap exists between the molecular two-photon transition and the $\t{LP}_2$ resonance (for the parameters here chosen). As a result, energy or amplitude transfer between polaritons and molecular doubly excited-states is more efficient when the LP is resonantly driven by the external field (see detailed discussion and connection to experiments \cite{xiang2019a} in Sections 3 and 4 of the main text). 

Note also that, for the parameters chosen to obtain Fig. \ref{fig:chi3ri}, the maxima of the nonlinear susceptibility obtained for the molecular system inside and outside of an optical cavity are of the same order of magnitude. However, we expect that if $\Omega_R$ is modified so that two-polariton states (LP$_2$ in this example) become nearly-resonant with molecular doubly excited-states, the molecular nonlinear susceptibility under strong coupling will likely undergo significant enhancement, since in this case, spectral overlap between LP$_2$ and molecular doubly excited-states will be large, and the latter will provide an efficient sink for energy disposal by the former (this is not the case for any of the scenarios shown in Fig. \ref{fig:chi3ri}).

\begin{figure}
\includegraphics[width=\columnwidth]{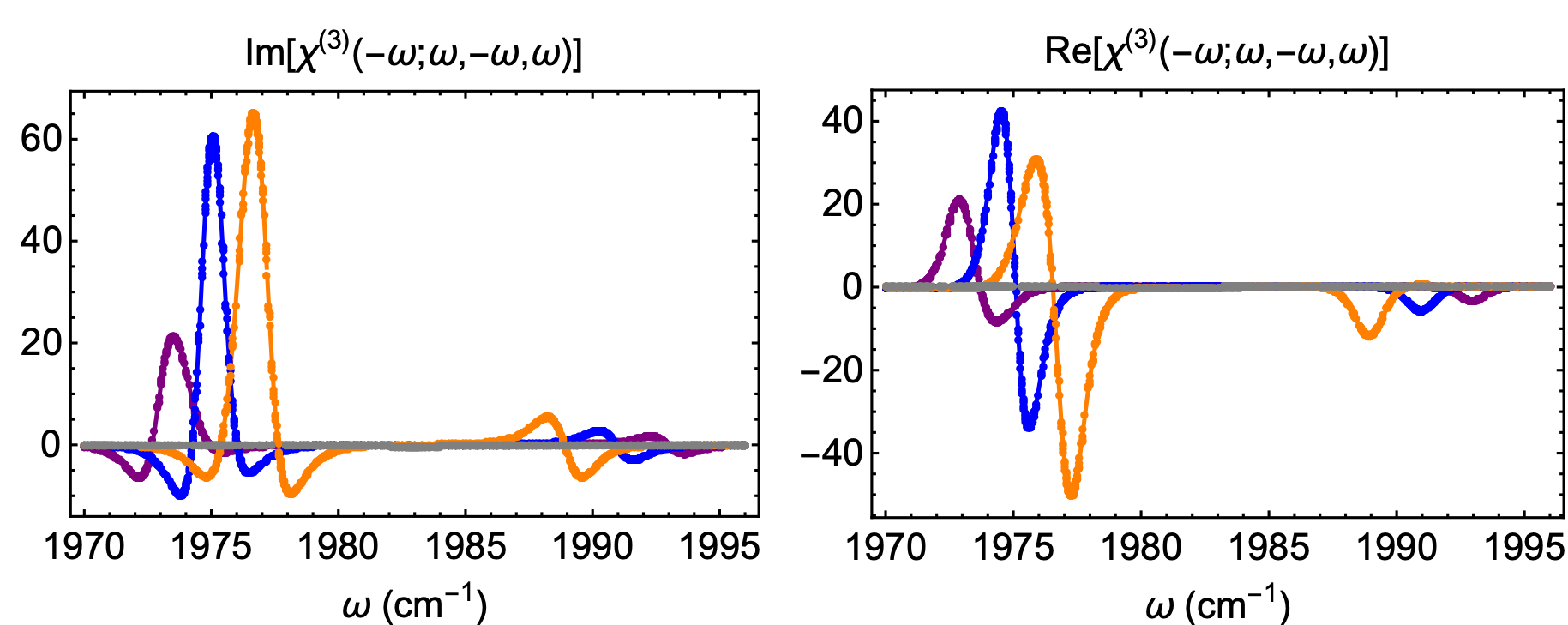}
\caption{Left (Right): Imaginary (Real) parts of $\chi^{(3)}(-\omega;\omega,-\omega,\omega)$ and $\chi_0(-\omega;\omega,-\omega,\omega)$ for a system with equal cavity and molecular fundamental frequencies and decay rates and varying Rabi splitting. The barely visible grey curve corresponds to results obtained for the bare molecular system,  whereas the purple, blue, and orange correspond to $\Omega_R = 20, 16,~\t{and}~12~\t{cm}^{-1}$.}\label{fig:chi3or}
\end{figure}

We conclude this section by presenting in Fig. \ref{fig:chi3or} the behavior of the strongly coupled molecular nonlinear susceptibility for $\Omega_R = 20, 16, ~\t{and}~12~\t{cm}^{-1}$ for a system with zero real and imaginary detuning ($\omega_c = \omega_0 = 1983~\t{cm}^{-1}$  and $\kappa = \gamma = 3~\t{cm}^{-1}$, respectively) and $\Delta = 8~\t{cm}^{-1}$. Our expectation of an enhanced molecular nonlinear susceptibility under strong coupling with a moderate quality cavity is now verified. Figure \ref{fig:chi3or} shows that as $\Omega_R - 2\Delta \rightarrow 0$, the nonlinear polarization of the molecular subsystem becomes larger, especially when the two-LP frequency $2\omega_0 -\Omega_R$ approaches the TPA resonance at  $ 2\omega = 2\omega_0 -2\Delta$. We can observe enhancement of both real and imaginary parts of $\chi^{(3)}$ relative to $\chi^{(3)}_0$ by two orders of magnitude at $\omega = \omega_0 - 2\Delta$ when the condition $\Omega_R = 2\Delta$ is satisfied. Note that $\t{Im}\left[\chi^{(3)}(-\omega; \omega,-\omega,\omega)\right]$ has absorptive lineshapes at the TPA transition. This feature suggests that the enhanced signal at $\omega = \omega_{\t{LP}}$ is due to two-LP decay into molecular doubly excited-states. This channel is discussed in detail in Secs. 3 and 4 of the main manuscript.
 
 \section{Energy eigenvalues and eigenstates of non-dissipative system}\label{sec:opt_spec}
 In this section, we obtain the optical spectrum of the hybrid system discussed in the main manuscript. For this purpose, we neglect the effects of dissipation, so that the obtained transition frequencies are real. In fact, the Hamiltonian of the hybrid system can be written in this case as:
\begin{align}
 	H = & \hbar \omega_{c}b^\dagger b + \sum_{i=1}^{N} \hbar\omega_0 a_i^\dagger a_i -\hbar \Delta \sum_{i=1}^N a_i^\dagger a_i^\dagger a_i a_i - \sum_{i=1}^N \hbar g \left(a_i^\dagger b+b^\dagger a_i\right), \end{align}
where $g$ is the single-molecule light-matter coupling constant. two conservation laws follow from the effective Hamiltonian given in the main text. First, the Hamiltonian is invariant under permutation of the molecules. Thus, the eigenstates of $H$ can be classified according to the irreducible representation fo the permutation group of $N$ symbols ($S_N$), and time-dependent evolution only allows transitions between states which belong to the same irrep.  Second, it follows from the RWA approximation to the light-matter interaction that the Hamiltonian evolution of the composite system preserves the \textit{total} number of excitations of the photonic and matter subsystems $M = \sum_{i=1}^N a_i^\dagger a_i+ b^{\dagger} b$. Therefore, the eigenstates of $H$ may also be classified according to the total number of excitations in the molecular and photonic subsystems. For instance, the ground-state of the system ($M = 0$) has all molecules in the ground-state, while the cavity field is in its vacuum state. The states with $M=1$ contain either a single excited vibration ($\ket{1_i}$ where $1\leq i \leq N$), or a single-photon ($\ket{1_0}$), etc.

Of the many irreps of $S_N$, only the totally-symmetric is relevant in our case. In the manifold of states with $M=1$, this feature is well-known: only the totally-symmetric superposition of states with a single excited molecule exchanges energy with the cavity field. The non-totally-symmetric states are dark and thus provide no contribution to the optical response of the hybrid system (in the studied ideal model). 

The lower and upper polariton states are denoted by $\ket{\t{LP}}$ and $\ket{\t{UP}}$. They can be written in terms of the local-mode basis states as follows: 
\begin{align} 
&\ket{\t{LP}} = -\t{sin}(\theta/2) \ket{1_0} +  \t{cos}(\theta/2)\ket{1_S}, \\
&\ket{\t{UP}} = \t{cos}(\theta/2) \ket{1_0} + \t{sin}(\theta/2)\ket{1_S},
\end{align}
where $2\theta = \t{tan}^{-1} \left[2g \sqrt{N}/(\omega_c-\omega_0)\right]$ and $\ket{1_S} = N^{-1/2} \sum_{i=1}^N \ket{1_i}$ is the molecular singly-excited bright state, and we denote by $g$ the single-molecule light-matter coupling.

The bright subspace of the doubly-excited state ($M=2$) manifold contains the four two-particle (hybrid) states which are totally-symmetric under permutation of the molecular labels. These states are the only which can be accessed via two-photon transitions in our model (dark modes are never accessed since they require molecular permutational symmetry-breaking operators which are disregarded in our treatment). They are given by:
\begin{align}
&\ket{2_0}, ~~\ket{1_01_m} = \frac{1}{\sqrt{N}} \sum_{a=1}^N \ket{1_0 1_a},  ~~ \ket{1_{m}1_{m'}}_{m\neq m'} = \sqrt{\frac{2}{N(N-1)}}\sum_{a>b} \ket{1_a1_b}, ~~\ket{2_m} = \frac{1}{\sqrt{N}} \sum_{a=1}^N \ket{2_a}.\label{eq:bs2}
\end{align}
Figure S1 illustrates how Hamiltonian evolution induces transitions between these states. From this figure, we can also see that these four states are the only which can be accessed from a two-photon initial state. In the subspace spanned by the priorly defined states, the total Hamiltonian is given by:
\begin{align}
H_2^{\t{B}}(N) = \begin{pmatrix}
2\hbar\omega_c & \hbar g\sqrt{2N} & 0 & 0 \\
\hbar g\sqrt{2N} & \hbar\omega_0+\hbar \omega_c & \hbar g\sqrt{2(N-1)}& \hbar g\sqrt{2} \\
0 & \hbar g \sqrt{2(N-1)} &  2\hbar \omega_0 & 0 \\
0 & \hbar g\sqrt{2} & 0 & 2\hbar\omega_{0}-2\hbar\Delta
\end{pmatrix},
\end{align}
where the matrix was ordered in the same way as the basis states in Eq. \ref{eq:bs2}. From now on, we will focus on the case where $\omega_c \approx \omega_0$ since this gives the simplest analytical results, and is also the most relevant.

 If the molecular oscillators were two-level systems, we would obtain the restriction of the Tavis-Cummings Hamiltonian to the $M=2$ Hamiltonian, which is given by: \begin{align}
H_{2\t{TC}}^{\t{B}}(N) = \begin{pmatrix}
2\hbar\omega_c & \hbar g\sqrt{2N} & 0 \\
\hbar g\sqrt{2N} & \hbar\omega_0+\hbar\omega_c & \hbar g\sqrt{2(N-1)}\\
0 & \hbar g\sqrt{2(N-1)} &  2\hbar\omega_0
\end{pmatrix}.
\end{align}
When $\omega_0 = \omega_c$, the TC eigenstates can be readily obtained since the secular equation can be written in the simple form:
\begin{align}
& (2\omega_0-\lambda) \left[\left(2\omega_0-\lambda\right)^2-2(N-1)g^2\right] - 2g^2N(2\omega_0-\lambda) = 0,
\end{align}
which has solutions:
\begin{align}
& \lambda_{\t{UP}_2}^{\t{TC}} = 2\omega_0 +2g \sqrt{N-1/2} \approx 2\omega_{\t{UP}} -\frac{g}{2\sqrt{N}}, \\
& \lambda_{\t{LP}_2}^{\t{TC}} = 2\omega_0 - 2g \sqrt{N-1/2} \approx 2\omega_{\t{LP}} +\frac{g}{2\sqrt{N}},\\
& \lambda_{\t{LU}}^{\t{TC}} = 2\omega_0,
\end{align}
where the approximate expressions resulted from taking the limit where $N \rightarrow \infty$. In terms of the bare states $\ket{2_0}$, $\ket{1_01_m}$ and $\ket{1_m 1_{m'}}$, the eigenstates corresponding to the above energies are given by:
\begin{align}
& \ket{\t{UP}_2^{\t{TC}}} = \sqrt{\frac{N}{4N-2}}\ket{2_0}+ \sqrt{\frac{1}{2}}\ket{1_0 1_m} +
\sqrt{\frac{N-1}{4N-2}}\ket{1_m 1_{m'}} \\
&\ket{\t{LP}_2^{\t{TC}}} = \sqrt{\frac{N}{4N-2}}\ket{2_0}- \sqrt{\frac{1}{2}}\ket{1_0 1_m} +
\sqrt{\frac{N-1}{4N-2}}\ket{1_m 1_{m'}}   \\
& \ket{{\t{LU}}^{\t{TC}}} = \sqrt{\frac{N-1}{2N-1}}\ket{2_0} - \sqrt{\frac{N}{2N-1}}\ket{1_m1_{m'}}.
\end{align}
Using these states along with the $\ket{2_i}$ as the new basis vectors for the totally-symmetric doubly-excited manifold of the system allowing two excitations, the Hamiltonian matrix (with the row and column indices in the order $\ket{\t{LU}^{\t{TC}}}, \ket{\t{UP}_2^{\t{TC}}}, \ket{\t{LP}_2^{\t{TC}}}, \ket{2_i}$), acquires the simple form:
\begin{align}
H_2^{\t{B}}(N) = \begin{pmatrix}
2\hbar\omega_0& 0 & 0 & 0 \\
0 &  \hbar\lambda_{\t{UP}_2}^{\t{TC}} & 0 &\hbar g \\
0 &  0 & \hbar\lambda_{\t{LP}_2}^{\t{TC}}& -\hbar g\\
0 & \hbar g  & -\hbar g  & 2\hbar\omega_{0}-2\hbar\Delta
\end{pmatrix}, ~~\hbar\omega_c = \hbar\omega_0.
\end{align}
From this, we can see that the state $\ket{\t{LU}^{\t{TC}}}$ is also an eigenstate of the complete Hamiltonian, and that despite its delocalization, the totally-symmetric doubly-excited molecular state $\ket{2_S}$ is only weakly-coupled to polaritons (the corresponding coupling constant is given by the single-molecule light-matter interaction energy $g$). If we take the single-molecule light-matter coupling to be very weak compared to the energy differences $\lambda_{\t{UP}_2}^{\t{TC}} - (2\omega_0 -2\Delta) = g\sqrt{4N-2} +2\Delta$ and $\lambda_{\t{LP}_2}^{\t{TC}} - (2\omega_0 -2\Delta) = -g\sqrt{4N-2} +2\Delta$, we can obtain reasonable approximate eigenstates and eigenvalues of $H_2^{\t{B}}(N)$. This will almost always be a valid assumption, even if $2\Delta$ is nearly equal to $g\sqrt{4N-2}$, since the single-molecule-light coupling constant $g$ is generally too small compared to the energy scale of vibrational motion, and there exists a large number of (non-totally symmetric) molecular doubly excited-states with energy $2\omega_0-2\Delta$ that provides an efficient decay channel for LP$_2$ states. In other words, the TC eigenstates will almost always be very good approximations to the eigenstates of  $H_2^{\t{B}}(N)$. The leading-order perturbatively-corrected eigenvalues are given by:
\begin{align}
&\omega_{\t{UP}_2} \approx 2\omega_0 +g\sqrt{4N-2}  +\frac{1}{2}\frac{2g^2}{g\sqrt{4N-2}+2\Delta}\approx 2\omega_{\t{UP}}-\frac{g}{2\sqrt{N}}+\frac{1}{2}\frac{2g^2}{g\sqrt{4N-2}+2\Delta}, \\
& \omega_{\t{LP}_2} \approx 2\omega_0 -g\sqrt{4N-2}  -\frac{1}{2}\frac{2g^2}{g\sqrt{4N-2}-2\Delta} \approx 2\omega_{\t{LP}}+\frac{g}{2\sqrt{N}} -\frac{1}{2}\frac{2g^2}{g\sqrt{4N-2}-2\Delta}, \\
&\omega_{2_S} \approx 2\omega_0 -2\Delta +\Delta \frac{g^2}{g^2(N-1/2)-\Delta^2}, \\
& \omega_{\t{LU}} =  2\omega_0,
\end{align}
where we included the exact eigenvalue of the $\ket{\t{LU}}$ state for completeness. The corresponding approximate eigenstates can be written as:\begin{align}
& \ket{\t{UP}_2}  \approx \sqrt{\frac{N}{4N-2}}\ket{2_0}+\sqrt{\frac{1}{2}}\ket{1_0 1_m} +
\sqrt{\frac{N-1}{4N-2}}\ket{1_m 1_{m'}} + \frac{g}{g\sqrt{4N-2}+2\Delta} \ket{2_m}, \\
& \ket{\t{LP}_2}  \approx \sqrt{\frac{N}{4N-2}}\ket{2_0}- \sqrt{\frac{1}{2}}\ket{1_0 1_m} +
\sqrt{\frac{N-1}{4N-2}}\ket{1_m 1_{m'}} + \frac{g}{g\sqrt{4N-2}-2\Delta} \ket{2_m}, \\
& \ket{2_S} \approx  \ket{2_m}  - \frac{g^2}{g^2(2N-1)-2\Delta^2}\left(\sqrt{N}\ket{2_0} +\sqrt{N-1}\ket{1_m1_{m'}}\right)+\frac{\sqrt{2}g\Delta}{g^2(2N-1)-2\Delta^2}\ket{1_01_m},\\
& \ket{\t{LU}} = \sqrt{\frac{N-1}{2N-1}}\ket{2_0} - \sqrt{\frac{N}{2N-1}}\ket{1_m1_{m'}}.
\end{align}

\begin{figure}	
\begin{tikzpicture}[scale=.9, transform shape]
\tikzstyle{bl} = [circle, fill=gray!30]
\tikzstyle{arrow} = [draw, <->]
\draw ++(0,0) node[bl] (a) {$\ket{2_0}$};
\draw ++(3.5,0) node[bl] (b){$\ket{1_0 1_m}$};
\draw ++(7.0,0) node[bl] (c) {$\ket{1_m 1_{m'}}$};
\draw++(3.5,-2.0) node[bl] (d) {$\ket{2_m}$};
\path[arrow] (a) -- node[above] {$g\sqrt{2N}$} (b) ;
\path[arrow] (b) --  node[above] {$g\sqrt{2(N-1)}$} (c);
\path[arrow] (b) --  node[left] {$g\sqrt{2}$}  (d);
\end{tikzpicture}
\caption{Scheme representing the bright (totally-symmetric matter and cavity states) two-particles states which play a role in the nonlinear spectroscopy of vibrational polaritons discussed here. Above each arrow connecting a pair of states we provide the corresponding Hamiltonian matrix elements (coupling constants).}
\end{figure}
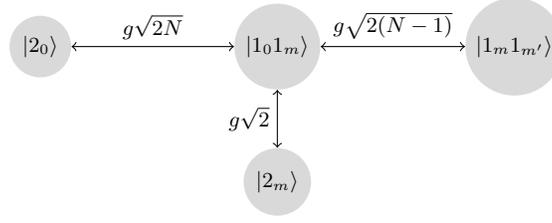

 \bibliography{lib}